\documentclass[useAMS,usenatbib,onecolumn]{mn2e}
\usepackage{graphicx}
\newcommand{\oversim}[2]{\protect{\mbox{\lower0.5ex\vbox{%
  \baselineskip=0pt\lineskip=0.2ex
  \ialign{$\mathsurround=0pt #1\hfil##\hfil$\crcr#2\crcr\sim\crcr}}}}}
\newcommand{\simgreat}{\mbox{$\,\mathrel{\mathpalette\oversim>}\,$}} 
\newcommand{\simless} {\mbox{$\,\mathrel{\mathpalette\oversim<}\,$}} 
\title[Influence of multiplicity on the IMF]{The influence of multiple
stars on the high-mass stellar initial mass function and age-dating of young
massive star clusters}
\author[C.~Weidner, P.~Kroupa and
  T.~Maschberger]{C.~Weidner$^{1,2}$\thanks{E-mail:
    Carsten.Weidner@st-andrews.ac.uk}, P.~Kroupa$^{3}$\thanks{E-mail:
    pavel@astro.uni-bonn.de} and
  T.~Maschberger$^{3,4}$\thanks{E-mail: tmasch@astro.uni-bonn.de}\\ 
$^{1}$Departamento de Astronom{\'i}a y Astrof{\'i}sica, Pontificia
      Universidad Cat{\'o}lica de Chile, Av. Vicu{\~n}a MacKenna 4860,\\
      Macul, Santiago, Chile\\
$^{2}$Scottish Universities Physics Alliance (SUPA), School of Pysics and
      Astronomy, University of St. Andrews, North Haugh,\\
      St. Andrews, Fife KY16 9SS, UK\\
$^{3}$Argelander-Institut f\"ur Astronomie (Sternwarte), Auf dem H{\"u}gel 71,
D-53121 Bonn, Germany\\
$^{4}$Institute of Astronomy, Madingley Road, Cambridge CB3 0HA, UK
}

\begin{document}
\bibliographystyle{mn2e}
\date{Accepted 2008 November 16. Received 2008 November 5; in original
  form 2008 September 2}

\pagerange{\pageref{firstpage}--\pageref{lastpage}} \pubyear{2008}

\maketitle

\label{firstpage}

\begin{abstract}
The study of young stellar populations has revealed that most stars
are in binary or higher order multiple systems. In this study the
influence on the stellar initial mass function (IMF) of large
quantities of unresolved multiple massive stars is investigated by
taking into account stellar evolution and 
photometrically determined system masses. The models where initial
masses are derived from the luminosity and colour of unresolved
multiple systems show that even under extreme circumstances (100\% binaries or
higher order multiples) the difference between the power-law index of the mass
function of all stars and the observed mass function is small ($\simless$
0.1). Thus, if the observed IMF has the Salpeter index $\alpha$ = 2.35 then
the true stellar IMF has an index not flatter than $\alpha$ =
2.25. Additionally, unresolved multiple systems may hide between 15 and 60\% of
the underlying true mass of a star cluster. While already a known
result, it is important to point out that the presence of a large
number of unresolved binaries amongst pre-main-sequence (PMS) stars
induces a significant spread in the measured ages of these stars even
if there is none. Also, lower-mass stars in a single-age binary-rich
cluster appear older than the massive stars by about 0.6 Myr.
\end{abstract}


\section{Introduction}
\label{se:intro}
Most late-type stars ($\simless 1 M_{\odot}$) are born in clusters of binary
systems with companion masses distributed randomly from the stellar initial
mass function \citep[IMF;][]{GK05,GKG07,GW07,KBG08}. Subsequent dynamical
encounters in the cluster evolve the stellar population to that observed in
the Galactic field \citep{DM91,Kr95a,Kr95b,Kr95c,Kr95d,KB03}. If massive
($\simgreat {\rm few} M_{\odot}$) stars were also to be born with companions
distributed randomly from the IMF, then the vast majority of O stars would
have M-dwarf companions while massive equal-mass binaries would be extremely
unlikely. For random sampling from a canonical IMF \citep{WK04} for any
O star\footnote{The percentages are valid for any given star but if
  the ``companion'' is more massive than the primary they would switch
  places and the ratios would change. If we constrain ourselves to O
  stars as primaries it does not matter if the companion is more
  massive than the primary because they would be both O stars.} it would
be expected that 0.2\% of the companions are O~stars
(m $\simgreat 16 M_{\odot}$), 2.2\% B~stars ($2.7 M_{\odot} \simless m
\simless 16 M_{\odot}$), 2.1\% A~stars ($1.7 M_{\odot} \simless m \simless 2.7
M_{\odot}$), 3.7\% F~stars ($1.1 M_{\odot} \simless m \simless 1.7
M_{\odot}$), 3.4\% G~stars ($0.85 M_{\odot} \simless m \simless 1.1
M_{\odot}$), 12.0\% K~stars ($0.5 M_{\odot} \simless m \simless 0.85
M_{\odot}$) and 76.3\% M~stars (0.08 $M_{\odot} \simless m \simless 0.5
M_{\odot}$). These M-dwarfs (as well as the K and at least the G stars,
altogether more than 90\% of all companions) would be more or less
invisible. However, the observed fraction of O stars in {\it massive} multiple
systems lies between at least 20 and 80\%
\citep{GCM80,GM01,DME06,KKK06,Lu06,ABK07,SGN07,KBP07,TBR08} and thus strongly
contradicts secondaries which are randomly taken from the IMF, at least for
massive primaries. Unfortunately, the mass ratio ($q=\frac{m_{\rm
secondary}}{m_{\rm primary}}$) distribution\footnote{The $q$-distribution
for random pairing is roughly flat, showing little preference for any
particular $q$-value \citep{Kr95d,GL06}.} for massive stars is observationally
not well explored for several reasons: Massive stars have very broad spectral
lines, making radial velocity studies a challenge. Even rather massive and
close companions are easy to miss. And young stellar nurseries are usually
deeply embedded in gas and dust. Furthermore, massive stars are predominately
found in massive clusters. So the issue of crowding and fast dynamical
evolution arises and therefore it is not clear if the $q$-distribution is
primordial. As massive stars either sink to the centre of star clusters
through energy equipartition \citep{Ge05,SBB06,FBL06,VMP06,ZCW06} or are
(exclusively?) formed there \citep{BBZ98,BBV03,BVB04}, dynamical processes in
the cluster centre can expel (massive) stars from the cluster and/or exchange
binary partners, thereby changing the $q$-distribution
\citep{PAK06,PO07}. This picture is strongly supported by the fact that
``field'' and runaway O stars have a far lower binary fraction than O stars in
clusters and associations \citep{TBR08,MHG08}. A recent study of triple and
quadruple system by \citet{To08} might also be used as an argument
against random pairing of stars as the author finds a surprisingly
large amount of these systems in which all stars are of similar
mass. Whether or not these results are applicable to this study is a
matter of debate. The \citet{To08} sample consists to about 80\% of
primary stars below 5 M$_{\odot}$ and it is not restricted to young
objects. If the author's claim is correct that properties like orbital
periods and eccentricities of old multiple systems are not consistent with
$N$-body calculations his study would be an important argument against
random pairing of multiple stars. It should be noted here that in a
cluster of the richness of the Orion Nebula cluster stellar dynamics may change
the mass function and binary properties in the core significantly in less than
$10^{6}$ yr \citep{PAK06,PO07}. The study of the $q$-distribution in the
present paper is of stars at the stage when they are observed and may thus be
representative of dynamically evolved distributions.\\

The influence of a substantial fraction of unresolved binary systems on
the low-mass stellar mass function (MF) has been well explored
\citep{SaRi91,KGT91,KTG93,Kr95a,Kr95b,MaZi01,TK07,TK08}. Beginning from the
lowest mass stars and brown dwarfs, \citet{TK07,TK08} corrected for
un-resolved companions in such systems and showed that the IMF is discontinuous
near 0.1 $M_{\odot}$. In an extensive study of late-type stars,
\citet{KGT91,KTG93} showed that the mass function of these stars is
significantly affected by unresolved systems. They used this result to
formulate the standard or canonical-IMF with little or no evidence for
systematic variations with star forming conditions \citep[][for more details
on the canonical IMF: see Appendix~\ref{app:IMF}]{Kr01}. Interestingly, there
seems now to be strong evidence for a universal companion mass function for
solar type stars \citep{MH08}. But only a very limited number of studies have
addressed how the large proportion of binaries affects IMF derivations for
massive stars. \citet{SaRi91} studied the MFs of five young star clusters in
the Large Magellanic Cloud (LMC) including the influence of optical and
physical binaries on the derived slope of the MFs in the range of 2--14
$M_{\odot}$ by Monte Carlo simulations of cluster and field populations with
random pairing over this mass interval. They found a very strong steepening of
the observed MF in the case of a binary fraction larger than
50\%. \citet{VB82} investigated the IMFs of primaries and single stars
analytically up to stars above 100 $M_{\odot}$ but did not explicitly discuss
any possible effects of unresolved binaries on the observed IMF. Only rather 
recently, the study by \citet{MA08} explores the biases on the IMF introduced
by unresolved binaries for massive stars. There a different colour is used
($U-V$) while in this contribution we resort to $B-V$, and only random pairing
is studied by \citet[][see \S~\ref{sub:pair} for details on binary
pairing]{MA08} while here we consider different viable $q$-distributions. This
contribution can therefore be seen as an extension of these previous studies
\citep{KGT91,KTG93,SaRi91,MaZi01,TK07,TK08} to higher masses as well as a
comparison to the \citet{MA08} study.

\section{The Model}
\label{se:model}
In what follows, $dN=\xi(m)\,dm$ is the number of stars in the mass
interval $m$ to $m+dm$.

For all methods the IMF, $\xi(m)$, described in Appendix~\ref{app:IMF} is used
as the input IMF. For every model the percentage of binaries, triples and
quadruples is assigned as an initial condition and then one of the following
pairing algorithms is used to obtain the masses of the stars. The primary,
secondary etc.~is always the most-massive, second massive, etc., companion.

\subsection{Pairing algorithms}
\label{sub:pair}
As the initial properties of (massive) binaries and higher-order
multiple systems are generally unknown, several different methods of
pairing stars were used. 
\begin{itemize}
\item Random pairing (RP): A given number of stars is taken fully randomly
  from the input IMF over the mass range of 0.08 to 150 $M_{\odot}$ and
  randomly assigned to pairs or higher-order multiple systems.
\item Special pairing (SpP): Again a given number of stars is taken
  randomly from the input IMF over the mass range of 0.08 to 150 $M_{\odot}$
  and randomly grouped together. But in the cases where the primary exceeds 2
  $M_{\odot}$ the secondary has to fulfil a minimum mass criterion as
  described in Appendix~\ref{app:secon}. If the secondary fails this test it
  is exchanged with another star from the random list until the minimum
  mass is reached or exceeded. In the case that the new secondary is more
  massive than the primary the two exchange places and the secondary becomes
  the new primary and the old one the new secondary. This new secondary still
  has to fulfil the minimum mass criteria or it is exchanged with another
  star. It is possible that no such secondary is available in the list
  anymore. Then the minimum mass is lowered by 20\% and if this is still
  insufficient, the star is kept as a single. In the numerical experiments
  presented here no such case was encountered. In the case of triple or
  quadruple systems, if the secondary component is also larger than 2
  $M_{\odot}$ the tertiary component is searched for with the same criterion
  as the secondary before (ie the secondary is treated as the ``primary'' in
  the secondary-tertiary sub-system). The same procedure is extended to the
  quartiary companions (ie.~the quartiary becomes the ``secondary'' of the
  tertiary). 
\item System pairing (SyP): A given number of {\it system} masses is randomly
  taken from an input (not the canonical) IMF with 0.16 $\le m_{\rm sys} \le$
  300 $M_{\odot}$ and with a change of slope from 1.3 to 2.35 at 1 $M_{\odot}$
  instead of 0.5 $M_{\odot}$. The lower and upper mass limit for {\it stars}
  is always kept at 0.08 and 150 $M_{\odot}$. If the system mass is below 10
  $M_{\odot}$ the system is randomly split into two, three or four components
  assuming a uniform mass-ratio distribution being equivalent to random
  pairing on this mass interval. But for systems with a mass that
  exceeds 10 $M_{\odot}$ the $q$-values used to split the systems into
  components are biased towards high-$q$-values as described in
  Appendix~\ref{app:fourth}. This model is motivated by a possible 
  merger-origin of massive stars but could also represent the outcome of the
  competitive accretion model \citep{BBZ98}. In order to have a smooth
  transition between the random mass ratios and the preferential
  formation of massive companions for the SyP model the change is
  necessary at 10 $M_{\odot}$ and not at 2 $M_{\odot}$ as in the case
  of the SpP model.
\end{itemize}

All pairing mechanisms preserve a lower mass limit of 0.08 $M_\odot$
for all components, because stars and brown dwarfs rarely pair up
\citep{TK07,TK08}.

After the stars are assigned into their systems, the mass function for all
stars, for the primaries, secondaries, tertiaries (if applicable) and
quartiaries (if applicable) is calculated along with the IMF for the
systems. The method used to measure the slopes is presented in
\S~\ref{sub:slope}. The direct results of these calculations are presented in
\S~\ref{sub:noevo} while the calculations additionally applying stellar
evolution (as described in \S~\ref{sub:evol}) are discussed in
\S~\ref{sub:evo}.

\subsection{Stellar evolution}
\label{sub:evol}
Stellar observations usually do not result directly in stellar masses and ages 
but magnitudes and colours. Therefore, the stars produced by the
pairing mechanisms described in Section~\ref{sub:pair} are assigned a
random age with a Gaussian distribution between 0.01 and 4.5 Myr with a
median of 2 Myr and a dispersion of 0.5 Myr. Stars in the same multiple system
are assumed to be coeval. Using a large grid of stellar
models\footnote{\citet{MM03} models for massive stars, \citet{HPT00} for the
evolution after the pre-main sequence (PMS) for stars below 9 $M_{\odot}$ and
\citet{BHS93}, \citet{BMH97}, \citet{DAM97}, \citet{DM98} and \citet{BM01} for
PMS evolution. For further details on the used models for stellar evolution
see \citet{WKG08}.} this age is then used to evolve the stars and obtain the
evolved masses, effective temperatures and radii. These values are further
used to calculate the Johnson V and B band magnitudes with the help of
\citet{Ku92} and \citet{BWB95} bolometric corrections using a program kindly
provided by J.~Hurley \citep[see][]{Hu03}. The luminosities of the
individual stars  are then combined to a single $V$ magnitude and
$B-V$ colour of the unresolved multiple and searched for in a large
table of $V$ and $B-V$ values made with the same models but for truly
single stars and with an age range between 0.01 and 100 Myr\footnote{A
larger age than the input age of 4.5 Myr is used to additionally show
the effect of unresolved multiples on the age determination of
stars.}. In this manner so-called {\it pseudo masses} and {\it pseudo
ages} for the unresolved multiples are derived as if they were single
stars. Note that the derived {\it pseudo masses} are the {\it
zero-age} masses.

Note that photometric mass and age determination is know to be inaccurate in
comparison to spectroscopic measurements \citep{Mass03}. But as spectra are
far more expensive in terms of observational time the photometric method is
still widely in use. Furthermore, for OB stars a spectroscopic analyses does
not automatically solve the problem of un-resolved multiples as O stars have
very broad lines which easily hide even massive companions. Only with elaborate
multi-epoch spectroscopic observations solid constrains can be put on the
multiplicity of O stars \citep{SGN07,SMS08}.

\subsection{A bias-free fitting method}
\label{sub:slope}

The common method to analyse power-law distributed data is to group
them in bins of constant logarithmic size and then perform a linear
regression to measure the power-law index $\alpha$. However, this method can
lead to strongly biased results \citep{MU05}. Therefore here the method of
\citet{MK08} is used. It is based on the estimates for $\alpha$ obtained with
the maximum likelihood method.

The maximum likelihood estimate of the power-law exponent
\citep[see][]{AMP06} is calculated by minimising the logarithmic
likelihood function of the data $\{X_i\}_{i=1}^{n}$,where $X_i = m_i$ is the
mass of star $i$. The estimated exponent ${\alpha}_{ML} $ is then the solution
of
\begin{eqnarray}
 -\frac{n}{1-{\alpha}_{ML} } + n \frac{Z^{1-{\alpha}_{ML} } \log Z -
  Y^{1-{\alpha}_{ML} } \log Y}{Z^{1-{\alpha}_{ML} } -
  Y^{1-{\alpha}_{ML} } } - T &=& 0, 
\end{eqnarray}
with the smallest data point, $Y= \min X_i$, the largest data point,
$Z = \max X_i$, and the sum of the logarithms of all data,
$T=\sum_{i=1}^n \log X_i$. 
The bias-free estimate follows with
\begin{eqnarray}
{\alpha} &=& \frac{n}{n-2} ( {\alpha}_{ML} - 1 ) + 1 .
\end{eqnarray}

The standard deviation and bias of the method were investigated by
using a Monte-Carlo sample of 1000 synthetic data sets, generated with
a lower limit of 1.3 $M_\odot$, an upper limit of 150 $M_\odot$,
different sizes of the data set and different exponents. 
For each data set the exponent was estimated and from the sample of
estimates the standard deviation and the bias (difference between the
average of the estimated exponents and the input value) were
calculated. 
The results are summarised in Table \ref{table_estimator}.
The bias is in all cases negligible.
The standard deviation decreases with increasing size of the data set
and increases with increasing exponent. 
For the sample sizes used in this work a conservative estimate of the
standard deviation is 0.05. 

\begin{table}
\caption{Standard deviation and bias for the estimation method of the
  exponent, for different sample sizes and output exponents.}
\label{table_estimator}
\begin{tabular}{llll}
\multicolumn{4}{l}{Standard deviation}\\
\# stars       & $\alpha=1.5$ & $\alpha=2.3$ & $\alpha=5.0$ \\
\hline
$10^3$        & 0.026        & 0.043        & 0.127 \\
$10^4$        & 0.009        & 0.014        & 0.042 \\
$10^5$        & 0.003        & 0.004        & 0.012 \\
&&&\\
\multicolumn{4}{l}{Bias}\\
\# stars       & $\alpha=1.5$        & $\alpha=2.3$         &
$\alpha=5.0$ \\
\hline
$10^3$        &  3 $\times\ 10^{-4}$ & 3 $\times\ 10^{-3}$  & 1 $\times\
10^{-2}$ \\ 
$10^4$        &  4 $\times\ 10^{-5}$ & 3 $\times\ 10^{-4}$  & 2 $\times\
10^{-3}$ \\ 
$10^5$        &  1 $\times\ 10^{-5}$ & 3 $\times\ 10^{-5}$  & 1 $\times\
10^{-4}$ \\ 
\end{tabular}
\end{table}

\section{Results}
\label{se:res}
In \S~\ref{sub:noevo} the results of simulating binaries, triples and
quadruples are described under the assumption that all stellar masses
are known perfectly. In \S~\ref{sub:evo} the stars are evolved with
the use of stellar evolution models and the combined magnitudes and
colours are used to model the process of observing stellar populations
containing unresolved multiple systems.

\subsection{Models without stellar evolution}
\label{sub:noevo}
Shown in Tab.~\ref{tab:noevo} are the parameters and results for a
series of models. A number of abbreviations are used in the
Table. ``100\% bin'' marks models made of 100\% binaries, ``100\%
trip'' are 100\% triples and ``100\% quad'' are 100\% quadruples. The
corresponding IMF slopes are: $\alpha_{\rm allstars}$\footnote{All IMF
slopes in this work are obtained by a bias-free fitting method
discussed in \S~\ref{sub:slope} for all stars above 1 $M_{\odot}$. In
the case of strong non-linear features in the MFs the line was fitted to
the stars above such features. The standard deviation in $\alpha$ is of the
order of 0.05 dex.} which accounts for all stars individually (ie.~the true
single-star IMF), $\alpha_{\rm systems}$ which is the mass function
for the combined masses of all multiple systems, $\alpha_{\rm
  primaries}$ which is the IMF of the most massive component of a
system, $\alpha_{\rm secondaries}$ for the second most
massive component of a system, $\alpha_{\rm tertiaries}$ for the third
massive object (only in triples and quadruplets) and finally
$\alpha_{\rm quartiaries}$ for the fourth component in a system if there is
any (only quadruplets).

All models are calculated without stellar evolution, meaning they are
treated as if all masses are known precisely. For each model $10^7$ stars are
generated, each model taking only about a few minutes of computational time on
a standard desktop PC.

As is seen in Tab.~\ref{tab:noevo} and the first three panels of
Fig.~\ref{fig:imf04} the models using random sampling (RP Model 1 to 3)
produce system IMFs which are somewhat steeper than the input IMF, and
very steep IMFs for the companions, while the primary IMFs are slightly
flatter. This can be understood as due to random pairing producing systems in
which most of the massive stars have low-mass companions. These do not change
the mass of the system significantly. But the low-mass systems tend to have
partners of very similar mass, therefore additionally depleting the total
number of low-mass systems and shifting them to higher masses. 

The special pairing mechanism explored here (SpP Model 4 to 6) has a different
effect. The system IMF is also steepened but only for the case of 100\% 
quadruples (SpP Model 6) and the companion star IMFs are
flattened below the Salpeter-value for the secondaries, tertiaries and
quartiaries for all three models. But for the primaries the IMF becomes
slightly steeper. In the panels 4, 5 and 6 of Figs.~\ref{fig:imf04} and
\ref{fig:imf08} this can be seen. This is expected since the SpP mechanism
searches for massive companions for massive stars and therefore depletes the
reservoir of massive stars remaining as primaries, and in doing so channels
large amounts of massive stars into the companion star IMFs.

A similar, but even more drastic result is obtained by system pairing (SyP
Models 7 to 9, panels 7 to 9 in Figs.~\ref{fig:imf08} and
\ref{fig:imf12}). Here the primary IMFs are steepened and the secondary IMFs
flattened in order to retain constant slopes for the system IMFs under the
condition of having increasingly higher $q$-values for more massive systems. 

The rather flat slopes of the companion mass functions for SpP and
SyP would be in agreement with the results of \citet{To08} that for a
large fraction of triple and quadruple systems the components are of rather
similar mass.

For reasons of comparison also an example of 100\% equal-mass binaries for
massive stars (for primaries $>$ 2 $M_{\odot}$) is shown (Model 10 in
Tab.~\ref{tab:noevo} and panel 10 in Fig.~\ref{fig:imf12}). Here in principle
all IMFs are parallel but again discontinuous features appear in the IMFs at
the borders of the random pairing for the low-mass stars and the $q=1$ pairing
for the more massive ones.

\begin{table*}
\caption{\label{tab:noevo} Properties of the models without stellar
  evolution. The pairing methods are described in
  Section~\ref{sub:pair}. All stars are in 100\% binaries, triples or
  quadruples, respectively. The slopes are measured by the bias-free
  method (see \S~\ref{sub:slope}) above 1.0 $M_{\odot}$ unless noted
  otherwise. For each model a total of $10^{7}$ stars are
  generated. Column \#1 denotes the model name, \#2 the pairing
method,\#3 if the systems are binaries, triples or quadruples, \#4
shows the slope for all stars, \#5 the slope of the systems, \#6 the
slope of the primaries, \#7 the slope of the secondaries, \#8
the slope of the tertiaries and \#9 the slope of the quartiaries.}
\begin{tabular}{ccccccccc}
Model&pairing method&multiple&$\alpha_{\rm all stars}$&$\alpha_{\rm
  systems}$&$\alpha_{\rm primaries}$&$\alpha_{\rm
  secondaries}$&$\alpha_{\rm tertiaries}$&$\alpha_{\rm quartiaries}$\\
\hline
1&RP&bin& 2.35&2.40$^1$&2.32&3.71&-   &-\\
2&RP&trip&2.35&2.47$^1$&2.28&3.64&5.14&-\\
3&RP&quad&2.35&2.54$^1$&2.26&3.60&4.96&6.06\\
\hline
4&SpP&bin& 2.35&2.34$^1$&2.36&2.04$^2$&-   &-\\
5&SpP&trip&2.35&2.38$^1$&2.34&2.13$^2$&1.93&-\\
6&SpP&quad&2.35&2.42$^1$&2.31&2.22$^2$&1.78$^2$&1.62$^2$\\
\hline
7&SyP&bin& 2.35&2.35&2.60$^1$&2.27&-   &-\\
8&SyP&trip&2.35&2.35&2.57$^1$&2.26&2.25&-\\
9&SyP&quad&2.35&2.35&2.55$^1$&2.25&2.25&2.25\\
\hline
10&q=1&bin&2.35$^1$&2.35$^1$&2.32&2.36$^1$&-&-\\
\hline
\end{tabular}

$^1$ slope measured above 10.0 $M_{\odot}$, 
$^2$ slope measured above 2.0 $M_{\odot}$
\end{table*}

\subsection{Models with stellar evolution}
\label{sub:evo}

\begin{figure*}
\begin{center}
\includegraphics[width=16cm]{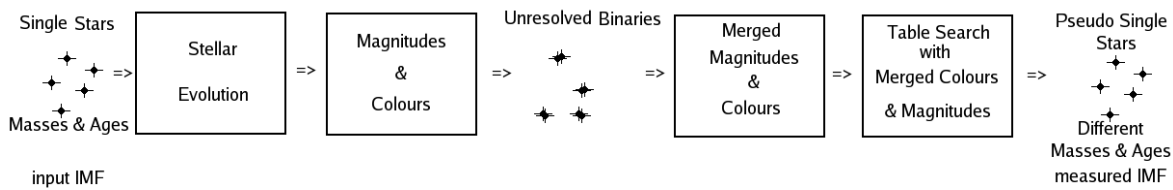}
\caption{Stylised flow chart how stellar evolution and unresolved binaries
  influence the observed mass function.}
\label{fig:flow}
\end{center}
\end{figure*}

In a second set of models the influence of the conversion of luminosities into
masses is studied. The systems have ages assigned from a Gaussian distribution
with a mean of 2 Myr and a standard deviation of 0.5 Myr and a cutoff at 0.01
and 4.5 Myr. With this age the stars are evolved according to the models
described in Section~\ref{sub:evol}. The resulting stellar parameters are used
to calculate luminosities and colours for each single star. These are
merged to form an unresolved multiple system ensuring coevality of all
companions. The resulting magnitudes and colours are searched for in a large
table of colours/magnitudes of single stars produced with the use of the same
models as in Section~\ref{sub:evol} in order to find a {\it pseudo zero-age
mass} and a {\it pseudo age} for the unresolved multiple as if it were a
single star. The table covers a mass range between 0.01 $M_{\odot}$ and 300
$M_{\odot}$ with ages between zero and 100 Myr. In Fig.~\ref{fig:flow}
the steps described above are visualised. As this procedure is rather
CPU time consuming only 1 million stars are generated for each model
in contrast to the 10 million in Section~\ref{sub:noevo}. Each model
needs about 27 hours of CPU time on a state-of-the-art PC.

The results and parameters for this second series of models are listed
in Tab.~\ref{tab:evo}. Additionally to the columns shown in
Tab.~\ref{tab:noevo} $\alpha_{\rm observed}$ is added: the IMF slope of the
mass function for all unresolved systems above 1 $M_{\odot}$ recovered by the
simulated observations. In some models strong non-linear features occur in the
output close to 1 $M_{\odot}$. In such cases a higher lower limit for the
slope determination is used as indicated in the Table. The slope is determined
in the same way as in \S~\ref{sub:noevo}. Furthermore, included is the column
``age fit'' in Tab.~\ref{tab:evo} which is the percentage of how many pseudo
stars are recovered with a pseudo age within 25\% of the real age of the
system. Finally, the last two columns contain the recovered total mass
fraction and the recovered mass fraction within the 25\% age limit. In
the first case the total mass of all recovered pseudo stars for one
model is divided by the true stellar mass within the model and in the
second case only the pseudo mass of the recovered pseudo stars with
pseudo ages within 25\% of their real ages is divided by the total
true mass. Furthermore, for reasons of comparison, Model Fit~0 is only
included in Tab.~\ref{tab:evo} but not otherwise used. This model is
similar to Model Fit~1 in all aspects but the age. Instead of an age
spread all stars have the same input age of 2 Myr.\\

Because of the fewer stars used here in comparison with \S~\ref{sub:noevo} the
errors in the slope determination are larger than in the previous experiments
without stellar evolution.

\begin{table*}
\caption{\label{tab:evo} Like Tab.~\ref{tab:noevo} but for the models
  with stellar evolution. For each model a total of 1 million stars are
  generated. Column \#1 (model) denotes the model name, \#2 (pairing
  method) the pairing method,\#3 (multiple) if the systems are
  binaries, triples or quadruples, \#4 ($\alpha_{\rm all stars}$)
shows the IMF slope for all stars above 1 $M_\odot$, \#5 ($\alpha_{\rm
  systems}$) the slope 
of the systems, \#6 ($\alpha_{\rm prim}$) the slope of the primaries,
\#7 ($\alpha_{\rm sec}$) the slope of the secondaries, \#8
($\alpha_{\rm tert}$) the slope of the tertiaries, \#9 ($\alpha_{\rm
  quart}$) the slope of the quartiaries, \#10 ($\alpha_{\rm
  observed}$) the slope of the recovered ``observed'' stars, \#11 (age
fit) marks the percentage for how many of the unresolved stars the
pseudo age of the system is recovered within 25\% of the original age,
\#12 (total mass recovered) shows the percentage of the input mass in
  stars which was recovered in ``observed'' stars and \#12 (age fit
  mass)  is the recovered mass in stars for which the pseudo ages are
  within 25\% of the real ages.  The differences in the values for the
  slopes of the primaries, 
  secondaries, tertiaries and quartiaries in comparison with
  Tab.~\ref{tab:noevo} are due to the lower number of stars used for the
  models with stellar evolution, as these lower numbers significantly increase
  the error of the slope determination.}
\begin{tabular}{ccccccccccccc}
Model&pairing&multiple&$\alpha_{\rm all stars}$&$\alpha_{\rm
  systems}$&$\alpha_{\rm prim}$&$\alpha_{\rm sec}$&$\alpha_{\rm
  tert}$&$\alpha_{\rm quart}$&$\alpha_{\rm observed}$&age fit&total mass&age fit\\
&method&&&&&&&&&&recovered&mass\\
\hline
Fit 0$^{1}$&RP&bin&2.35&2.40$^2$&2.32&3.71&-&-&2.36$^2$&50\%&84\%&59\%\\
\hline
Fit 1&RP&bin& 2.35&2.40$^2$&2.32&3.70&-   &-   &2.35$^3$&50\%&85\%&43\%\\
Fit 2&RP&trip&2.35&2.47$^2$&2.29&3.63&5.11&-   &2.34$^3$&34\%&73\%&43\%\\
Fit 3&RP&quad&2.35&2.54$^2$&2.25&3.62&4.98&6.08&2.33$^3$&25\%&65\%&33\%\\
\hline
Fit 4&SpP&bin& 2.35&2.36$^2$&2.37&2.07$^3$&-   &-   &2.43$^3$&49\%&80\%&44\%\\
Fit 5&SpP&trip&2.35&2.36$^2$&2.34&2.15$^3$&1.98&-   &2.42$^3$&32\%&68\%&30\%\\
Fit 6&SpP&quad&2.35&2.40$^2$&2.32&2.22$^3$&1.83$^3$&1.67$^3$&2.43$^3$&23\%&60\%&23\%\\
\hline
Fit 7&SyP&bin& 2.35&2.35&2.60$^2$&2.26&-   &-   &2.41$^2$&43\%&70\%&25\%\\
Fit 8&SyP&trip&2.35&2.35&2.56$^2$&2.25&2.24&-   &2.46$^2$&22\%&53\%&10\%\\
Fit 9&SyP&quad&2.35&2.35&2.54$^2$&2.25&2.25&2.23&2.43$^2$&19\%&43\%&5\%\\
\hline
Fit 10&q=1&bin&2.35$^2$&2.35$^2$&2.31&2.37$^2$&-&-&2.31$^2$&45\%&68\%&10\%\\
\hline
\end{tabular}

$^{1}$ This model is included for comparison only. It is identical to
  Fit 1 but all stars have a fixed input age of 2 Myr instead of the
  Gaussian distributed ages for the other models.\\
$^2$ slope measured above 10.0 $M_{\odot}$, 
$^3$ slope measured above 2.0 $M_{\odot}$

\end{table*}

\begin{figure*}
\begin{center}
\includegraphics[width=8cm]{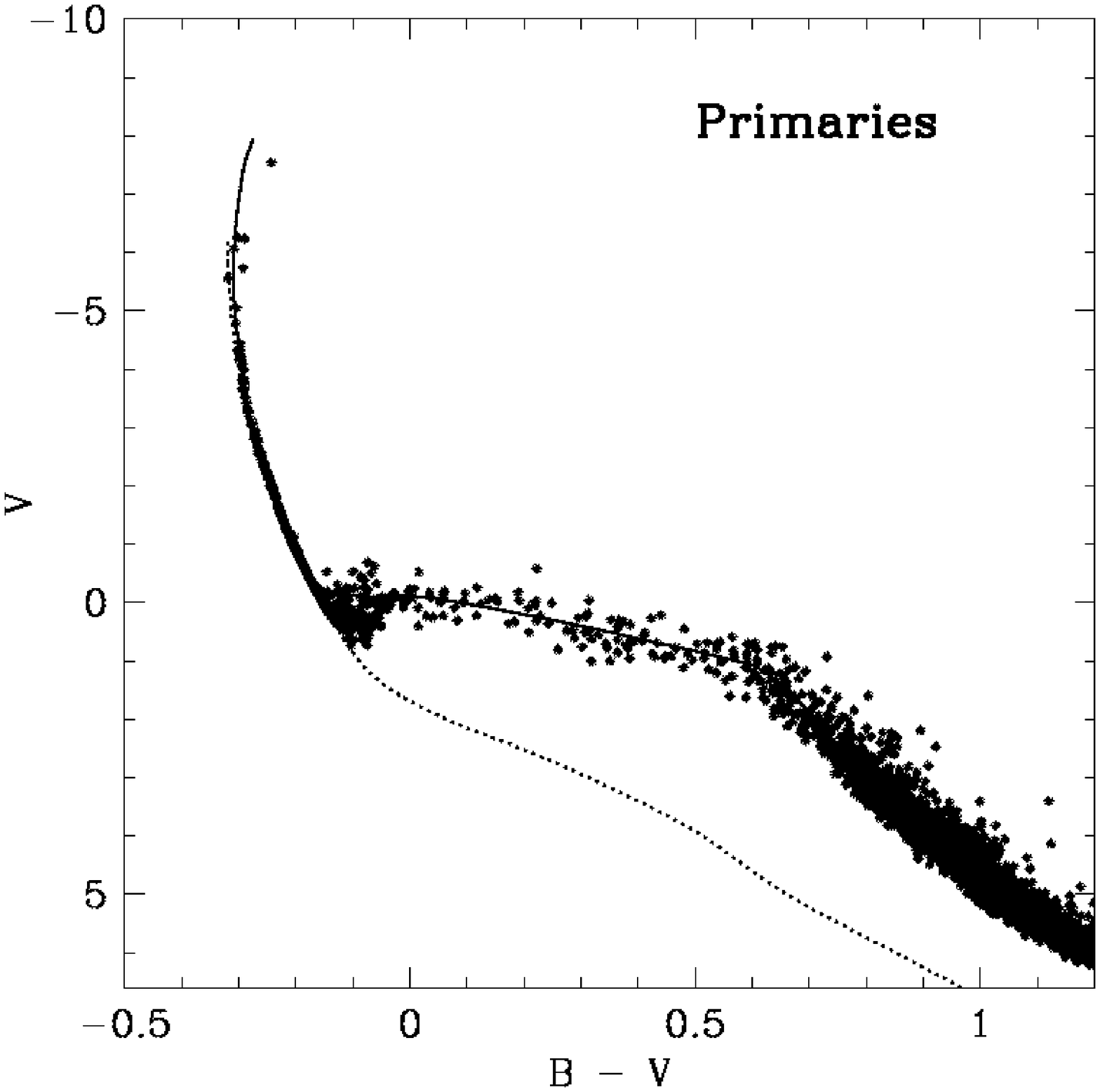}
\includegraphics[width=8cm]{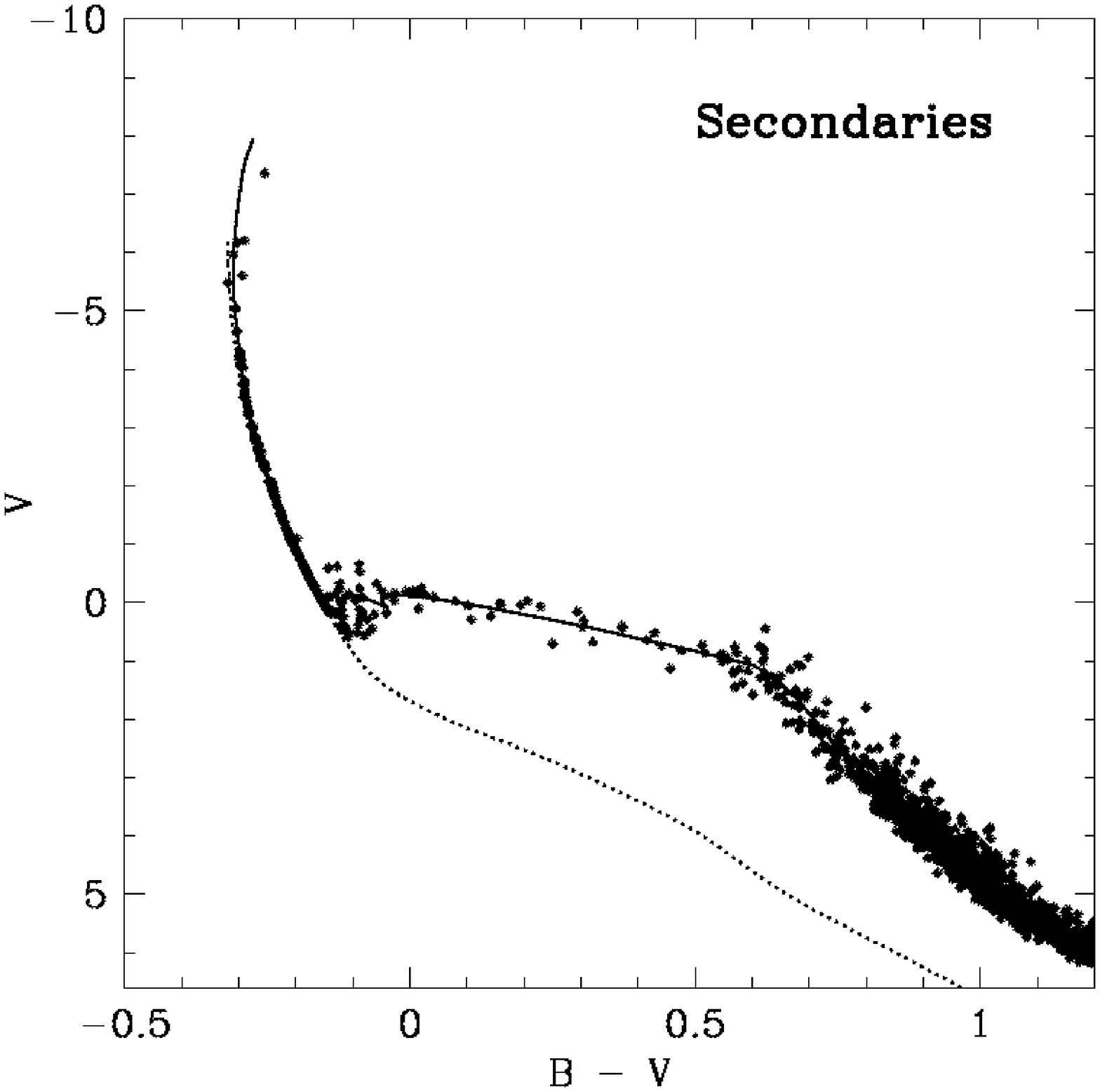}
\includegraphics[width=8cm]{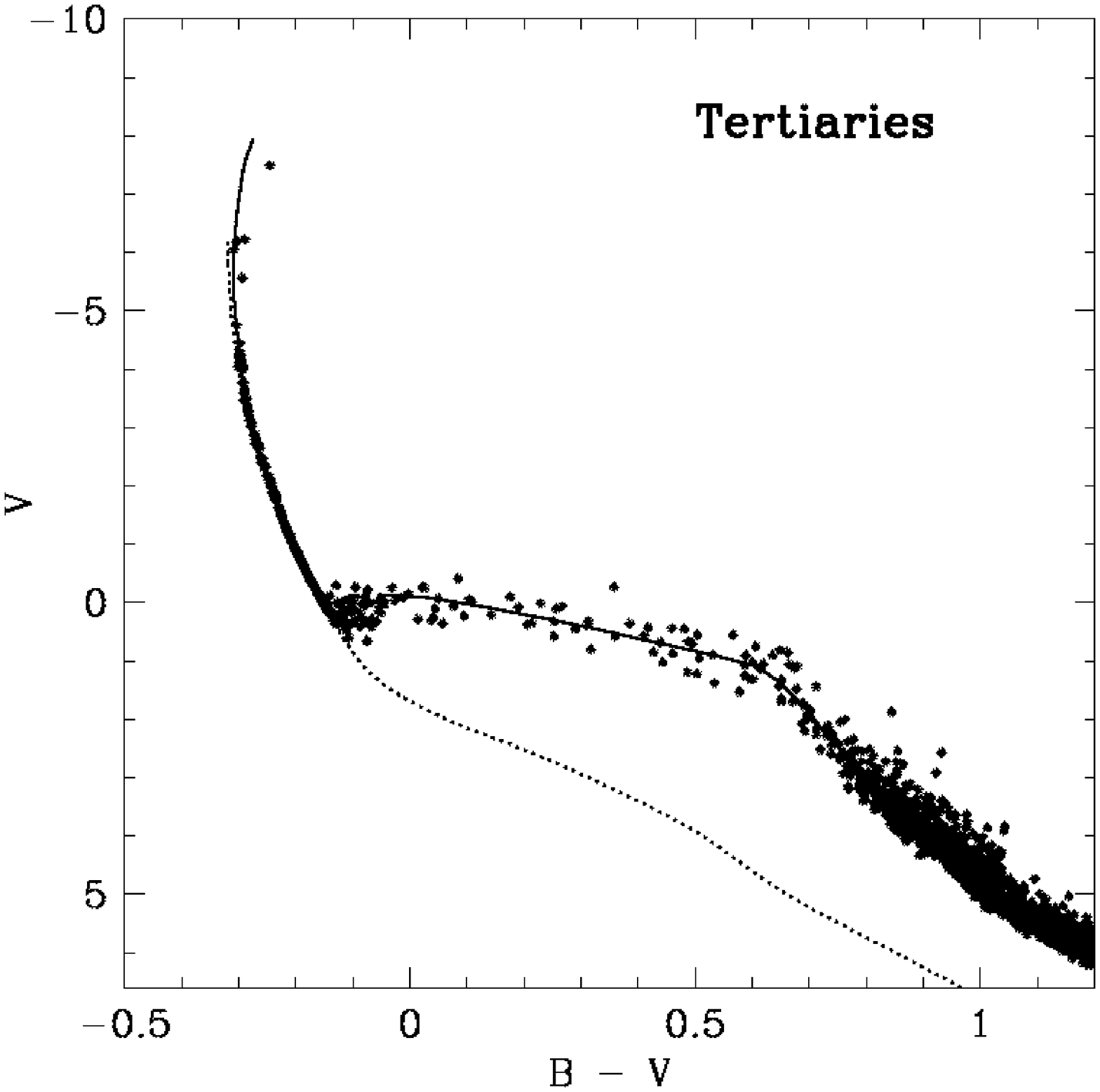}
\includegraphics[width=8cm]{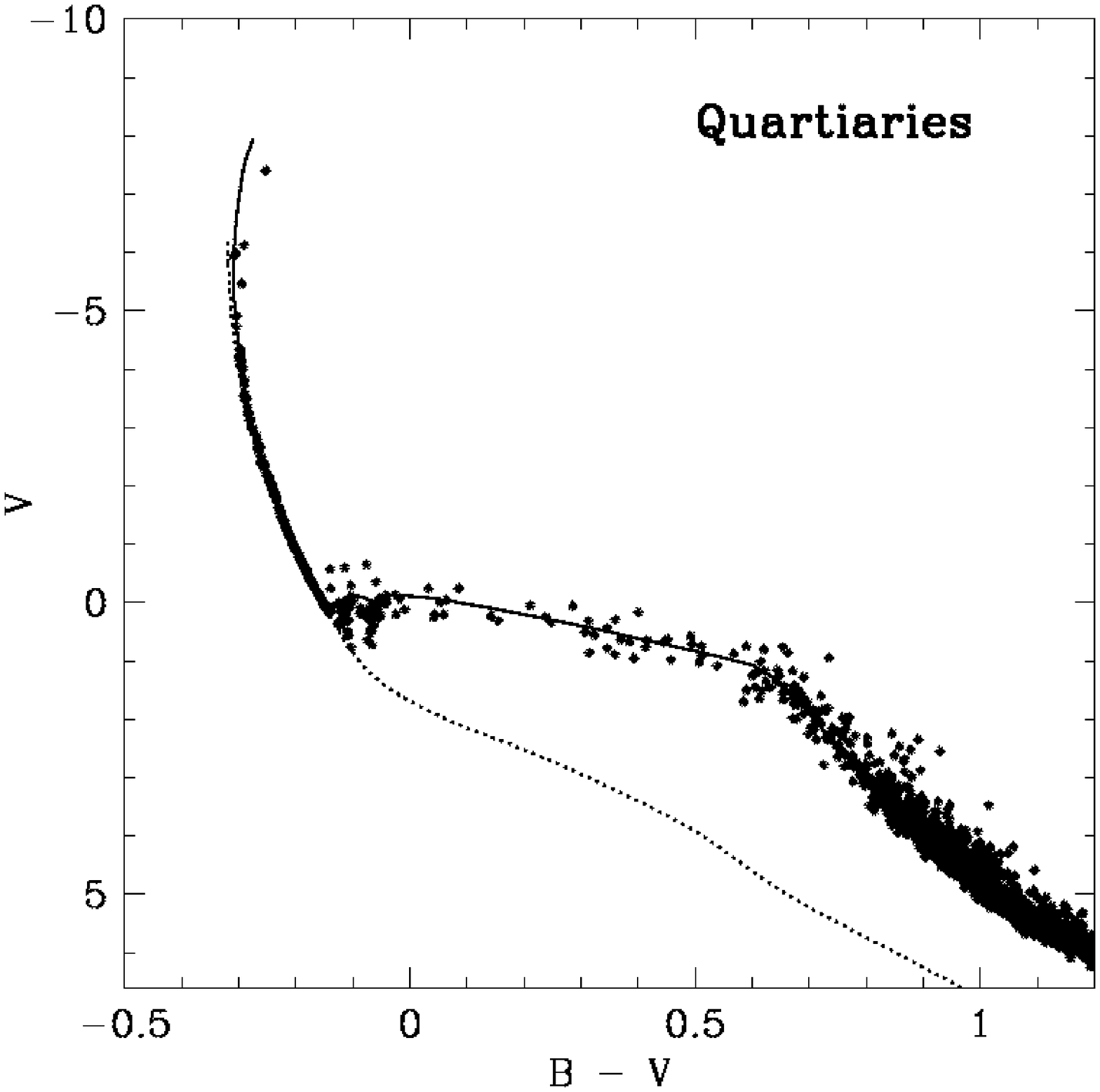}
\vspace*{-2.0cm}
\caption{HR-Diagrams for the primaries, secondaries, tertiaries and quartiaries
  stars for 17000 stars from model Fit 9. The {\it solid line} is a 2 Myr
  isochrone while the {\it dotted line} is a zero-age main-sequence.}
\label{fig:HRDF9a}
\end{center}
\end{figure*}

\begin{figure*}
\begin{center}
\includegraphics[width=8cm]{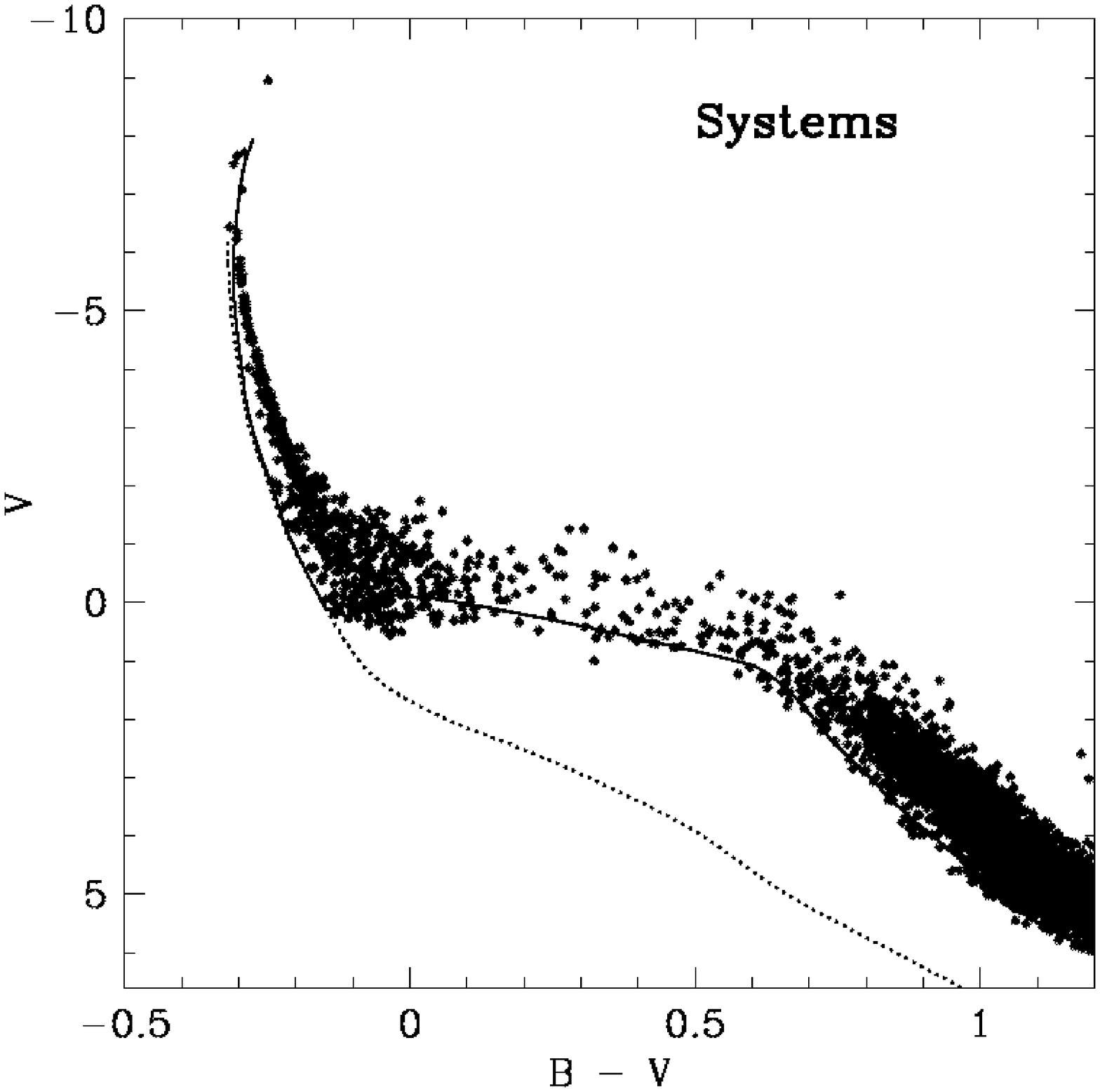}
\includegraphics[width=8cm]{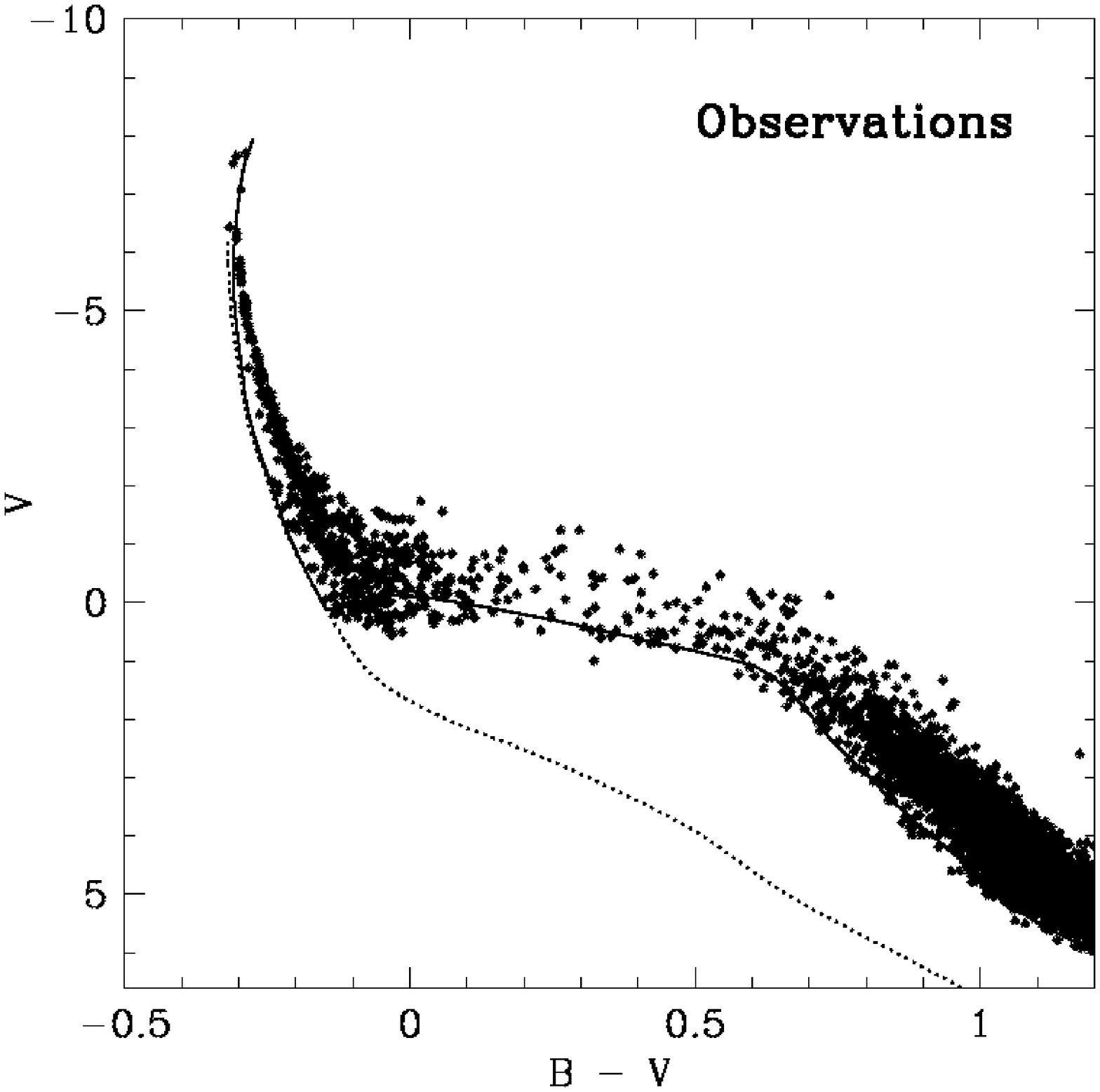}
\vspace*{-2.0cm}
\caption{HR-Diagrams for the systems and the observed pseudo stars for 17000
  stars from model Fit 9. The {\it solid line} is a 2 Myr isochrone while the
  {\it dotted line} is a zero-age main-sequence.}
\label{fig:HRDF9b}
\end{center}
\end{figure*}

In the first panel of Fig.~\ref{fig:Fit1to4} (Model Fit 1, random
pairing, 100\% binaries, see also the exemplary HR-Diagrams for this model in
Fig.~\ref{fig:HRDF9a}) the input, primary, secondary and system IMF
are the same as for Model 1 from Section~\ref{sub:noevo} but with a
larger scatter because of the smaller number of generated stars. The
derived observed IMF does not follow the (slightly) steeper system IMF
but follows instead the input IMF of all stars. The steeping effect seen in
the system IMF in comparison to the IMF of all stars is thus fully
negated. This is due to the fact that while in the model without stellar
evolution all masses are 100\% accounted for, in this model with
evolution the low-mass companions of massive stars are ``lost'' because the
primary dominates the light and only its mass is recovered. In the low-mass
regime all stars are of rather similar mass and therefore the unresolved
binaries change the luminosity of the system but not the ($B-V$) colour. This
makes them appear as younger PMS stars, usually rather close to the mass of
the primary. Both model series with mass-dependent mass-ratios (SpP and SyP,
panels 4 to 9 in Figs.~\ref{fig:Fit4} and \ref{fig:Fit3to4}) have slightly
steeper observed IMFs with a slope of about 2.45. But such a small deviation
from the Salpeter-slope is within the error margin of measured high-mass IMF
slopes. The final Model Fit 10 in which all stars above 2 $M_{\odot}$ reside
in equal mass binaries influences the slope of the primary IMF and the
observed IMF slightly. But such small differences ($\Delta \alpha$ = 0.04) are
still within the errors of the here used slope determinations.

In conclusion, the analysis presented here quite clearly shows that the
observed IMF for massive stars remains mostly indistinguishable from the
underlying true stellar IMF, which may be flatter by at most $\Delta \alpha =
0.1 $ even if all massive stars reside in quadruple systems. This would thus
indicate that measured IMF indices are quite robust and that the
Salpeter/Massey slope, $\alpha$ = 2.35, is not significantly affected
by unresolved multiple stars above a few $M_{\odot}$.

\subsubsection{Age spreads}
\label{subsub:age}
\begin{figure}
\begin{center}
\includegraphics[width=8cm]{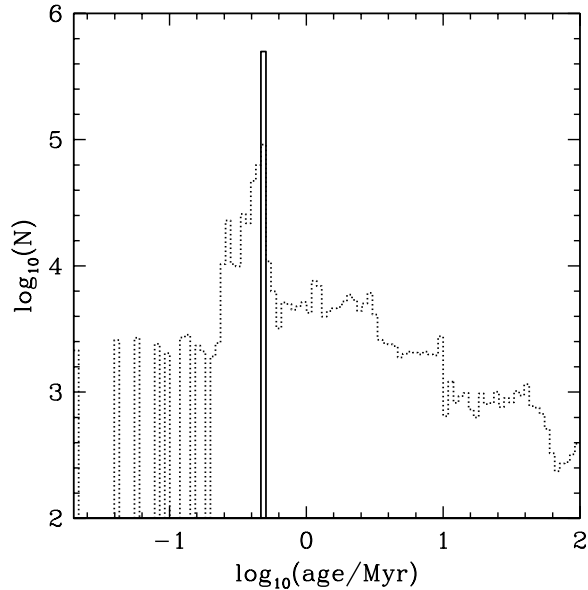}
\vspace*{-2.0cm}
\caption{Logarithm of the number of stars versus logarithm of the
  age for one million stars in 100\% unresolved RP binaries (500000
  systems). {\it Solid histogram}: The input age of 0.5 Myr. {\it Dotted
  histogram}: The recovered output ages for stars of all masses.}
\label{fig:agebin0.5Myr}
\end{center}
\end{figure}

\begin{figure}
\begin{center}
\includegraphics[width=8cm]{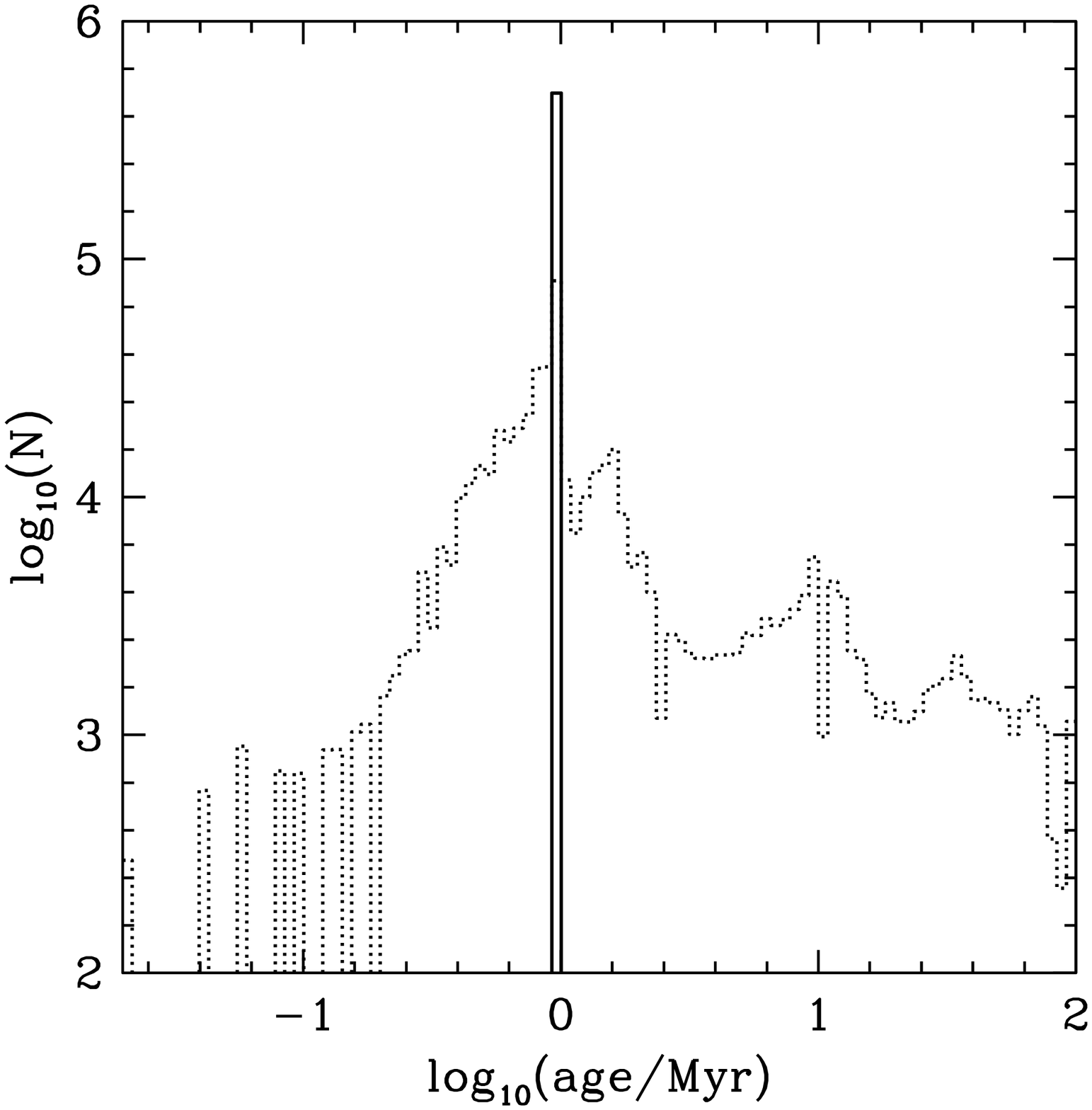}
\vspace*{-2.0cm}
\caption{Like Fig.~\ref{fig:agebin0.5Myr} but with an input age of 1 Myr.}
\label{fig:agebin1Myr}
\end{center}
\end{figure}

\begin{figure}
\begin{center}
\includegraphics[width=8cm]{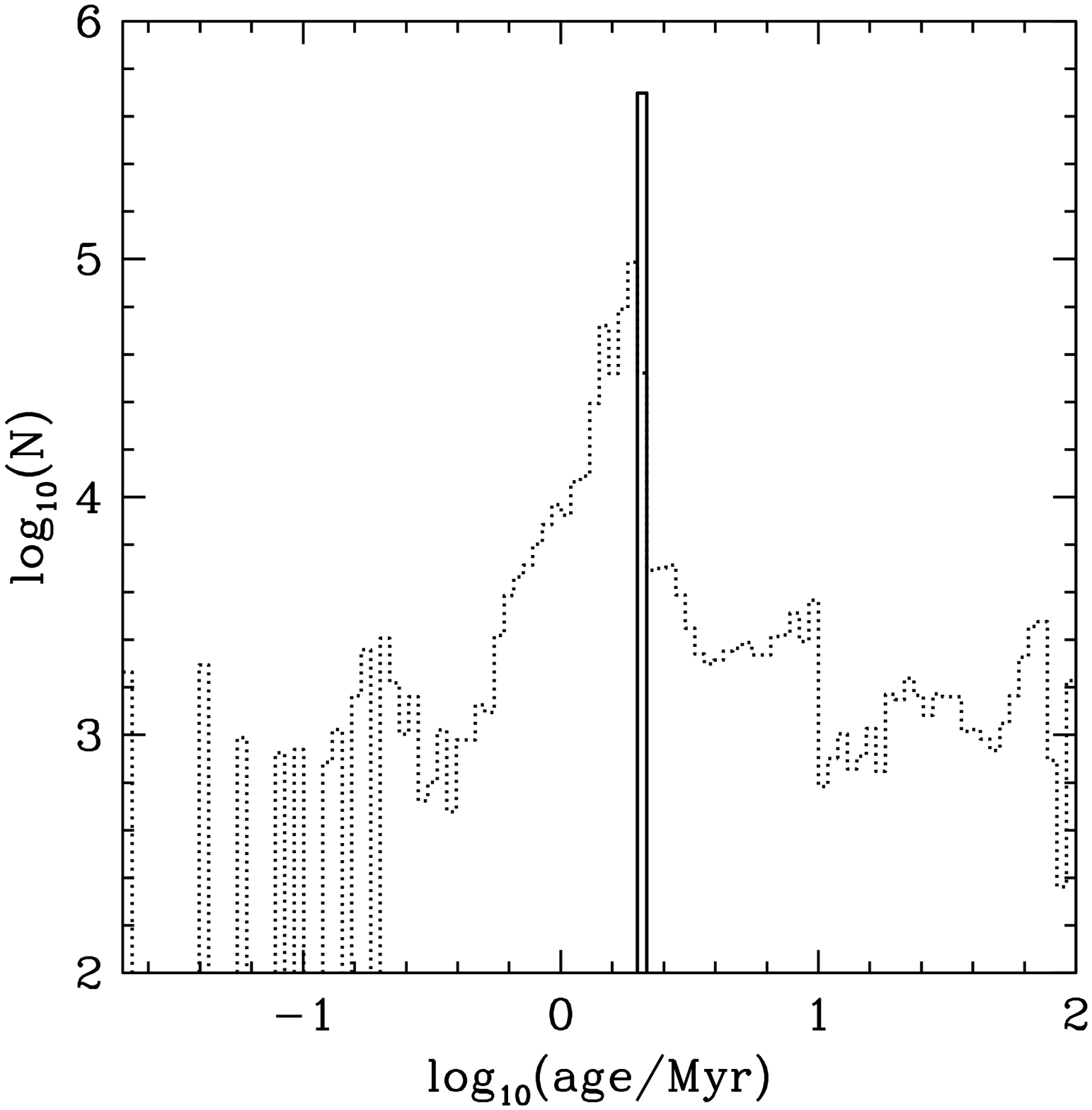}
\vspace*{-2.0cm}
\caption{Like Fig.~\ref{fig:agebin0.5Myr} but with an input age of 2 Myr.}
\label{fig:agebin2Myr}
\end{center}
\end{figure}

\begin{figure}
\begin{center}
\includegraphics[width=8cm]{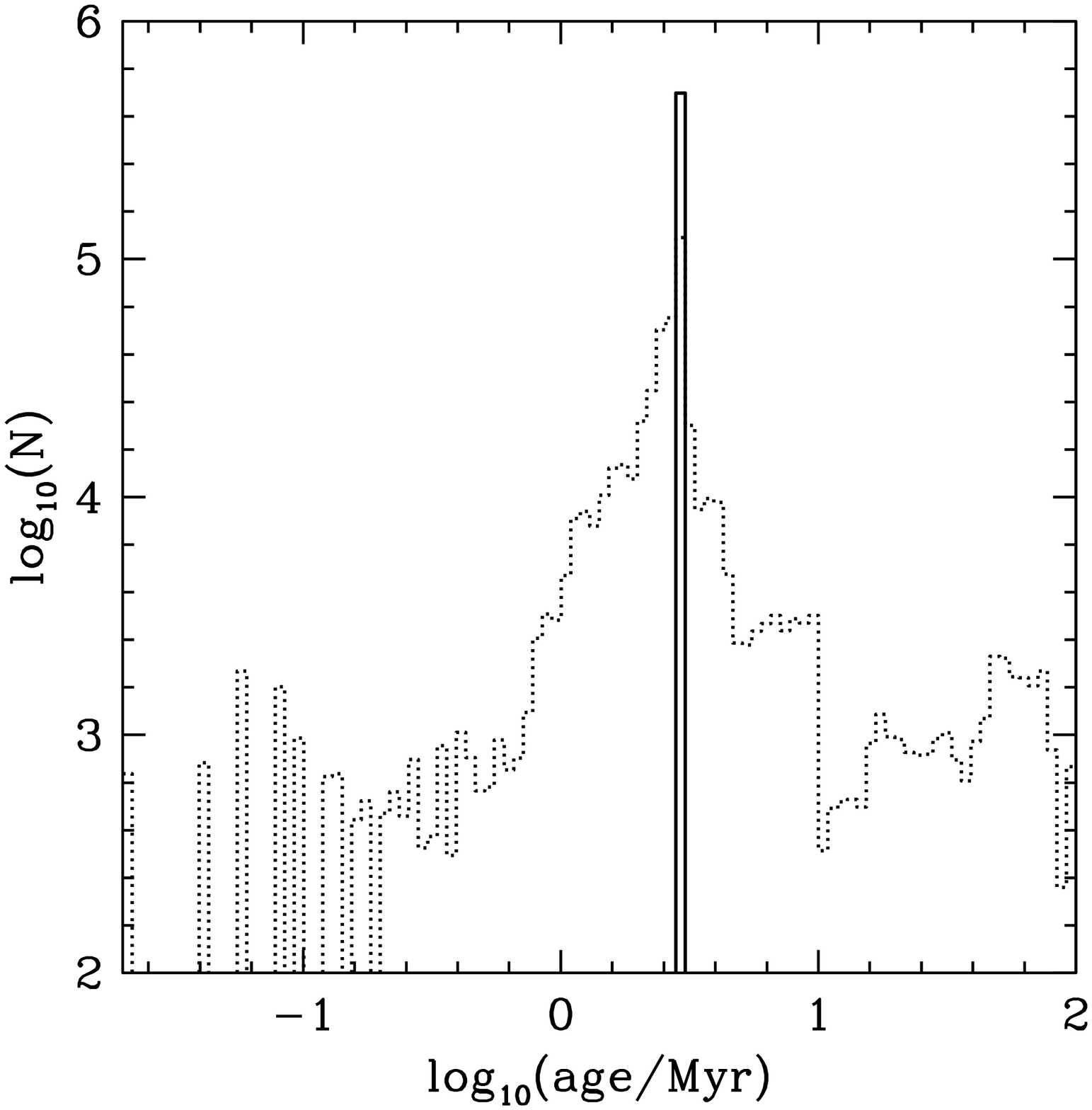}
\vspace*{-2.0cm}
\caption{Like Fig.~\ref{fig:agebin0.5Myr} but with an input age of 3 Myr.}
\label{fig:agebin3Myr}
\end{center}
\end{figure}

Interestingly, as is seen in the ninth column of Tab.~\ref{tab:evo},
only about 40 to 50\% of the pseudo stars are recovered with the right pseudo
age (with an error of less than 25\% compared to the true age). Actually, the
discrepancy between assigned age and fitted age can be rather
large. Therefore, large numbers of binaries (regardless of their masses) can
mimic a wide spread of formation ages. In order to further explore this
interesting result a series of models is calculated with $5 \cdot 10^{5}$
binaries (0.08 to 150 $M_{\odot}$ for the stars), random pairing and the same
input age. The distribution of the resulting recovered ages (using the same
method as before in \S~\ref{sub:evo}) for stars of all masses can be seen in
Figs.~\ref{fig:agebin0.5Myr} to \ref{fig:agebin3Myr}. 
Fig.~\ref{fig:agebin0.5Myr} features an input age of 0.5 Myr, 
Fig.~\ref{fig:agebin1Myr} 1 Myr, Fig.~\ref{fig:agebin2Myr} 2 Myr and
Fig.~\ref{fig:agebin3Myr} 3 Myr.

Using such a single input age (e.g.~1 Myr) as in
Fig.~\ref{fig:agebin1Myr} the distribution of the recovered output
ages is severely distorted ({\it dotted histogram} in
Fig.~\ref{fig:agebin1Myr}). In Tab.~\ref{tab:spread} are shown the
percentages of recovered ages in several age bins. Besides the models
with single input ages also the Models Fit 1, Fit 4, Fit 7 and Fit 10 are
included as well as a model with only 50\% binaries and the same age range as
the Models Fit 1 to Fit 10.

\begin{table*}
\caption{\label{tab:spread} Percentages of recovered ages in different
age bins for a number of models.}
\begin{tabular}{crrrrr}
Model&$<$ 0.5&0.5 - 1.5&1.5 - 5.0&5.0 - 10.0&10$^{+}$\\
&Myr&Myr&Myr&Myr&Myr\\
\hline
random ages (like Fit 1 to Fit 10),&&&&&\\
50\% bin RP&1.5&19.7&74.1&1.6&3.1\\
0.5 Myr, 100\% bin RP&58.5&22.1&11.5&3.5&4.4\\
1.0 Myr, 100\% bin RP&8.0&65.2&10.3&9.0&7.5\\
2.0 Myr, 100\% bin RP&4.8&28.9&53.0&4.3&9.0\\
3.0 Myr, 100\% bin RP&3.0&10.3&75.0&4.8&7.0\\
\hline
Fit 1&4.4&31.8&50.0&4.6&9.1\\
Fit 4&4.4&31.8&49.3&4.9&9.5\\
Fit 7&5.3&38.0&41.9&3.8&11.1\\
Fit 10&4.7&33.9&45.0&4.7&11.5\\
\hline
\end{tabular}
\end{table*}

\begin{figure}
\begin{center}
\includegraphics[width=8cm]{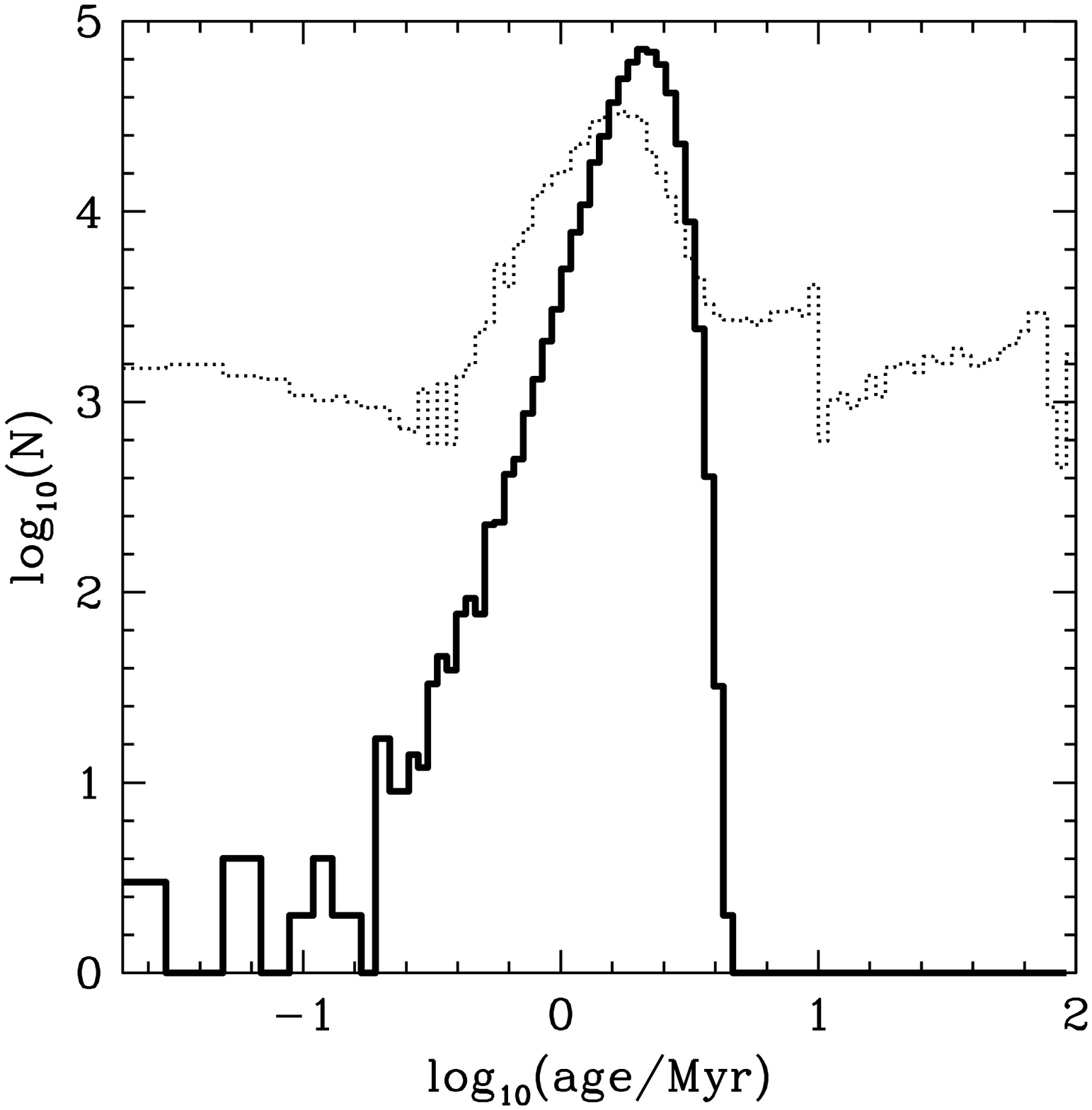}
\vspace*{-2.0cm}
\caption{Logarithm of the number of stars versus logarithm of the
  age for Model Fit 1. {\it Solid histogram}: The assigned input
  age. {\it Dotted histogram}: The recovered output age of all stars.}
\label{fig:agebin}
\end{center}
\end{figure}

In a real star cluster it is unlikely that all stars are truly of the
same age. Therefore in Fig.~\ref{fig:agebin} a distribution of input
ages is used. Fig.~\ref{fig:agebin} shows as an example the age
distribution for Model Fit 1. While the {\it solid histogram} is the
Gaussian input age distribution (distorted because of the logarithmic
ordinate), the {\it dotted histogram} is the recovered age distribution. The
mean of the recovered ages of all stars is slightly lower than the true mean
age (1.6 Myr instead of 2 Myr). This is consistent with the findings by
\citet{BZ97} and \citet{PZ99} that un-resolved binaries let a population
appear younger than it is. Additionally, the wings of the recovered
distribution are much wider - suggesting the presence of stars even up to an
age of 100 Myr, despite the maximal input age of about 4.5 Myr. Variability of
these PMS stars and photometric errors make the age derivation even more
difficult \citep{BPS07,HBW07}.

\begin{figure}
\begin{center}
\includegraphics[width=8cm]{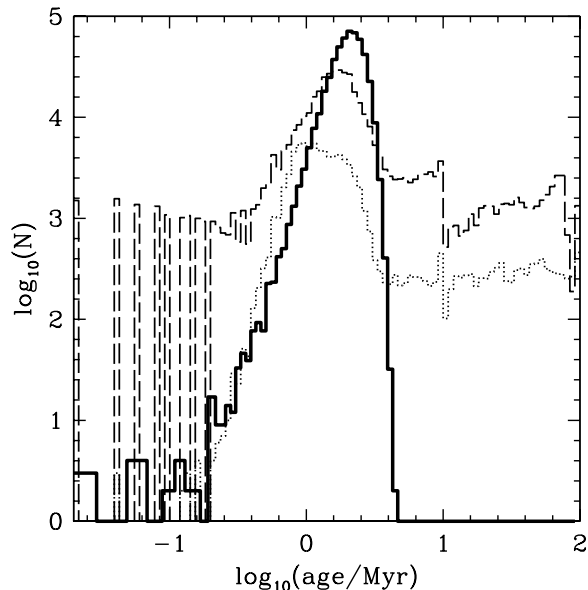}
\vspace*{-2.0cm}
\caption{Like Fig.~\ref{fig:agebin} but split for high- and low-mass
  stars. The {\it dotted line} shows the histogram of output ages for
  stars above 1 $M_{\odot}$ and the {\it long-dashed} one for stars
  below that limit.}
\label{fig:agebin3}
\end{center}
\end{figure}

In Fig.~\ref{fig:agebin3} the same recovered output ages as in
Fig.~\ref{fig:agebin} are shown but split into two, for stars above
1 $M_{\odot}$ ({\it dotted histogram}) and below ({\it long-dashed
histogram}). It can be seen that the tail towards old ages is
prominent in both populations but the tail on the left hand side,
towards young ages, is more dominant for low-mass stars. However, the
actual peak of the high-mass distribution is moved from 2 Myr to 1 Myr
while the low-mass stars peak in between at about 1.6 Myr. The bulk
low-mass stellar population therefore {\it appears older} by 0.6 Myr than the
massive stars.

So, it is important to note that unresolved binaries easily lead to the
impression of extended star-formation in a cluster and to the low-mass stars
appearing older. Therefore, any age spread encountered in young stellar
clusters needs to be interpreted very carefully, as there are also other
processes which may masquerade an age spread, e.g.~older stars captured
during the formation of the star cluster \citep{FKE06,PAK06b} and/or
unresolved multiple stars as shown here.

\begin{figure}
\begin{center}
\includegraphics[width=8cm]{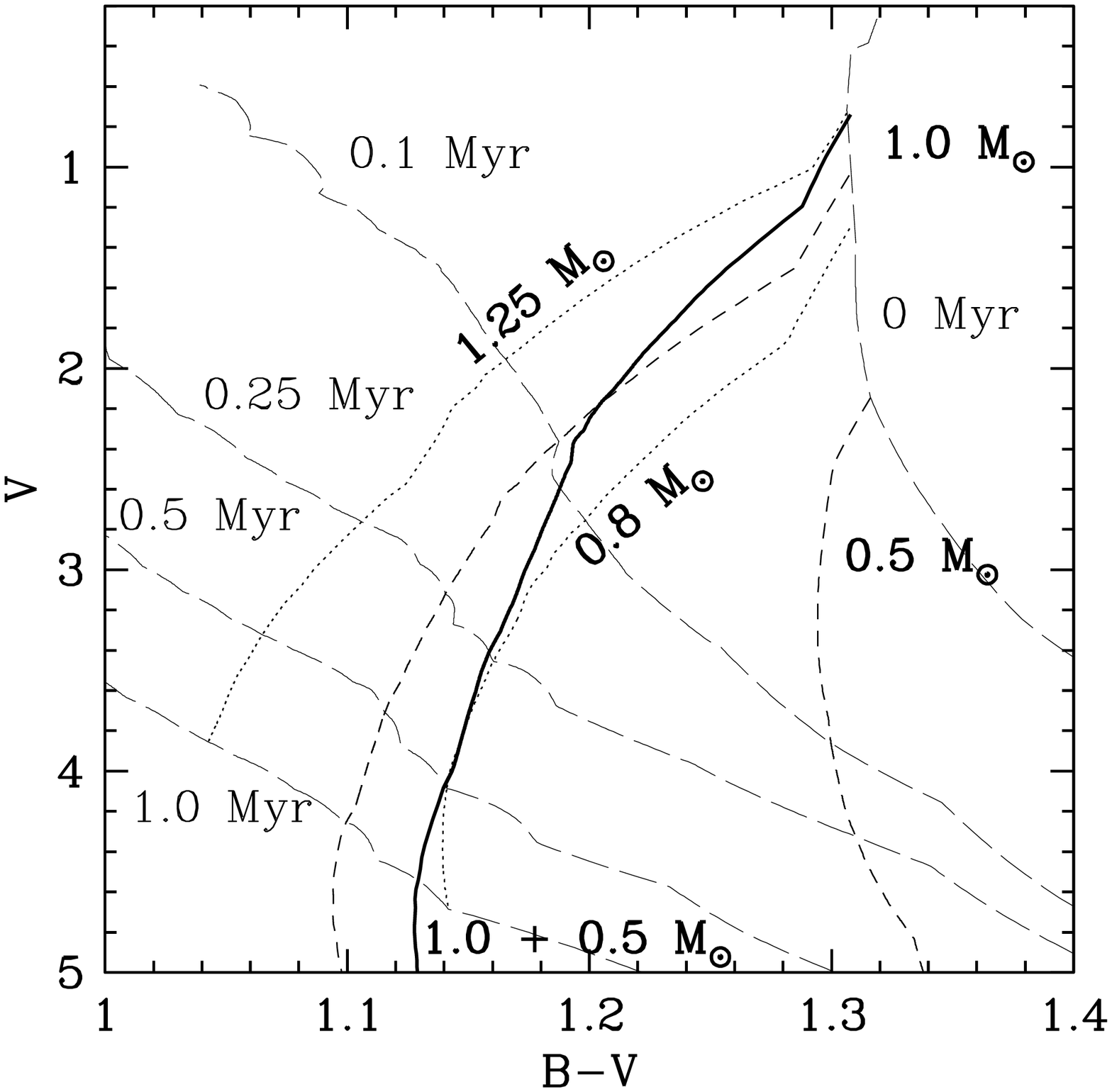}
\vspace*{-2.0cm}
\caption{The evolution of some young stars through the
  colour-magnitude diagram along the Hayashi contraction tracks for
  the first 3 Myr. The {\it dotted} and {\it short-dashed lines}
  indicate the evolution of single stars, while the {\it thick solid
  line} shows a binary made of a 1 $M_{\odot}$ and a 0.5 $M_{\odot}$
  star. The {\it thin long-dashed lines} represent isochrones for
  single stars with ages as indicated in the figure. For more details see
  the text in \S~\ref{subsub:age}.}
\label{fig:ageHRD1}
\end{center}
\end{figure}

\begin{figure}
\begin{center}
\includegraphics[width=8cm]{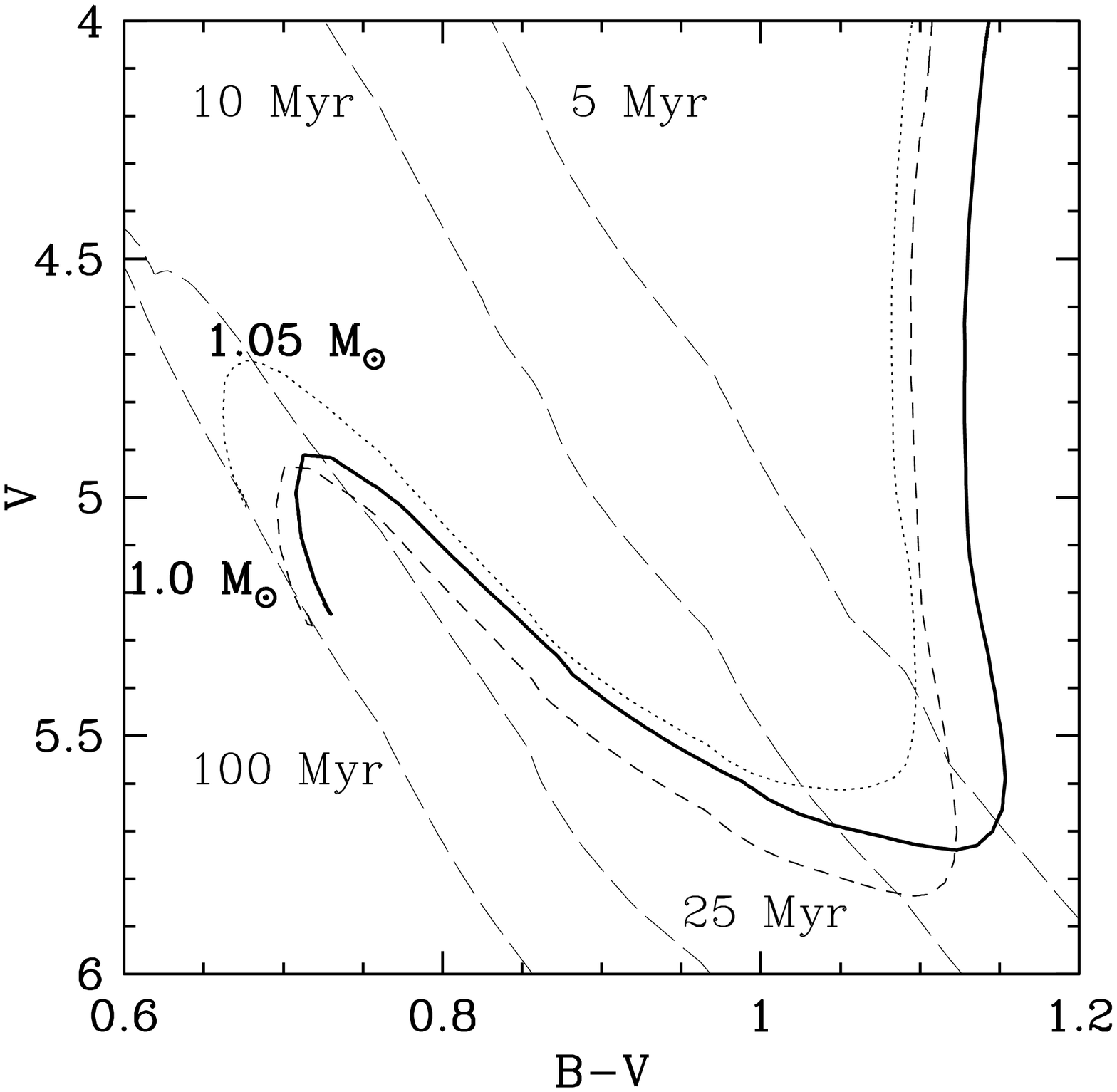}
\vspace*{-2.0cm}
\caption{Like Fig.~\ref{fig:ageHRD1} but following the PMS evolution
  until an age of 100 Myr. For details see the text in \S~\ref{subsub:age}.}
\label{fig:ageHRD2}
\end{center}
\end{figure}

Figs.~\ref{fig:ageHRD1} and \ref{fig:ageHRD2} show some details how
the difference in the age determination comes about. The solid line in
both figures refers to the evolution of an unresolved binary composed of
equal age 1 $M_{\odot}$ and 0.5 $M_{\odot}$ stars using the PMS and
main-sequence (MS) tracks as described in Section~\ref{sub:evol}. In
Fig.~\ref{fig:ageHRD1} the first few Myr are shown. Marked in the
figure are also the evolutionary tracks for four different single
stars. A 1 $M_{\odot}$ star, one of 1.25 $M_{\odot}$, a 0.8
$M_{\odot}$ and a 0.5 $M_{\odot}$ star. Within its first 1.4 Myr the
binary first looks like a 1.25 $M_{\odot}$ star until it evolves
parallel for some time to a 0.8 $M_{\odot}$ star. The measured age of the 
binary is 0.11 Myr when it crosses the 1 $M_{\odot}$ track, while the
true age of the two companion stars at that moment is 0.05 Myr.
Between 0.3 and 1.4 Myr the binary looks almost identical to a 0.8
$M_{\odot}$ star but this star evolves through this part of its track
at an age between 0.16 and 0.66 Myr. Therefore, the binary of 1.5
$M_{\odot}$, interpreted as a single star, looks for about 1 Myr like
a 0.8 $M_{\odot}$ star of only half its age.

In Fig.~\ref{fig:ageHRD2} the final contraction of the
pre-main-sequence stars to the zero-age-main-sequence (ZAMS) and the
first Myr of the later MS evolution up to an age of 100 Myr is
shown\footnote{The 1.05 $M_{\odot}$ star needs about 40 Myr to reach
  the ZAMS while the process takes roughly 80 Myr for the 1
  $M_{\odot}$ star.}. Here the binary evolves for some time parallel
to the track of a 1.05 $M_{\odot}$ star. But while the 1.05
$M_{\odot}$ object moves through this phase when 9.5 to 22 Myr old the
binary is about 5 Myr older, with a true age between 14 and 25 Myr.

\subsection{Recovered total stellar mass}
As not all stars are recovered when observing unresolved multiple
stars because they are either absorbed in pseudo stars or appear with
a pseudo age too different to the bulk age, not only the ages are
effected but also the deduced amount of mass in stars in a 
cluster. The two last columns in Tab.~\ref{tab:evo} show the fraction of
recovered mass in the different models in proportion to the initial mass of
the $10^6$ original stars. First, all the ``pseudo'' masses for all
recovered stars are added up and than divided by the sum of the real
masses of all stars in the calculation. And second, only the
``pseudo'' masses of the stars with pseudo ages recovered within 25\% of the
real age are added up and compared with the real total mass of all
stars. This second comparison is made because in a real observed sample
stars with ``wrong'' ages might not be counted as cluster
members but as field stars appearing superposed on the cluster.
As can be seen in the second last column between 15 and nearly
60\% of the true mass might be missed due to unresolved multiples. If
only the stars are considered which are well within the true age of
the systems (last column of Tab.~\ref{tab:evo}) between 41 and 95\% of
the actual mass of the star cluster can be missed. It should be noted
here that, as pointed out in \S~\ref{se:intro}, the dynamics of massive
stars in star clusters might change the mass-ratios of massive
multiple systems towards more massive companions. This would result in
an increasing amount of ``hidden'' mass in the cluster even if all stars
were randomly paired in the beginning. This can be seen in
Tab.~\ref{tab:evo} by the fact that both methods, ``SpP'' and ``SyP'',
have lower recovered total masses than the ``RP''
method. Additionally, the unresolved multiple stars can influence
dynamical mass estimates of star clusters by biasing radial velocity
measurements \citep{KG08}.

\section{Discussion and Conclusions}
\label{se:disc}

A large numerical study has been carried out to determine whether unresolved
multiple stars might hide a different underlying massive-star IMF in
the observational data. In order to do so three different pairing
mechanisms have been considered on order to produce the unresolved
systems.\\

The main difference between these three pairing mechanisms can be
summarised as follows: Random Pairing (RP) keeps the MF of all stars
and the MF of the primaries constant. As the MF of secondaries is
constrained by the input IMF RP changes the system IMF. The ``Special
Pairing'' (SpP) method with its mass-dependent lower mass limit for
companions of massive stars keeps all MFs (above $1 M_{\odot}$) 
parallel but changes the system MF in a way similar to RP. As ``System
Pairing'' (SyP) keeps the system MF as a constant input and puts constrains on
the mass-ratio for massive systems, the primary and companion MFs are changed
to fulfil these constrains.\\

\begin{figure}
\begin{center}
\includegraphics[width=8cm]{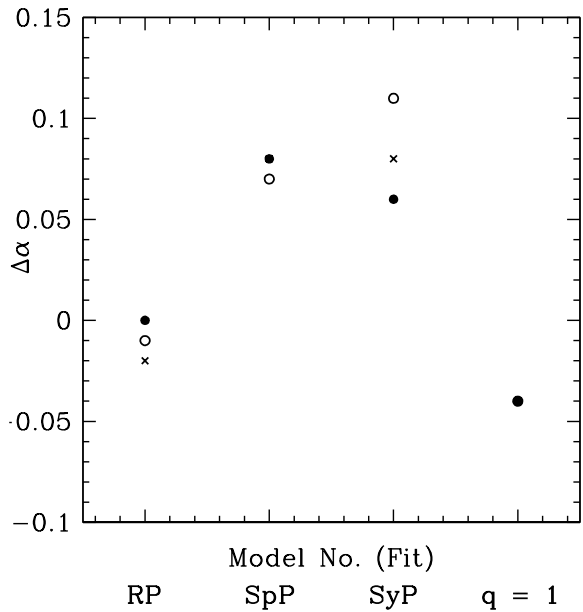}
\vspace*{-1.5cm}
\caption{Change in slope ($\Delta \alpha = \alpha_{\rm observed} -
  \alpha_{\rm allstars}$ ) between the observed slope and a slope for
  all stars of the models with stellar evolution (see
  Tab.~\ref{tab:evo}). {\it Filled black dots} indicate models with 100\%
  binaries, {\it open circles} are the models with 100\% triples and the {\it
  ``x''}-symbols show models with 100\% quadruples. Note that for Model Fit 4
  ($q$ = 1) only the 100\% binary case is shown and that for Model Fit 2 (SpP)
  the values for the binaries and the quadruples are too close together to be
  distinguishable.}
\label{fig:slope1}
\end{center}
\end{figure}

A first set of models (\S~\ref{sub:noevo}) has been studied under the
assumption that all the masses of the stars and systems can be
recovered exactly from the observations. While in the second set
(\S~\ref{sub:evo}) the stars are evolved according to stellar
evolution models, then merged following the different pairing
algorithms and retrieved as if they were single stars. The results of
this study of the effect of large numbers of unresolved multiple stars
can be summarised as follows:

\begin{itemize}
  \item For random pairing and special pairing the system IMF is always
  slightly steeper above 1 $M_\odot$ than the IMF for all stars
  (Tab.~\ref{tab:noevo}). 
  \item This, however, does not automatically lead to a steeper observed mass
  function.
  \item In the case of random sampling there is no effect on the
  observed IMF above 1 $M_\odot$ whatsoever (Tab.~\ref{tab:evo}). This
  seems to be in contrast with some of the results by
  \citet{MA08}. For certain combinations of lower and upper mass
  limits \citet{MA08} find differences in the slopes of up to $\Delta
  \alpha = \pm$ 0.2. But when using a similar upper mass limit (120
  $M_\odot$) as in this study (150 $M_\odot$) he finds only very
  little changes in the slope. He concludes that ``in most cases the
  existence of unresolved binaries has only a small effect on the
  massive-star IMF slope''. His conclusions are different for chance
  superpositions of stars which are not considered in this
  contribution.
  \item For special pairing and system pairing the observed IMF is slightly
  steeper ($\Delta \alpha \approx$ 0.1 dex) than the IMF for all stars (see
  Fig.~\ref{fig:slope1}). Therefore, an in reality observed IMF slope
  of 2.35 therefore might have a true underlying IMF for all stars of
  2.25 above a few $M_\odot$.
  \item The effect is in general smaller than the error bars on observational
  slopes of IMFs.
  \item Nonetheless, unresolved multiple systems can hide between 15\% and up
  to 95\% of the mass of a star cluster, depending on the pairing mechanism
  and the amount of higher order unresolved systems. 
  \item Furthermore, age dating in star clusters can by affected severely:
  \begin{itemize}
    \item Even a single-age population will appear to have an age
      spread if the fraction of unresolved multiples is large.
    \item Lower mass stars can appear 0.6 Myr older than massive stars
      in this case.
  \end{itemize}
\end{itemize}

Based on the broad range of models presented here we conclude that an
observed Salpeter/Massey IMF slope ($\alpha_\mathrm{obs}$ = 2.35) can
only mask a slightly flatter IMF for all stars with $\alpha_{\rm true}
\approx 2.25$.

It should be noted here that \citet{MA08} too finds little evidence
for effects of unresolved, random paired, binaries on the slope of the
IMF of massive stars despite using a different colour ($U-V$ instead
of $B-V$), different stellar evolution models, different routines to
calculate the colours and another method to measure the slopes of the 
mass functions. But he further concludes that chance
superpositions of stars in star clusters also introduce a bias on the slope 
determination, and that the slope-fitting method via histogram construction
additionally biases slope determinations \citep[see also][]{MU05}. Due to our
novel approach by using a bias-free maximum likelihood method (see
\S~\ref{sub:slope}) to measure the slopes of the mass functions we do not
suffer from the disadvantages of the histogram fitting method as described in
\citet{MU05} and \citet{MA08}.\\

\section*{Acknowledgements}
We thank Christopher Tout for helpful discussions and Jarrod Hurley
for kindly providing us with his FORTRAN routines for the conversion
of luminosities and effective temperatures into magnitudes and
colours. Part of this work was financially supported by the Chilean FONDECYT
grand 3060096 and the European Commission Marie Curie Research Training Grant
CONSTELLATION (MRTN-CT-2006-035890).

\bibliography{mybiblio}

\begin{appendix}
\section{The stellar initial mass function}
\label{app:IMF}
The following multi-component power-law IMF is used throughout the paper:

{\small
\begin{equation}
\xi(m) = k \left\{\begin{array}{ll}
k^{'} \left(\frac{m}{m_{\rm H}} \right)^{-\alpha_{0}}&\hspace{-0.25cm},m_{\rm
  low} \le m < m_{\rm H},\\
\left(\frac{m}{m_{\rm H}} \right)^{-\alpha_{1}}&\hspace{-0.25cm},m_{\rm
  H} \le m < m_{0},\\
\left(\frac{m_{0}}{m_{\rm H}} \right)^{-\alpha_{1}}
  \left(\frac{m}{m_{0}} \right)^{-\alpha_{2}}&\hspace{-0.25cm},m_{0}
  \le m < m_{1},\\ 
\left(\frac{m_{0}}{m_{\rm H}} \right)^{-\alpha_{1}}
    \left(\frac{m_{1}}{m_{0}} \right)^{-\alpha_{2}}
    \left(\frac{m}{m_{1}} \right)^{-\alpha_{3}}&\hspace{-0.25cm},m_{1}
    \le m < m_{\rm max},\\ 
\end{array} \right. 
\label{eq:4pow}
\end{equation}
\noindent with exponents
\begin{equation}
          \begin{array}{l@{\quad\quad,\quad}l}
\alpha_0 = +0.30&0.01 \le m/{M}_\odot < 0.08,\\
\alpha_1 = +1.30&0.08 \le m/{M}_\odot < 0.50,\\
\alpha_2 = +2.35&0.50 \le m/{M}_\odot < 1.00,\\
\alpha_3 = +2.35&1.00 \le m/{M}_\odot.< m_{\rm max}\\
          \end{array}
\label{eq:imf}
\end{equation}}
\noindent where $dN = \xi(m)\,dm$ is the number of stars in the mass
interval $m$ to $m + dm$. The exponents $\alpha_{\rm i}$ represent the
standard or canonical IMF \citep{Kr01,Kr02}. The advantage
of such a multi-part power-law description is the easy integrability
and, more importantly, that {\it different parts of the IMF can be
changed readily without affecting other parts}. Note that this form is
a two-part power-law in the stellar regime, and that brown dwarfs
contribute about 4 per cent by mass only and need to be treated as a
separate population such that the IMF has a discontinuity near $m_{\rm
H}$ = 0.08 $M_\odot$ with $k^{'} \sim \frac{1}{3}$
\citep{KBD03,TK07,TK08}. A log-normal form below 1 $M_{\odot}$ with a
power-law extension to high masses was suggested by \citet{Ch03} but
is indistinguishable from the canonical form \citep{DHK08} and does
not cater for the discontinuity. The observed IMF is today understood
to be an invariant Salpeter/Massey power-law slope
\citep{Sal55,Mass03} above $0.5\, M_\odot$, being independent of the
cluster density and metallicity for metallicities $Z \simgreat 0.002$ 
\citep{MH98,SND00,SND02,PaZa01,Mass98,Mass02,Mass03,WGH02,BMK03,PBK04,PAK06}.

The basic assumption underlying our approach is the notion that all stars
in every cluster are drawn from this same universal parent IMF, which is
consistent with observational evidence \citep{Elme99,Kr01}.

It should be noted here that, while not indicated in eq.~\ref{eq:imf},
there is evidence of a maximal mass for stars \citep[$m_{\rm
    max*}\,\approx\,150\,M_{\odot}$,][]{WK04}, a result confirmed
by several independent studies \citep{OC05,Fi05,Ko06}, and that a
maximal-star-mass--star-cluster-mass relation, $m_{\rm max}(M_{\rm
  ecl}) \le m_{\rm max*}$, exists \citep{WK05b}.

\section{Special treatments to produce secondary stars}
\subsection{Special Pairing (SpP)}
\label{app:secon}
The following description has been developed in order to produce
multiple stars for which the mass-ratio increases with increasing
primary mass. Brown dwarfs ($m\,<\,0.08\,M_{\odot}$) are not included
as they are most likely to be treated as a separate population
\citep{TK07,TK08}.

\begin{figure}
\begin{center}
\includegraphics[width=8cm]{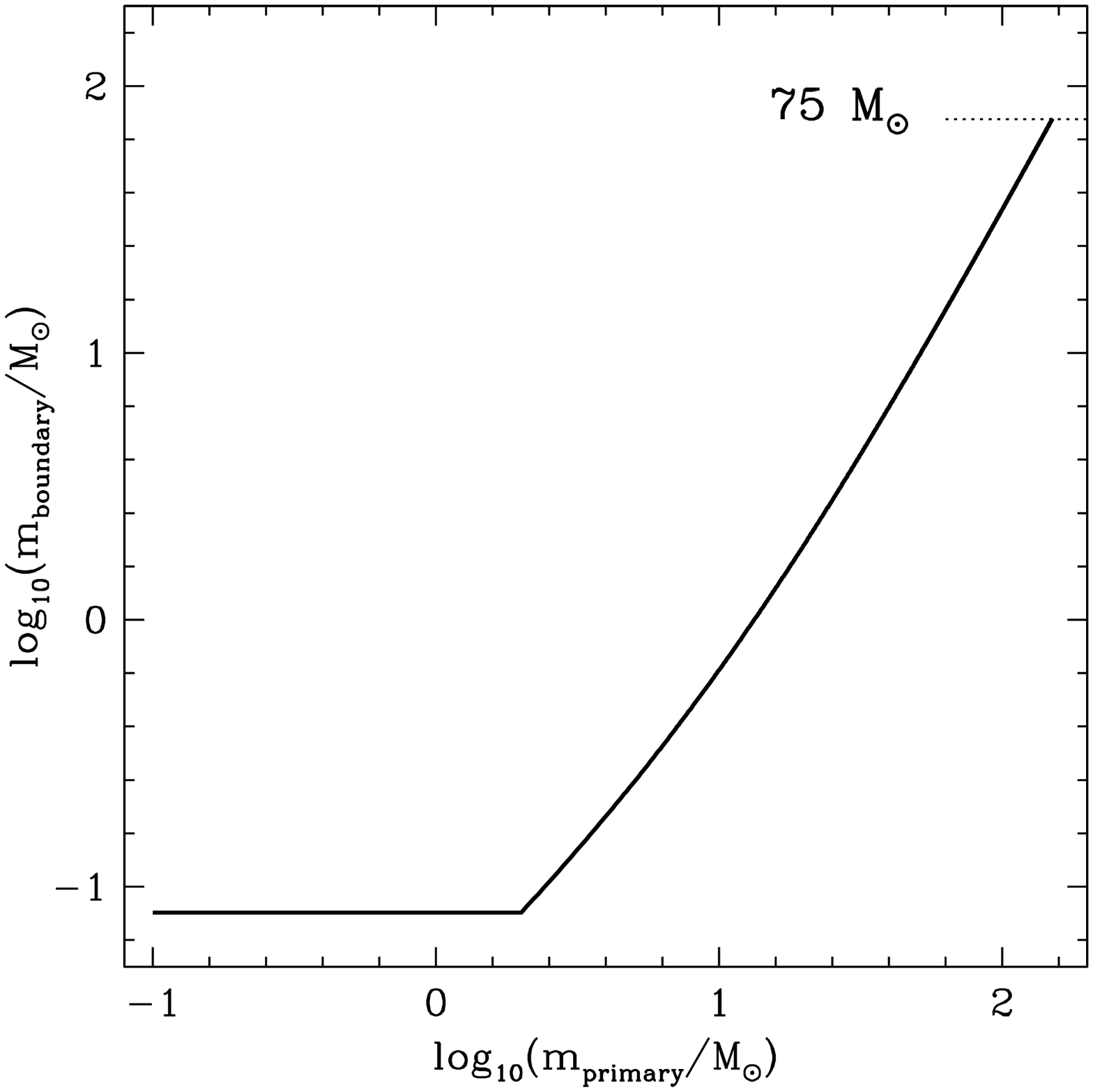}
\vspace*{-2.0cm}
\caption{The minimum possible secondary mass ($m_{\rm boundary}$) in
dependence of $m_{\rm primary}$, both in logarithmic units
(eq.~\ref{eq:boundhigh}). Indicated by a {\it dotted line} is the lower limit
for the secondary mass for a primary of 150 $M_{\odot}$.}
\label{fig:mseclow}
\end{center}
\end{figure}

In the regime above 2 $M_{\odot}$ the following quadratic relation was
chosen to define the range of mass ratios as a fraction of the primary star
mass, $m_{\rm primary}$: the least massive allowed companion, $m_{\rm
  boundary}$,

\begin{equation}
\label{eq:boundhigh}
  m_{\rm boundary} = 3.1 \cdot 10^{-3} \times
  m_{\rm primary}^{2} + 0.034 \times m_{\rm primary}.
\end{equation}
Companions are paired randomly within the so defined mass range. This
eq.~assures that primaries of up to 2 $M_{\odot}$ have a least-massive
companion of 0.08 $M_{\odot}$, while the least-massive companion is 75
$M_{\odot}$ for the most massive stars considered here (150
$M_{\odot}$). Fig.~\ref{fig:mseclow} shows the minimum possible secondary mass
($m_{\rm boundary}$) in dependence of $m_{\rm primary}$ resulting from
eq.~\ref{eq:boundhigh}.

\begin{figure}
\begin{center}
\includegraphics[width=8cm]{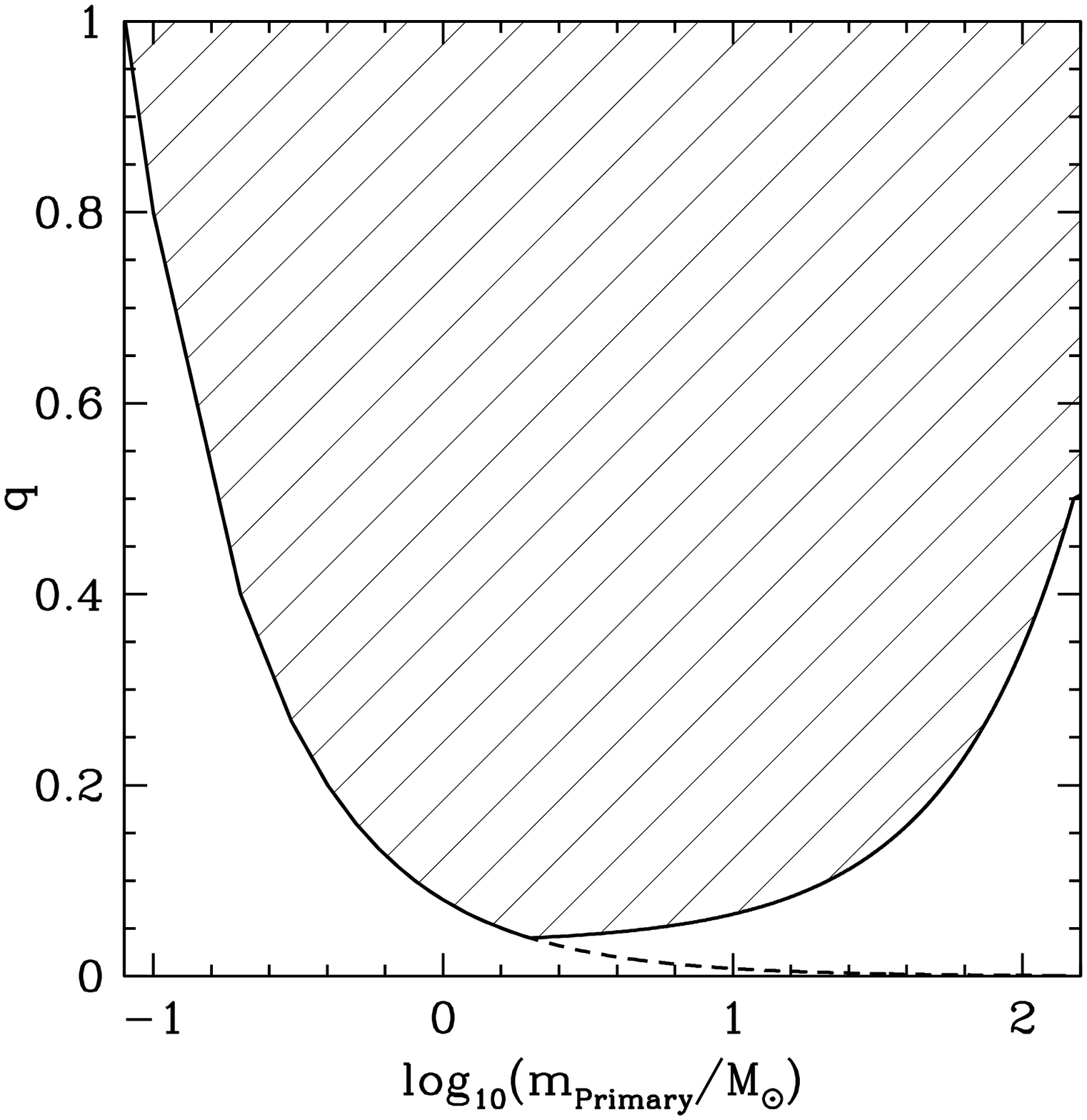}
\vspace*{-2.0cm}
\caption{The possible $q$-values ({\it shaded region}) in dependence of
  the primary mass (in logarithmic units) produced by the SpP procedure
  (Appendix~\ref{app:secon}). Inside the {\it shaded region} the
  companions are randomly drawn from the IMF. The {\it dashed line}
  shows the minimum possible $q$-value for random sampling over all
  masses with a companion minimum mass of 0.08 $M_{\odot}$.}
\label{fig:qenvelope}
\end{center}
\end{figure}

Stellar masses are created by randomly drawing all stars from the
IMF. Binaries are then made from this set of stars as described in
\S~\ref{sub:pair}.

The combined range of possible $q$-values for eq.~\ref{eq:boundhigh} and
random pairing for stars below 2 $M_{\odot}$ is shown in
Fig.~\ref{fig:qenvelope}. The {\it solid line} marks the minimum
allowed mass ratio. The companion masses lie within the {\it shaded region}.

\begin{figure}
\begin{center}
\includegraphics[width=8cm]{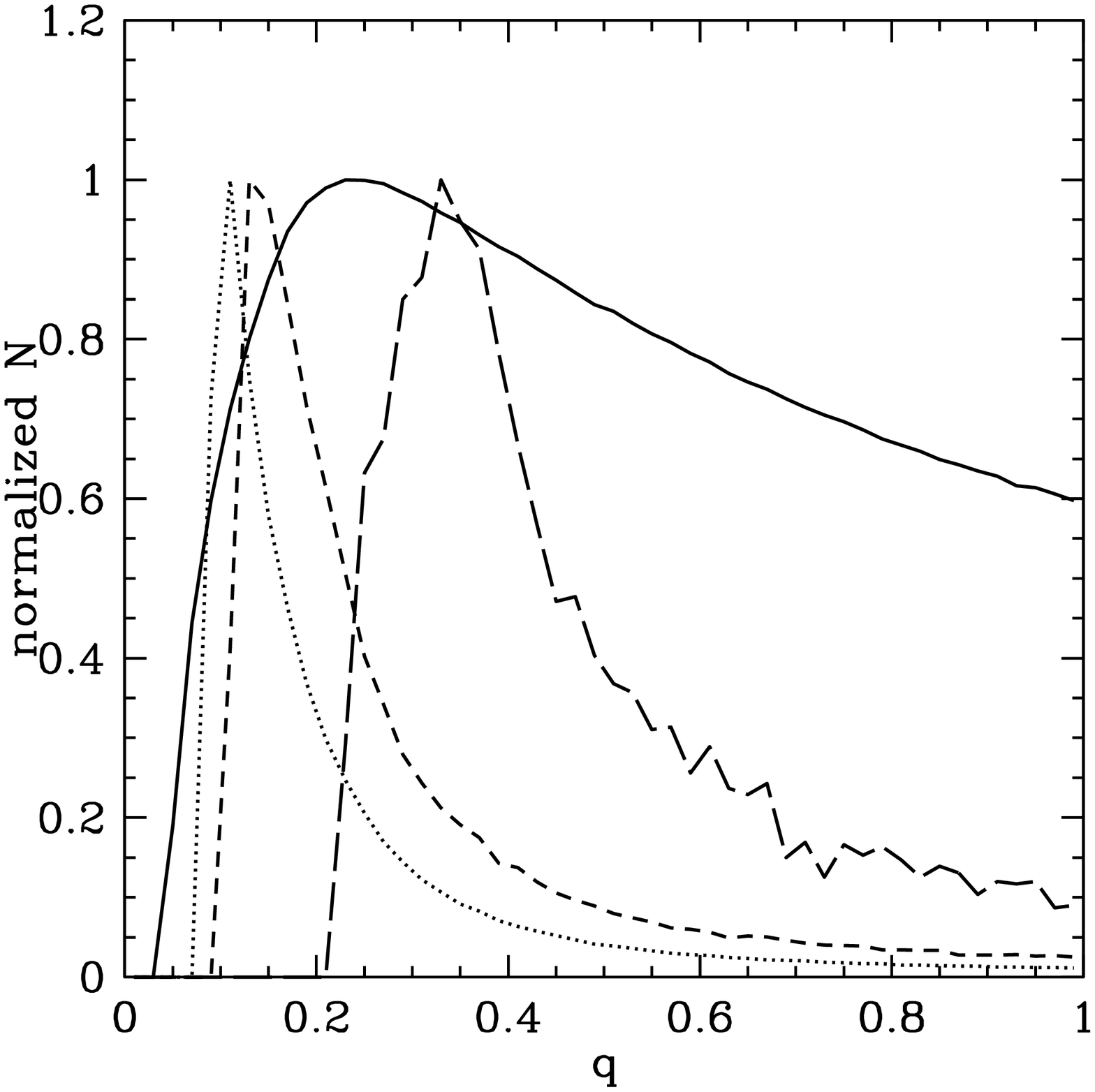}
\vspace*{-2.0cm}
\caption{The mass-ratio ($q$) distributions for primaries in several
  mass intervals for the SpP model. {\it Solid line:} $m_{\rm
    primary}$ from 0.08 to 2 $M_\odot$, {\it dotted line:} 2 - 10
  $M_\odot$, {\it short-dashed line:} 10 - 50 $M_\odot$ and {\it
    long-dashed line:} 50 - 100 $M_\odot$.}
\label{fig:qspecialII}
\end{center}
\end{figure}

Fig.~\ref{fig:qspecialII} shows the resulting $q$-distribution for 4
different mass bins. While the lowest mass bin (0.08 - 2 $M_\odot$) shows a
flat distribution resulting from random pairing the other three bins are
peaked towards the lower edge of allowed mass-ratios. This is due to the steep
power-law slope of the canonical IMF, which makes it highly likely that the
chosen companion for a massive star is close to the lower limit $m_{\rm
  boundary}$.

\subsection{System Pairing (SyP)}
\label{app:fourth}
In order to obtain systems with components which are strongly biased towards
equal mass systems, the SyP scheme is motivated by the possible merger origin
of massive stars and may also by applicable for the competitive accretion
scenario. System masses are drawn randomly from the adopted fixed system IMF
and then split into companions as follows:\\

For $m_{\rm system} > 10 M_{\odot}$:
\begin{eqnarray}
q&=& \frac{m_{\rm system} \times R}{10};\\
{\rm if~q}~>~{\rm 0.9:} q&=& (0.1 \times R) + 0.9;\\
m_{\rm primary}&=&\frac{m_{\rm system}}{1 + q};\\
m_{\rm companion}&=& m_{\rm system} - m_{\rm primary},
\end{eqnarray}
where $R$ is a random number between 0 and 1. This formulation biases the mass
ratio distribution for massive stars strongly towards more massive companions
while still allowing for occasional low-mass secondaries. Systems less-massive
than 10 $M_\odot$ are split randomly in two or more halves with the
limitation that the lower limit for the stellar mass of the companions is 0.08
$M_\odot$.

The IMF for the systems is the same as in \S~\ref{app:IMF} but with different
mass intervals. The lowest {\it system mass} allowed is 0.16 $M_\odot$ and the
highest 300 $M_\odot$ and the exponent changes at 1 $M_\odot$ instead of 0.5
$M_\odot$. But for the {\it stars} produced from these systems the lower and
upper mass limits are still 0.08 and 150 $M_\odot$.

\begin{figure}
\begin{center}
\includegraphics[width=8cm]{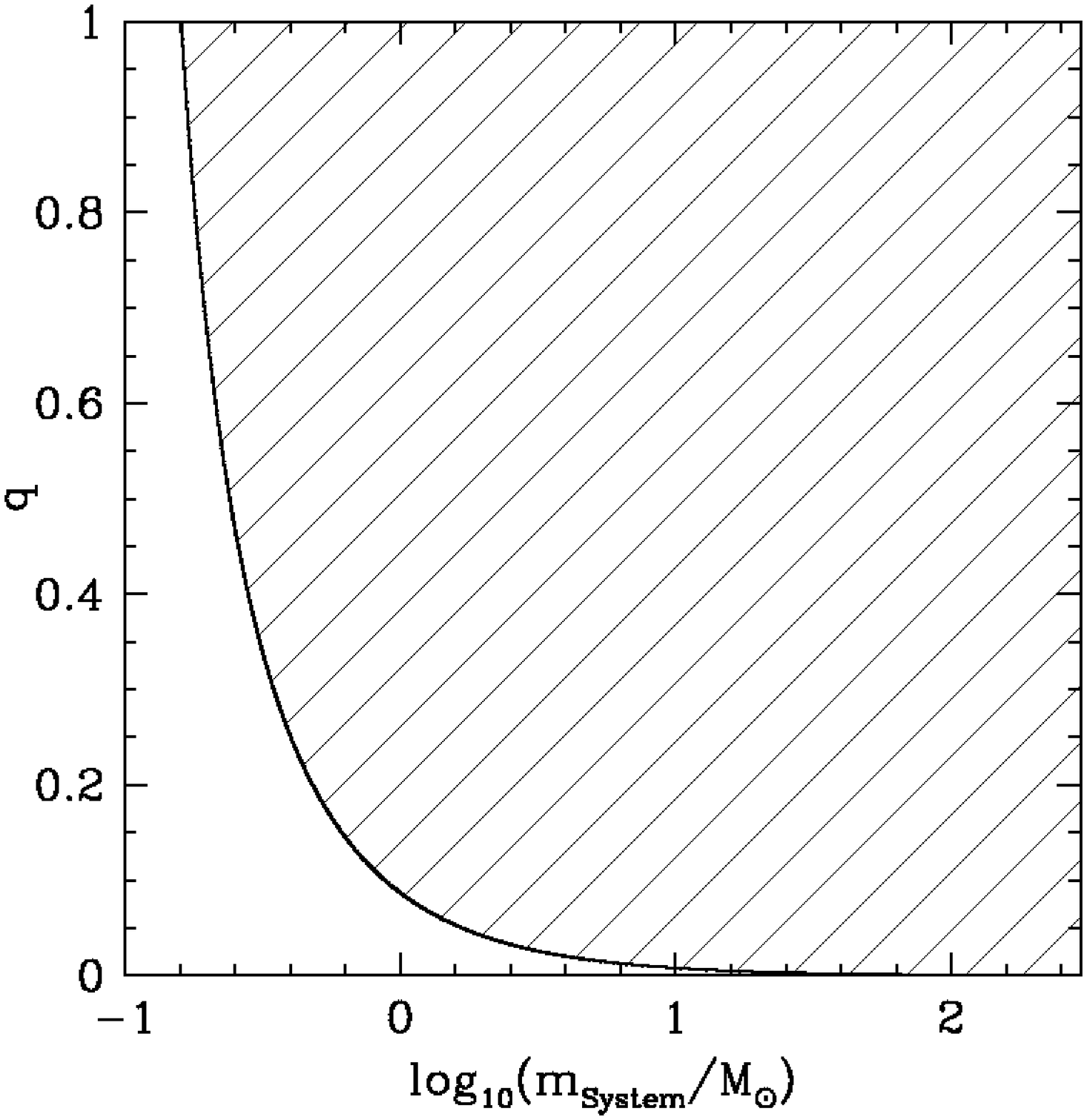}
\vspace*{-2.0cm}
\caption{The possible $q$-values ({\it shaded region}) in dependence of
  the system mass (in logarithmic units) produced by the SyP
  procedure (Appendix~\ref{app:fourth}). But within the {\it shaded 
  region} the $q$-values are only randomly distributed for $m_{\rm
  system} < 10 M_{\odot}$. Above this mass the $q$-values are from
  distributions as shown in Fig.~\ref{fig:qsystemII}.}
\label{fig:qenvelopeII}
\end{center}
\end{figure}

\begin{figure}
\begin{center}
\includegraphics[width=8cm]{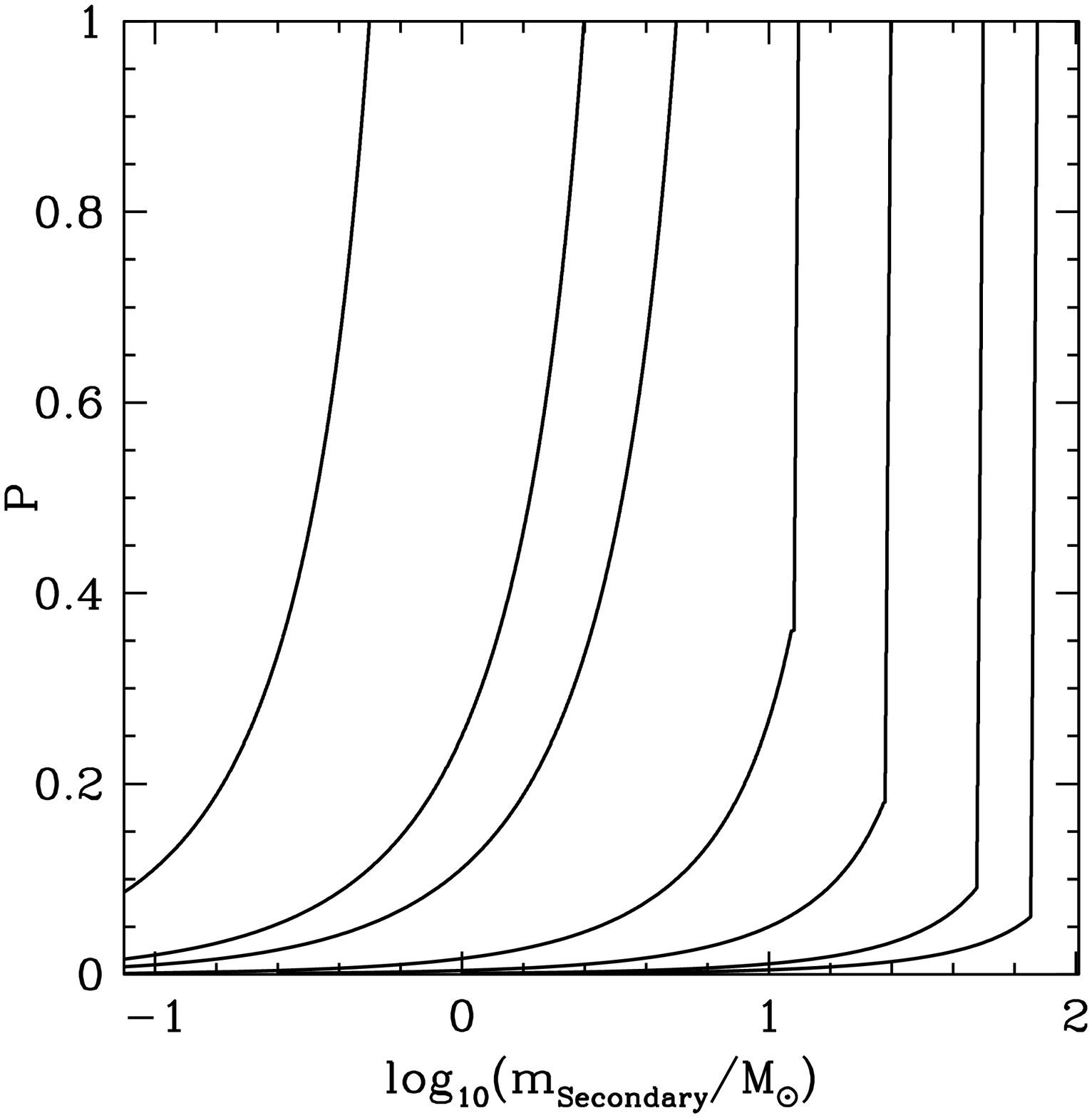}
\vspace*{-2.0cm}
\caption{The probability $P$ to pick the companion mass between
the lower limit (0.08 $M_{\odot}$) and half of the system mass for 
system masses of (from left to right ) 1, 5, 10, 25, 50, 100 and 150
$M_{\odot}$.}
\label{fig:psysII}
\end{center}
\end{figure}

The combined range of possible $q$-values is shown in
Fig.~\ref{fig:qenvelopeII} while the probabilities for the companion 
masses for a few examples are shown in Fig.~\ref{fig:psysII}. These
probabilities are chosen randomly from a Gaussian distribution between $m_{\rm
boundary}$ and the maximum possible mass for the secondary, $m_{\rm
sec, max}$, which is $\frac{m_{\rm system}} {2}$. The mean of the
Gaussian is $m_{\rm sec, max}$ and its dispersion is given by $\sigma
= m_{\rm sec, max} / (0.064 m_{\rm system} + 3.29)$.

\begin{figure}
\begin{center}
\includegraphics[width=8cm]{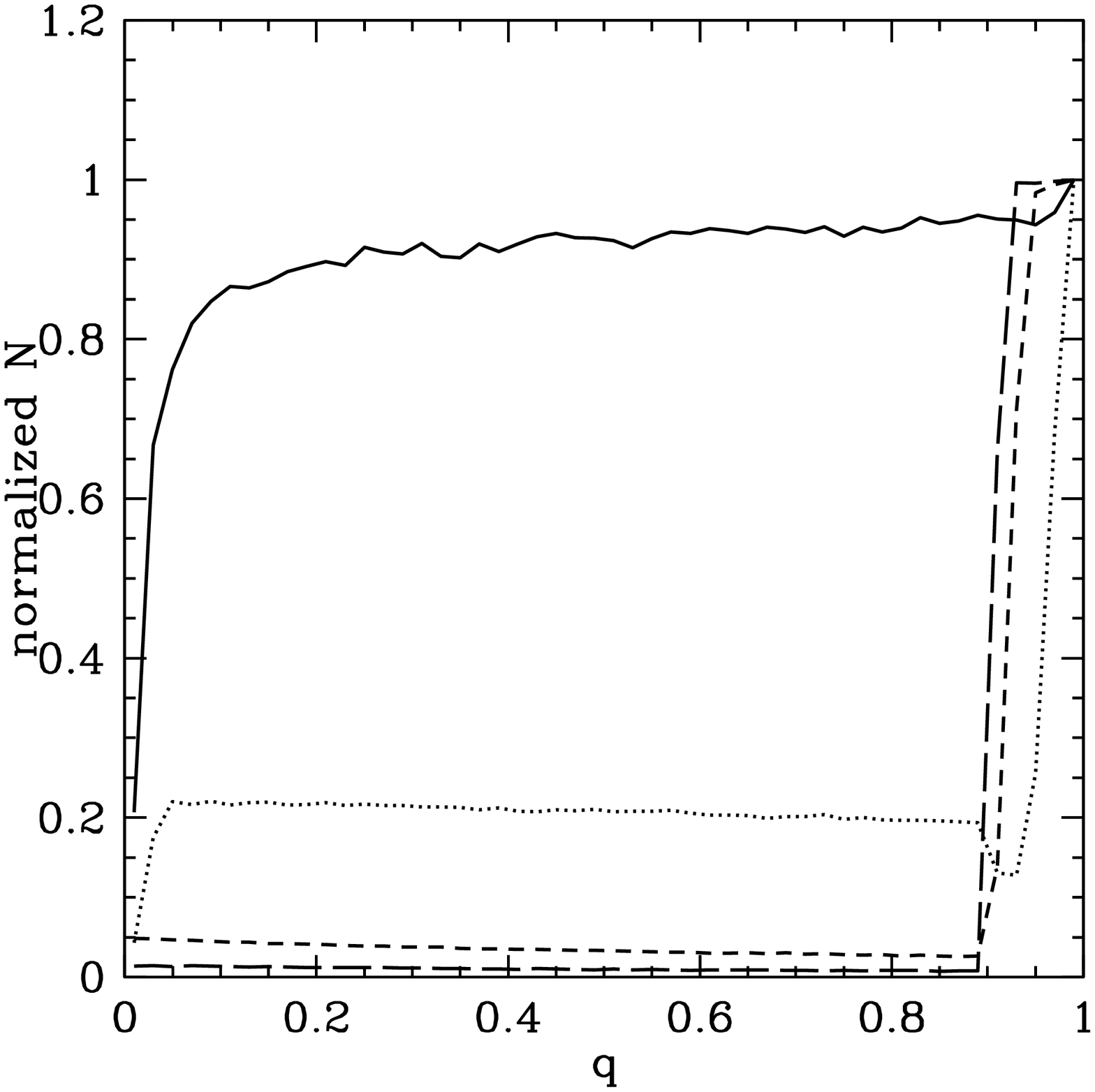}
\vspace*{-2.0cm}
\caption{The mass-ratio ($q$) distributions for primaries in several
  mass intervals (Appendix~\ref{app:fourth}). {\it Solid line:}
  $m_{\rm primary}$ from 0.08 to 2 $M_\odot$, {\it dotted line:} 2 -
  10 $M_\odot$, {\it short-dashed line:} 10 - 50 $M_\odot$ and {\it
  long-dashed line:} 50 - 100 $M_\odot$. The mass-ratio distribution for
  low-mass stars ({\it solid line}) is not completely flat due to the lower
  mass limit for stars of 0.08 $M_{\odot}$.}
\label{fig:qsystemII}
\end{center}
\end{figure}

In Fig.~\ref{fig:qsystemII} the $q$-distribution is shown
for 4 different mass bins. The three bins with stars above 10 $M_{\odot}$ peak
strongly above $q$ = 0.8, indicating the strong tendency to produce nearly
equal mass binaries with this algorithm.

\clearpage
\section{IMFs without stellar evolution}
\label{app:imfnoevo}
\begin{figure*}
\begin{center}
\includegraphics[width=8cm]{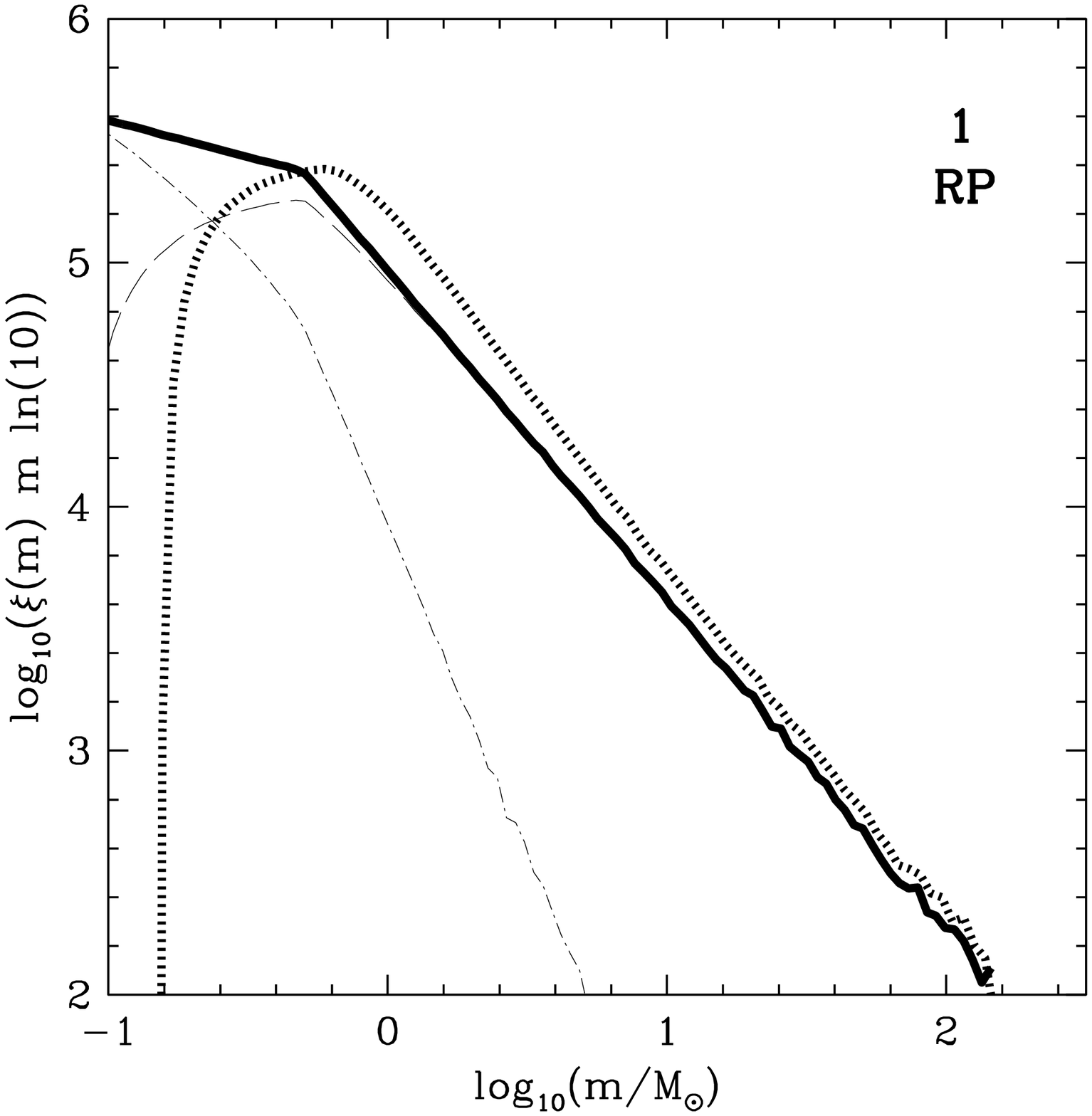}
\includegraphics[width=8cm]{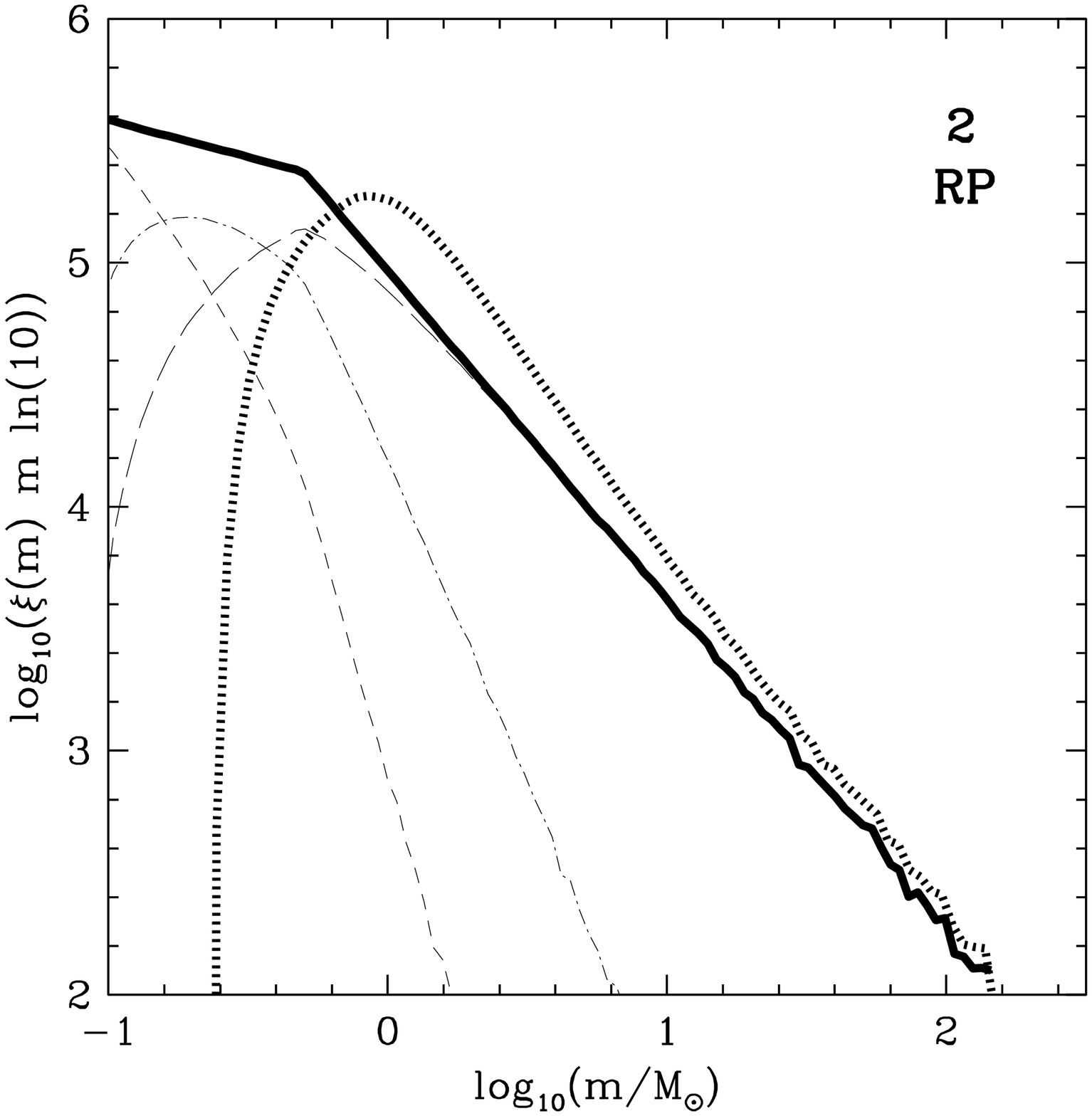}

\vspace*{-2.5cm}

\includegraphics[width=8cm]{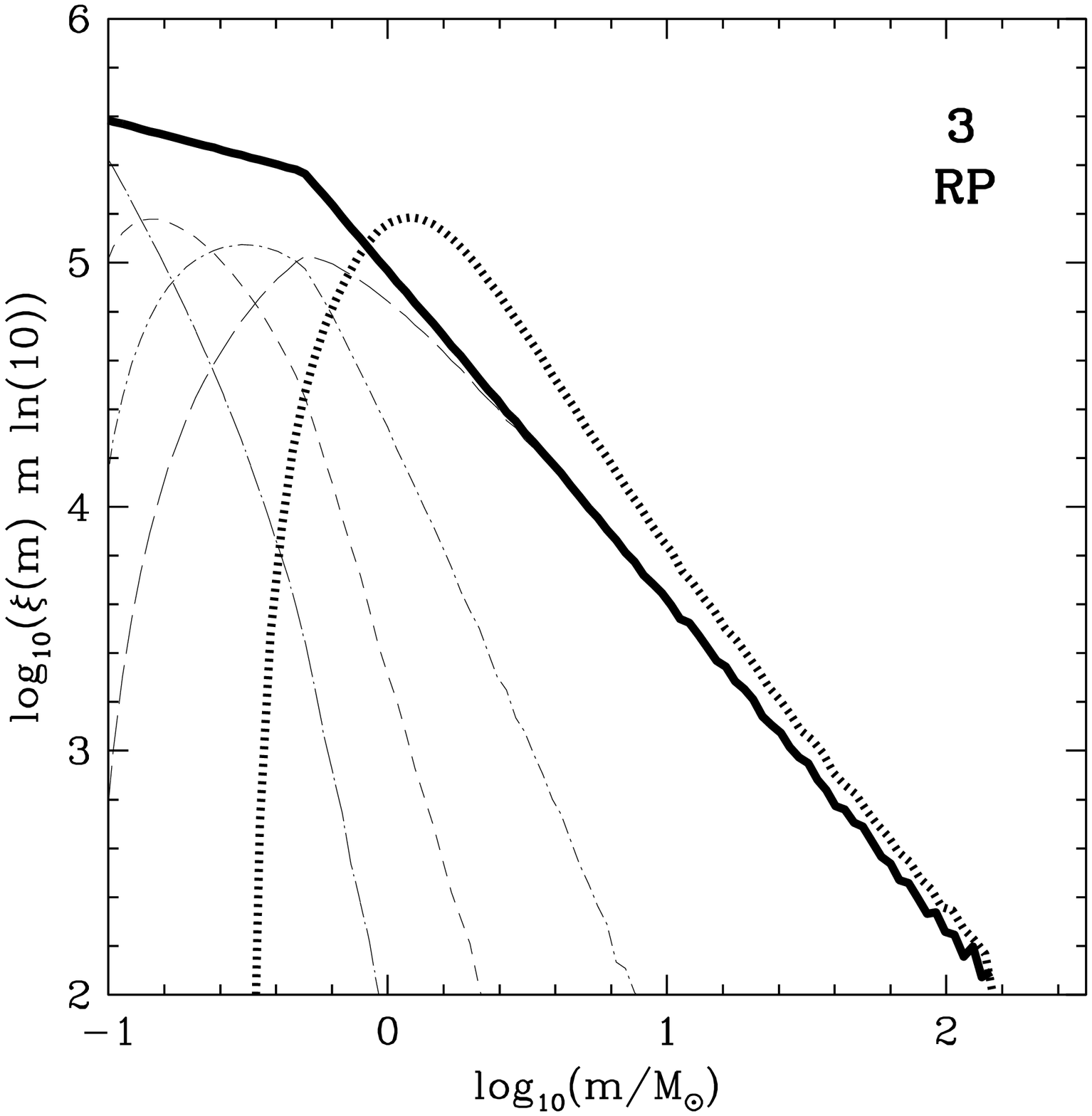}
\includegraphics[width=8cm]{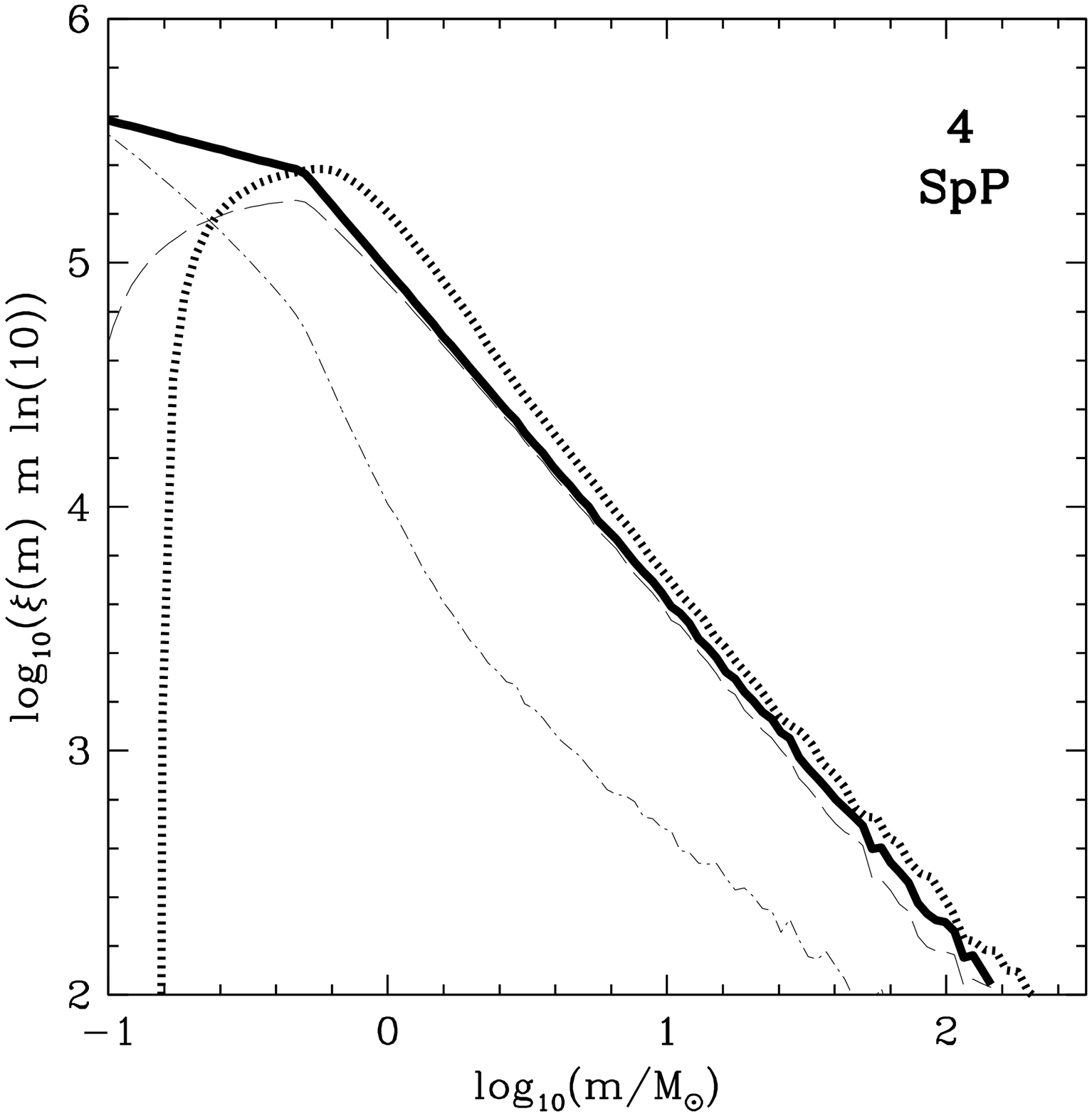}
\vspace*{-2.0cm}
\caption{IMFs of Models 1 to 4. The {\it thick solid lines} are the
  IMFs of all stars, the {\it thin solid lines} are the input IMFs
  (identical to the IMF of all stars for models 1 to 6), the {\it
  thick dotted lines} the system IMFs and the {\it thin lines}
  represent the mass functions for the individual components. The {\it
  thin long-dashed lines} are the primary star IMFs, the {\it thin
  dash-dotted lines} are the secondary star IMFs, the {\it thin
  short-dashed lines} are the tertiary star IMFs (only panels 2, 3, 5,
  6, 8 and 9) and the {\it thin long-dash-dotted lines} are the
  quartiary star IMFs (only panels 3, 6 and 9). The ``spikes'' in the
  system IMFs ({\it thick dotted lines}) in panels 4 to 10 are a
  result of changing the $q$-distribution at 2 $M_{\odot}$ .} 
\label{fig:imf04}
\end{center}
\end{figure*}

\begin{figure*}
\begin{center}
\includegraphics[width=8cm]{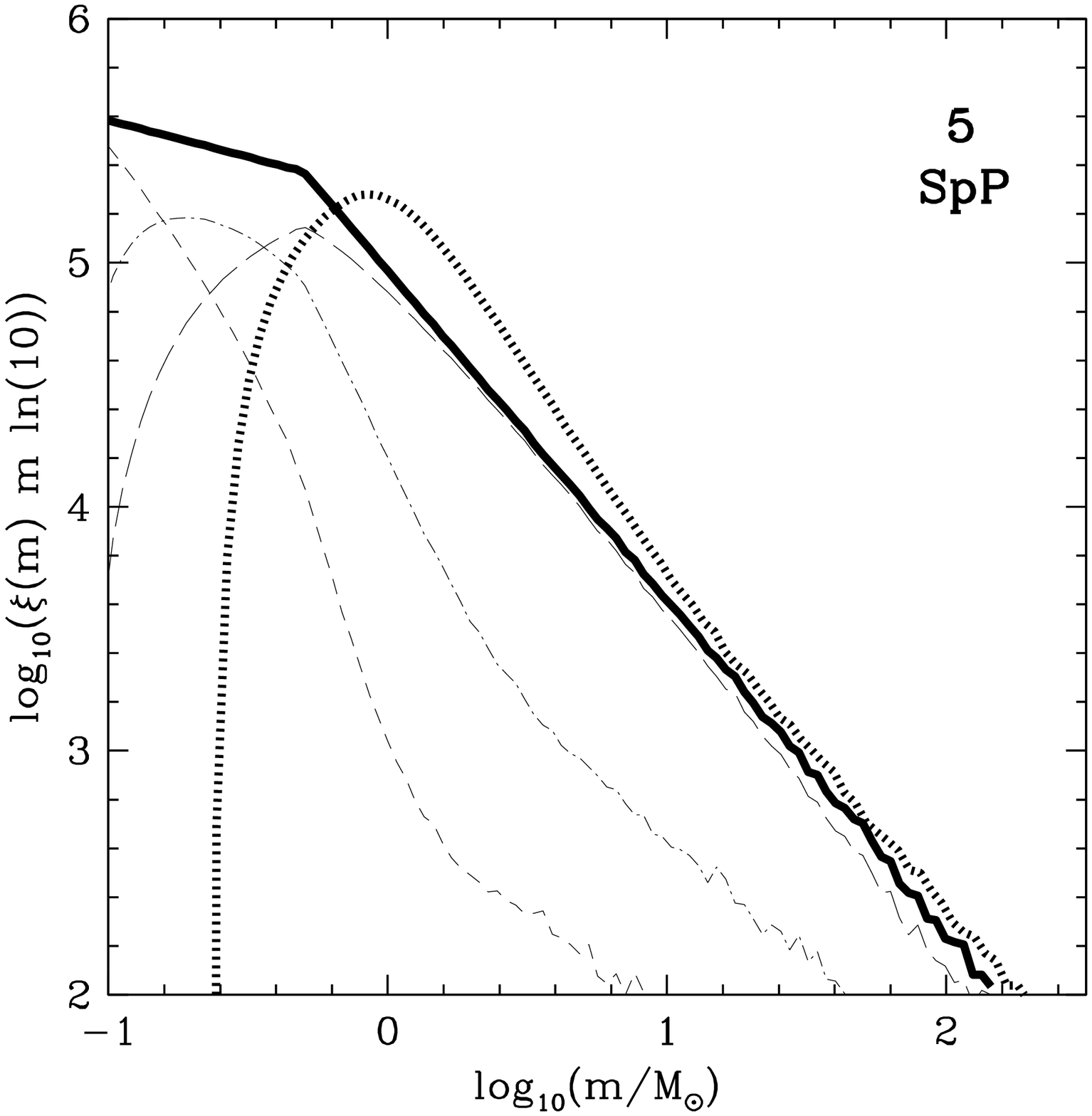}
\includegraphics[width=8cm]{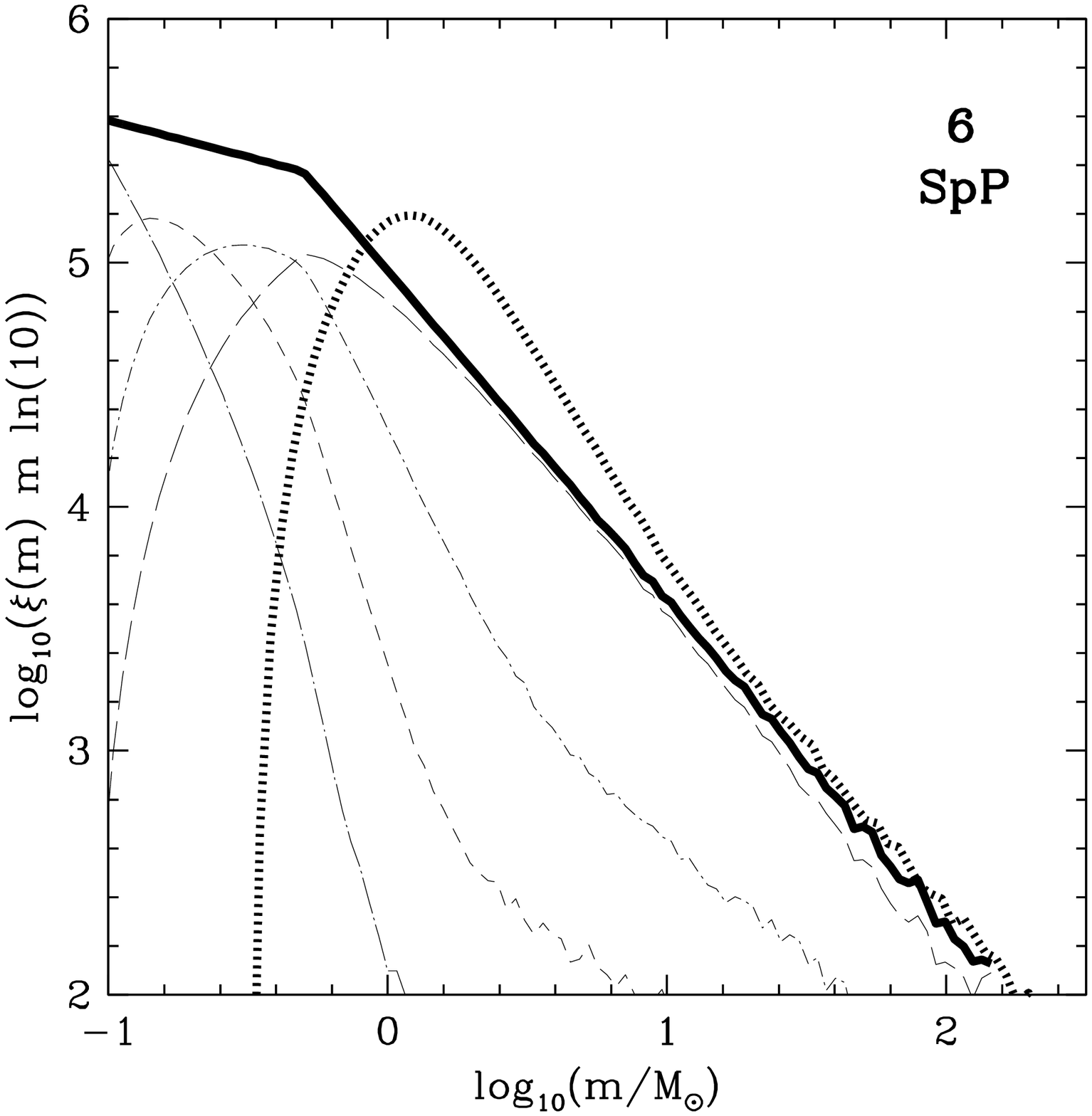}

\vspace*{-2.5cm}

\includegraphics[width=8cm]{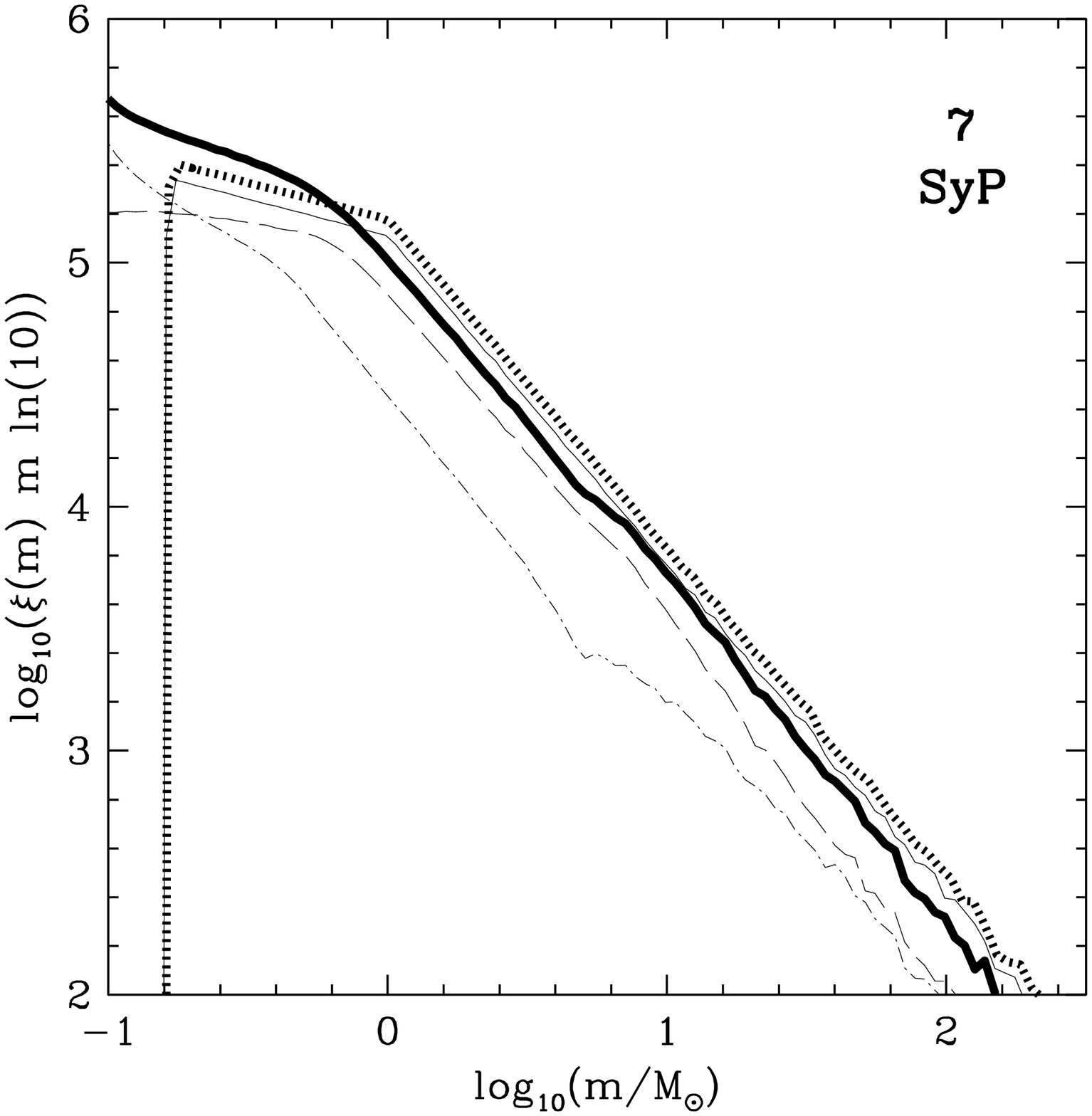}
\includegraphics[width=8cm]{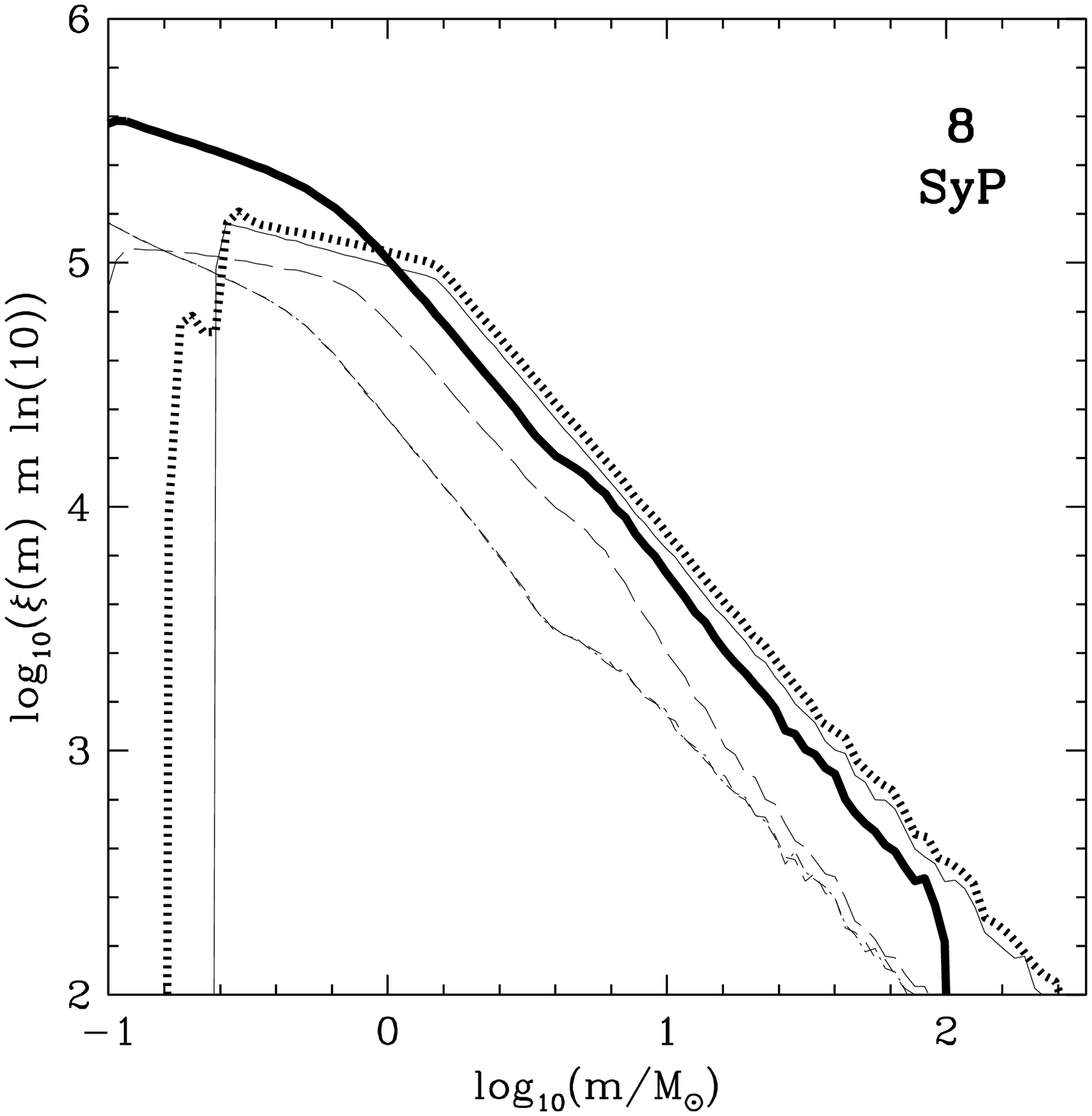}
\vspace*{-2.0cm}
\caption{Like Fig.~\ref{fig:imf04} but for the Models 5 to 8 from
  Table~\ref{tab:noevo}.}
\label{fig:imf08}
\end{center}
\end{figure*}

\begin{figure*}
\begin{center}
\includegraphics[width=8cm]{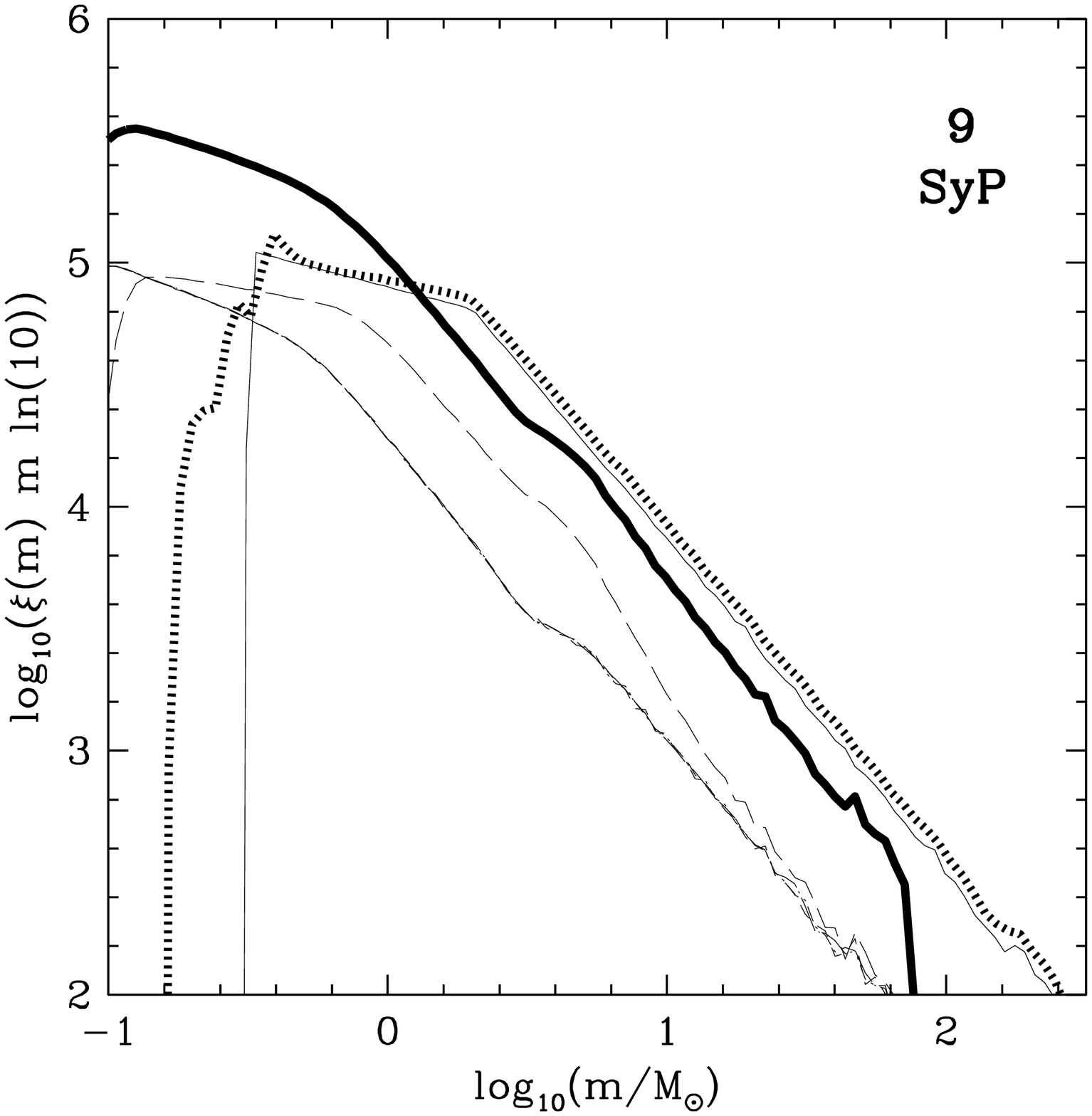}
\includegraphics[width=8cm]{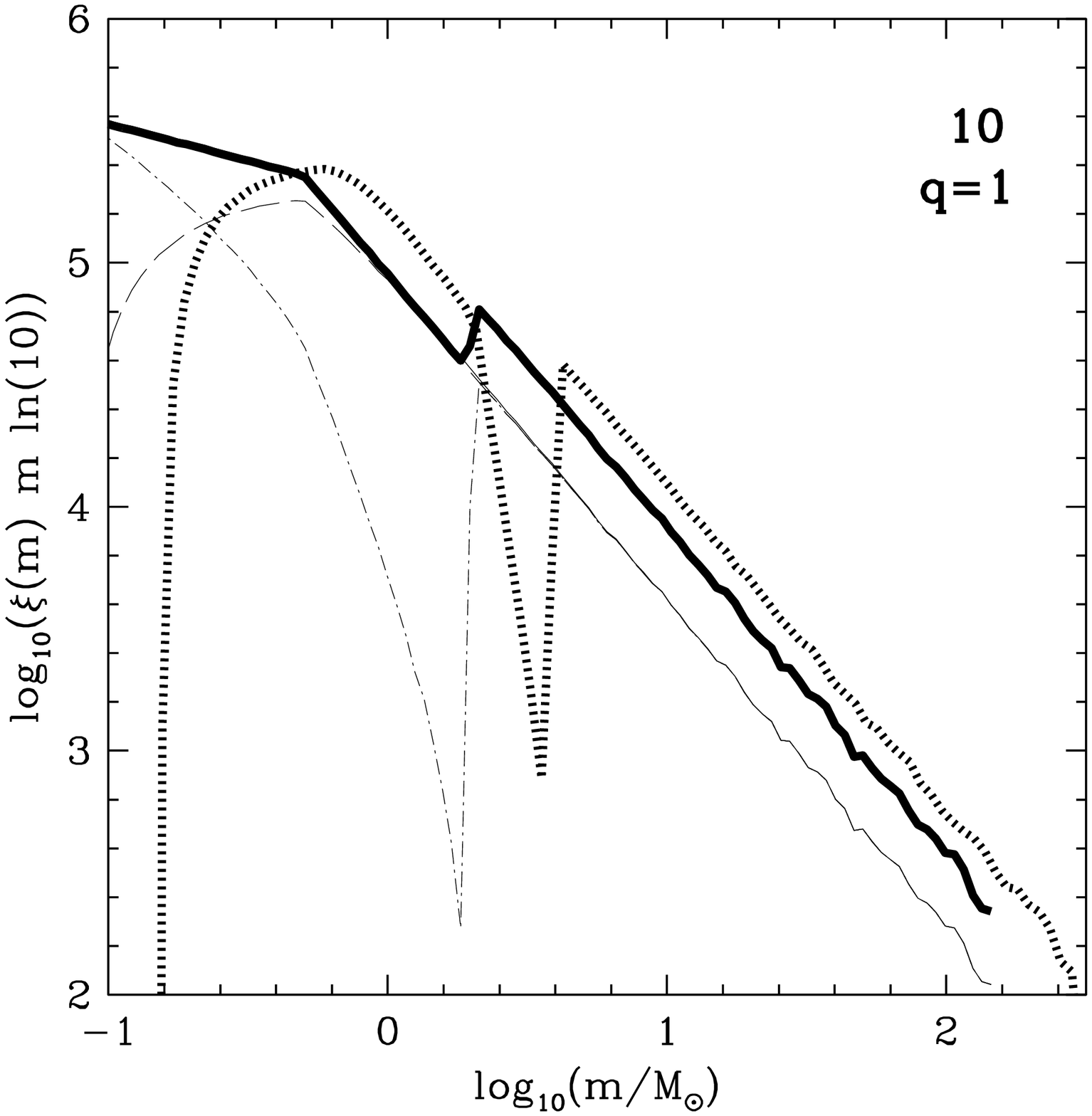}
\vspace*{-2.0cm}
\caption{Like Fig.~\ref{fig:imf04} but for the Models 9 and 10 from
  Table~\ref{tab:noevo}.}
\label{fig:imf12}
\end{center}
\end{figure*}

\clearpage
\section{IMFs with stellar evolution}
\label{app:imfevo}
\begin{figure*}
\begin{center}
\includegraphics[width=8cm]{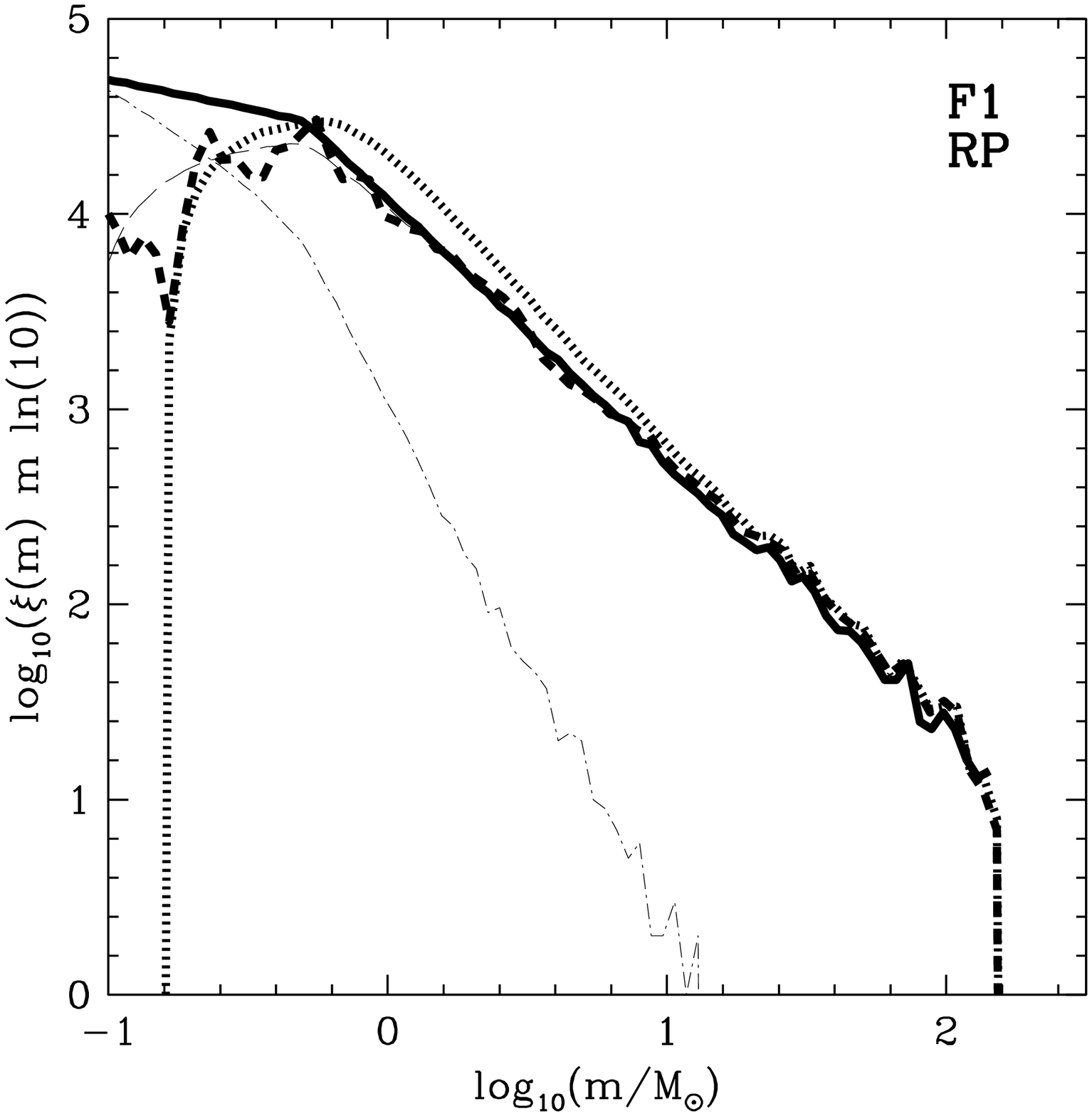}
\includegraphics[width=8cm]{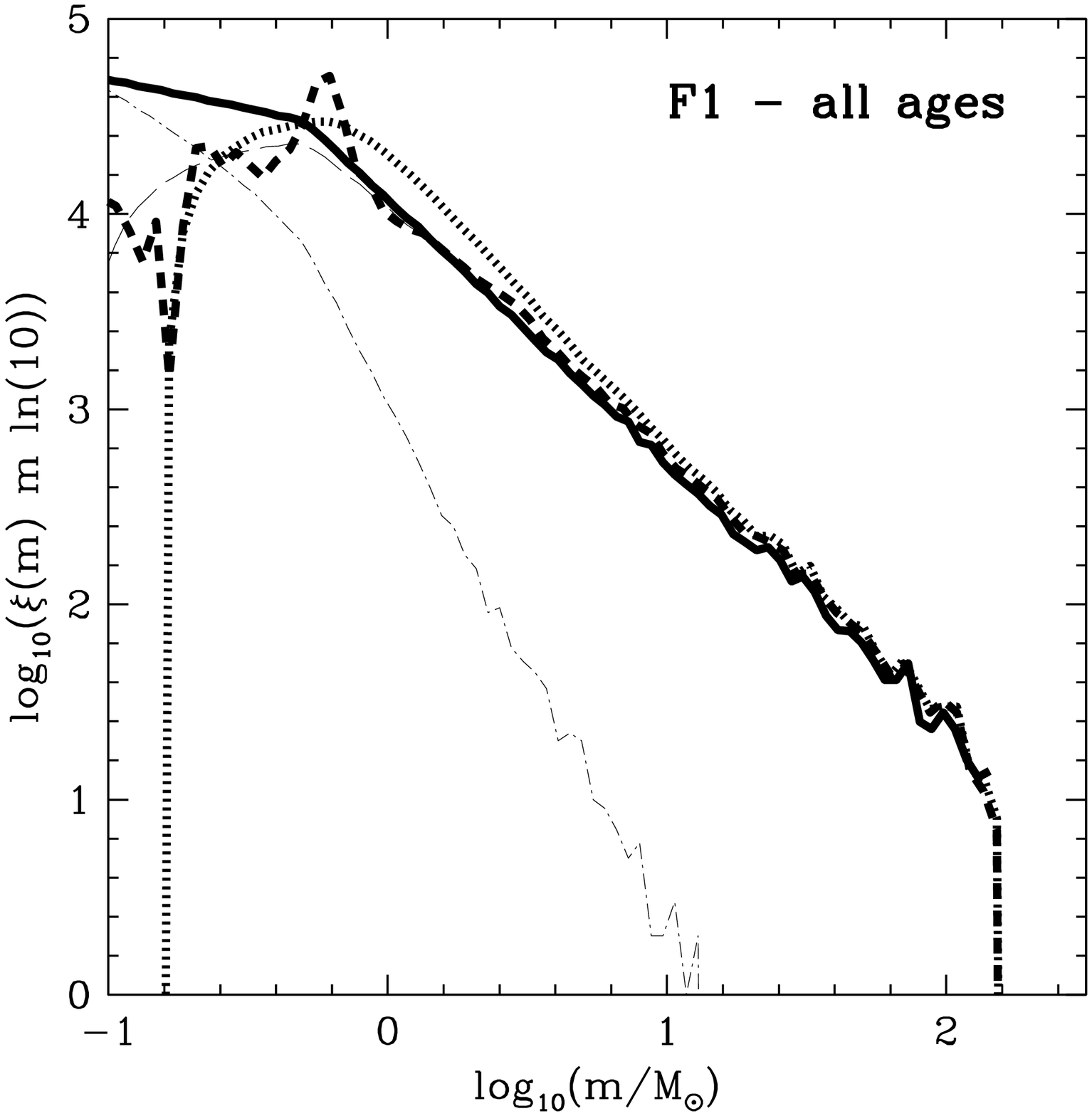}

\vspace*{-2.5cm}

\includegraphics[width=8cm]{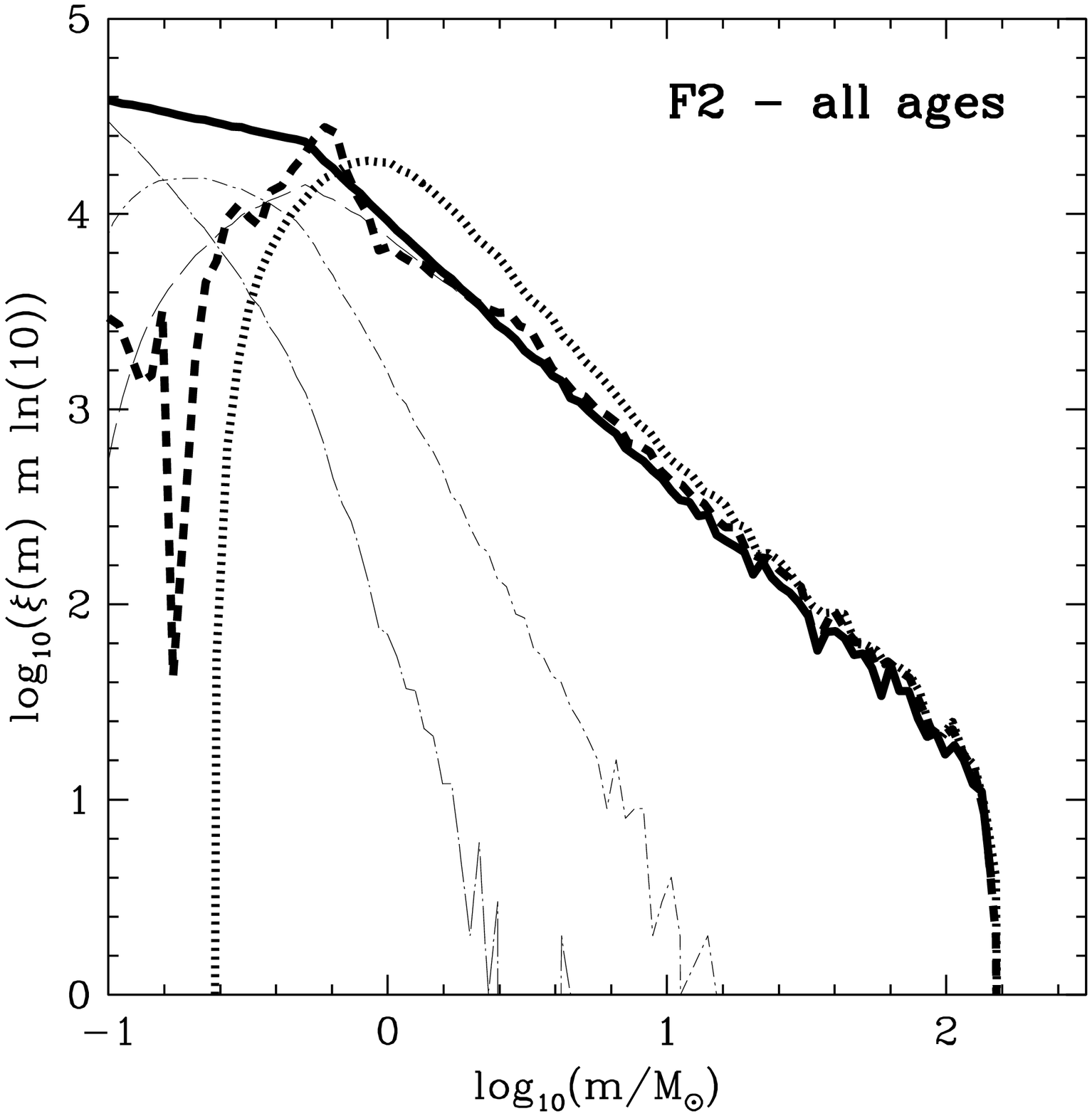}
\includegraphics[width=8cm]{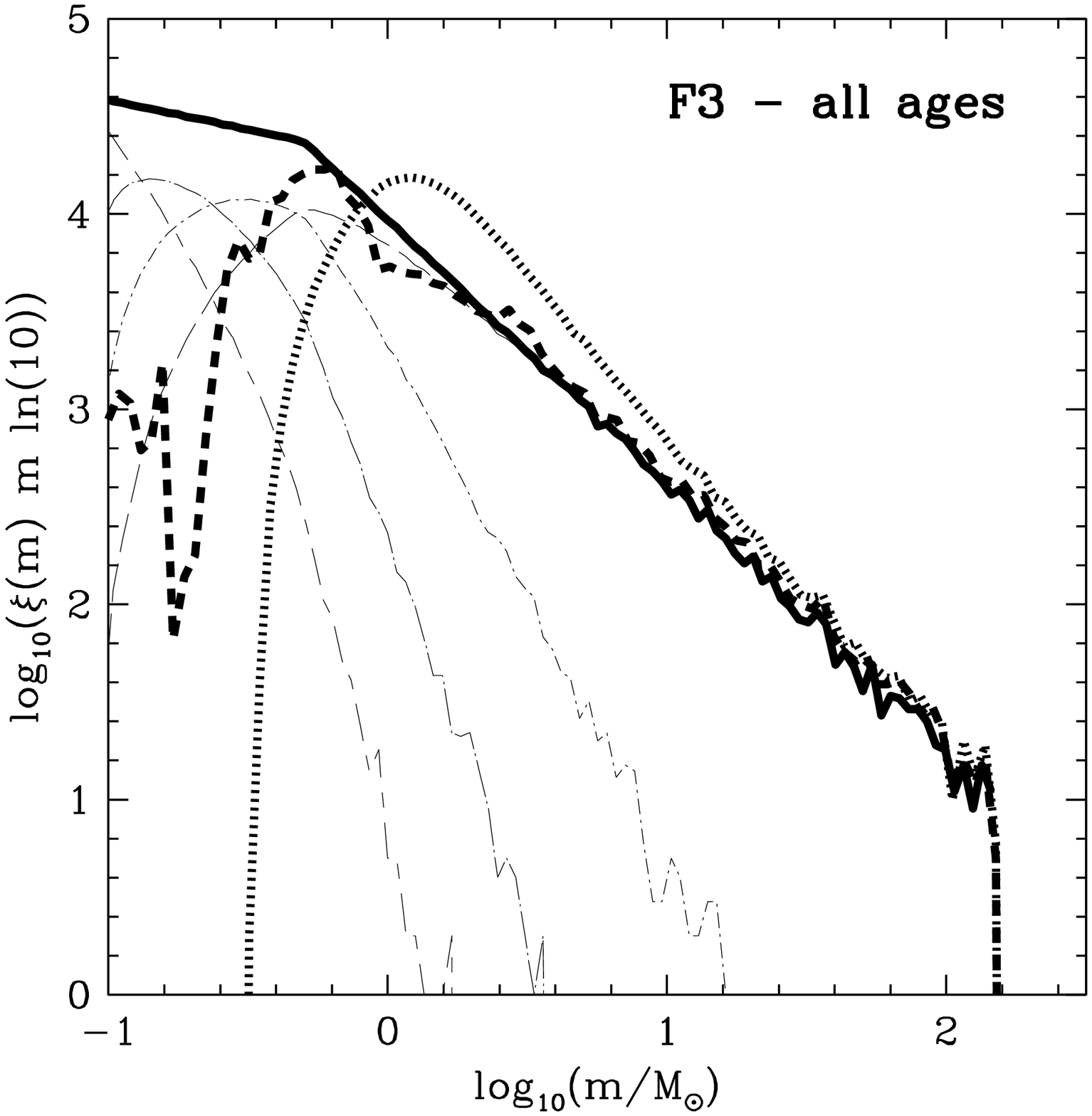}
\vspace*{-2.0cm}
\caption{IMFs of Models Fit 1 to Fit 3 from Table~\ref{tab:evo}. {\it
  Thick solid lines} are the input IMFs, the {\it thick dotted lines}
  are the system IMFs without stellar evolution, the {\it thick
  short-dashed lines} are the ``observed'' IMFs, the {\it long-dashed
  lines} are the primary star IMFs, the {\it dash-dotted lines}
  are the secondary star IMFs, the {\it thin short-dashed lines} are the
  tertiary star IMFs (only Models Fit 2, 3, 5, 6, 8 and 9) and the {\it thin
  long-dash-dotted lines} are the quartiary star IMFs (only Models Fit 3, 6
  and 9). For the first panel (``F1 RP'') the ``observed'' IMF is created only
  from recovered stars with pseudo ages less than 5 Myr. For each of the other
  three panels all recovered stars regardless of their ages are used. These
  panels are marked by ``all ages''. Note the dip in all models around
  $\log_{10}(m) \approx$ -0.8 (about 0.16 $M_\odot$). It is the result of
  differences in the PMS tracks from different authors used here. These
  differences shift some of the stars between 0.15 and 0.3 $M_\odot$ to lower
  masses but have no influence on the results of this study as we restrain
  ourselves only to the high-mass end of the IMF.}
\label{fig:Fit1to4}
\end{center}
\end{figure*}

\begin{figure*}
\begin{center}
\includegraphics[width=8cm]{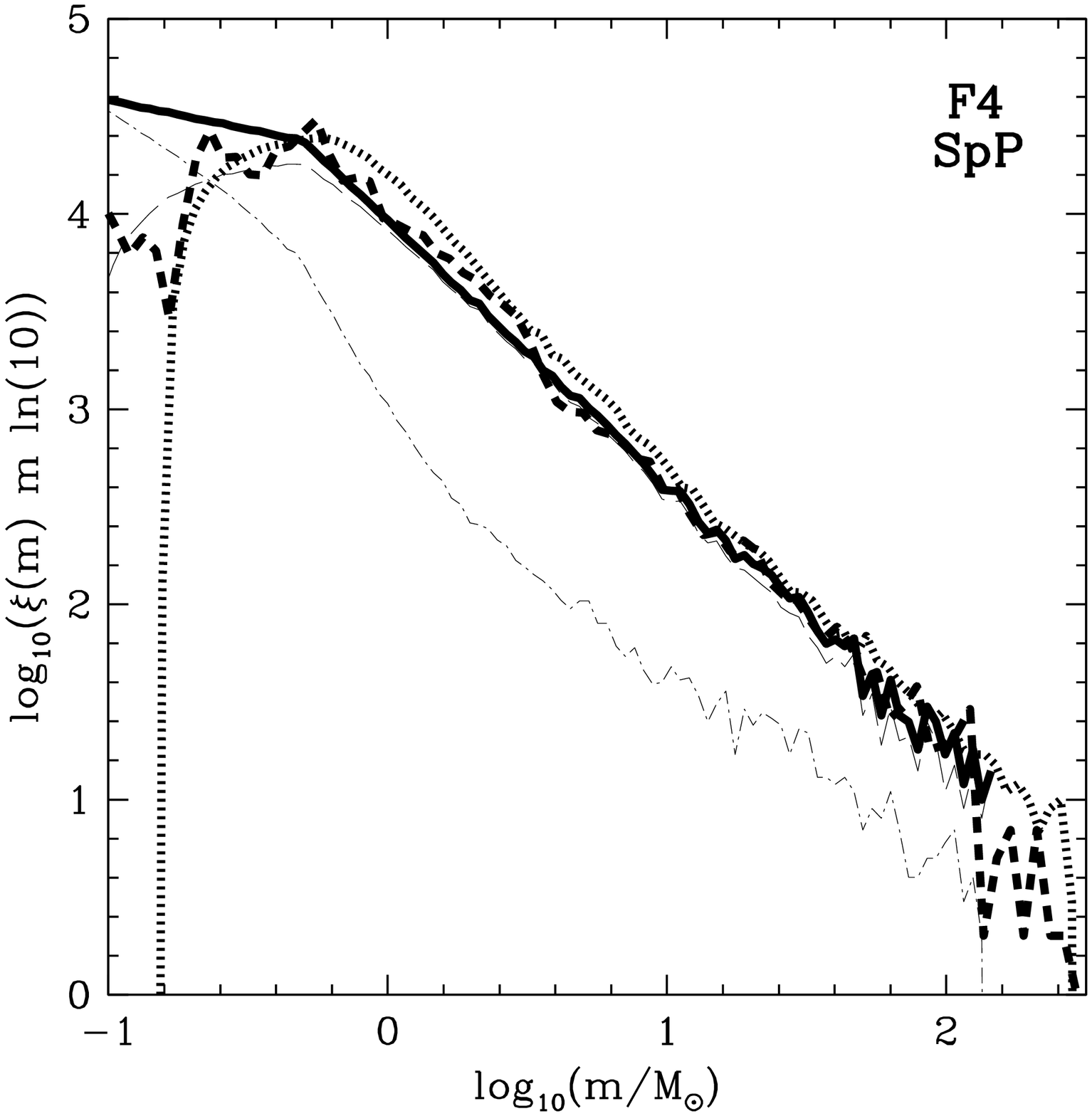}
\includegraphics[width=8cm]{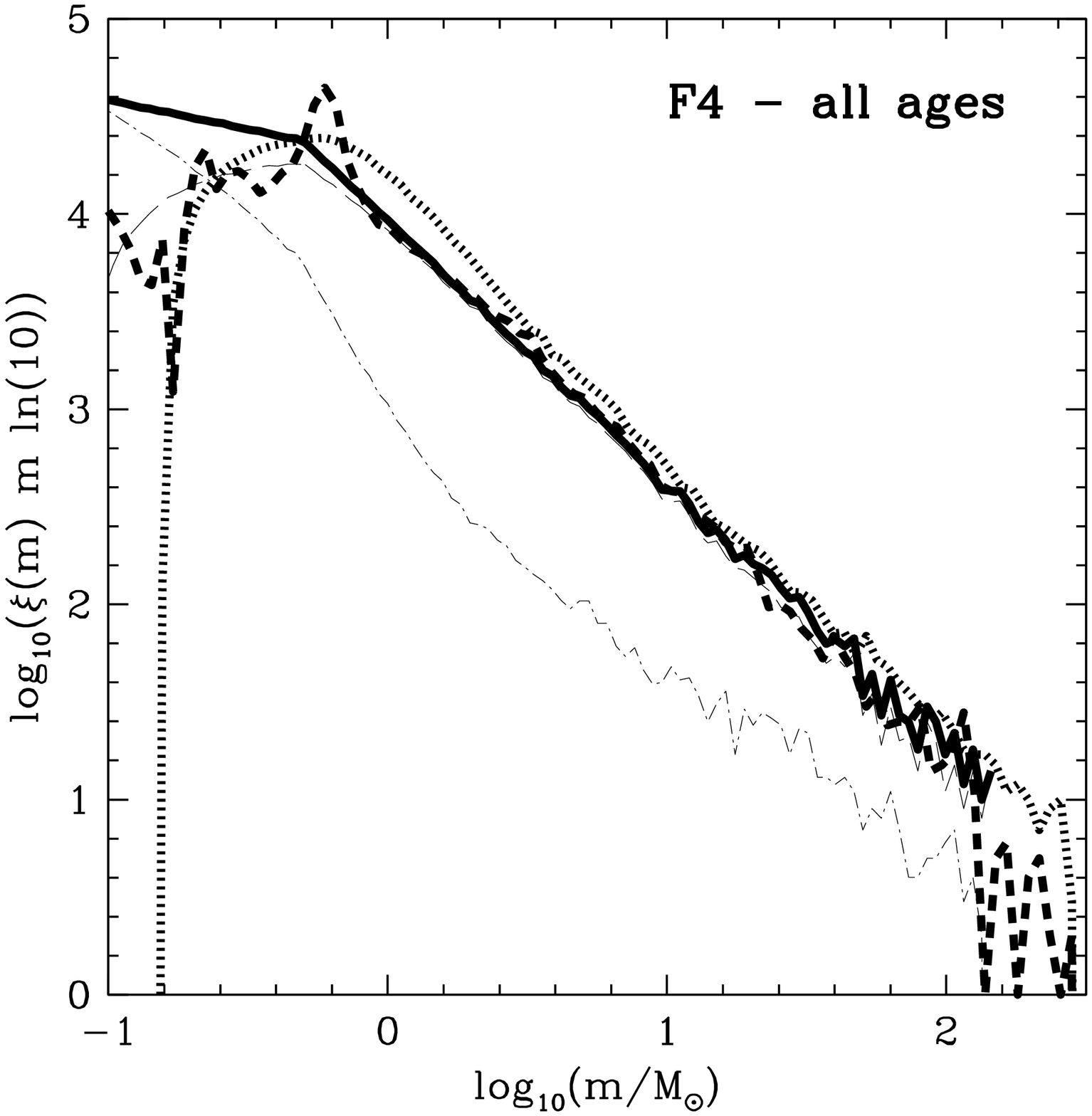}

\vspace*{-2.5cm}

\includegraphics[width=8cm]{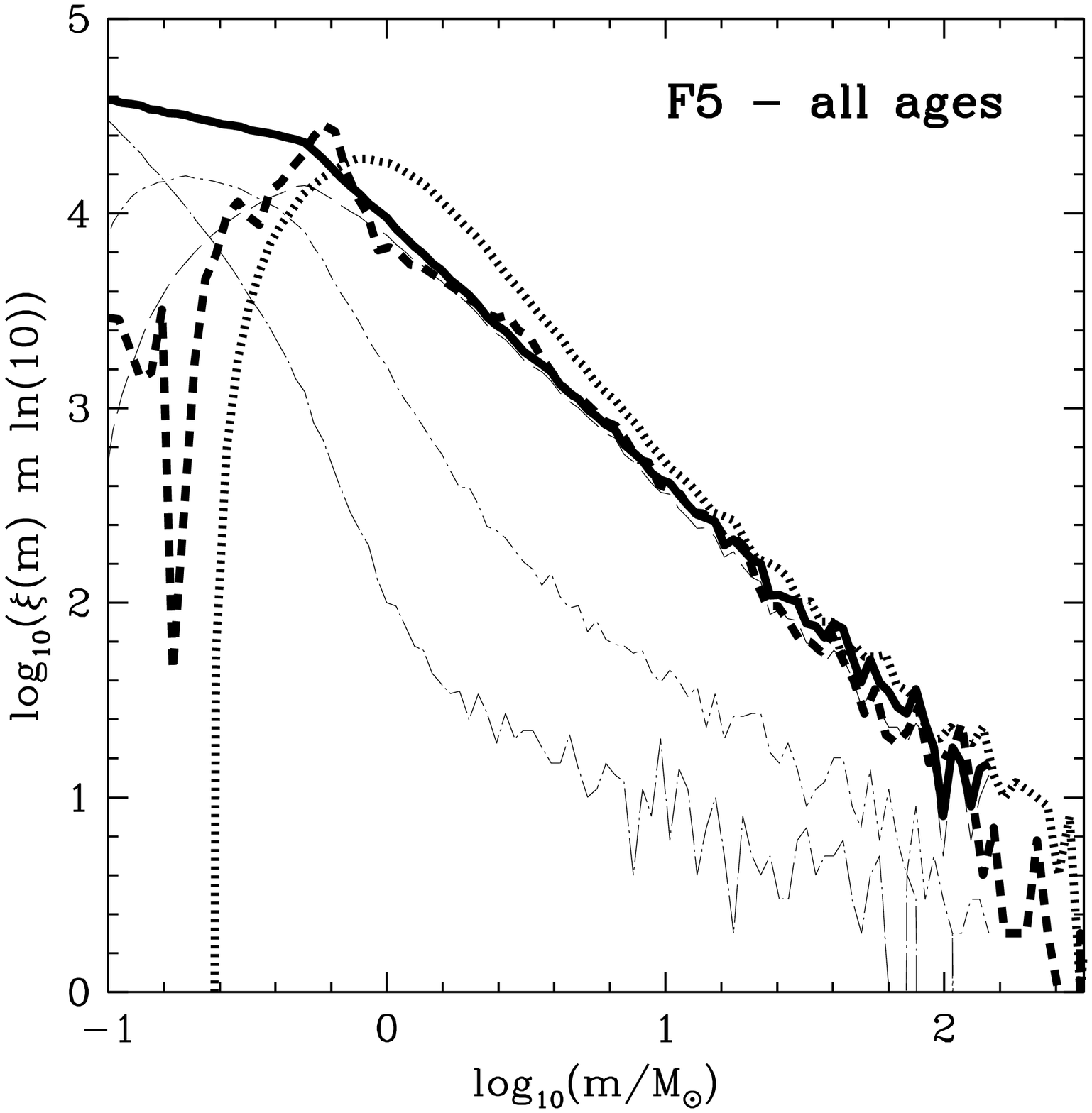}
\includegraphics[width=8cm]{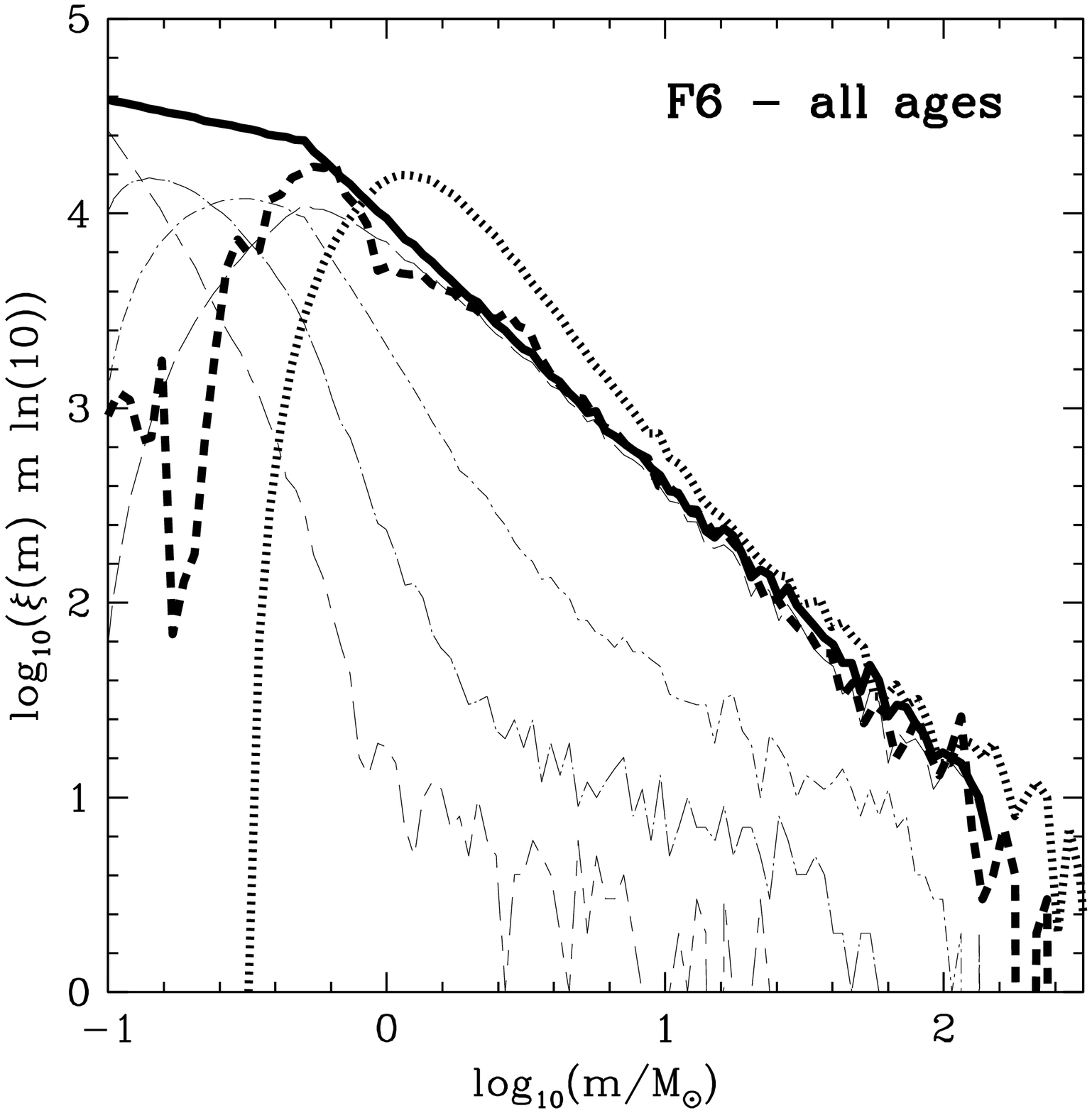}
\vspace*{-2.0cm}
\caption{Like Fig.~\ref{fig:Fit1to4} but for the Models Fit 4 to Fit
  6 from Table~\ref{tab:evo}.}
\label{fig:Fit4}
\end{center}
\end{figure*}

\begin{figure*}
\begin{center}
\includegraphics[width=8cm]{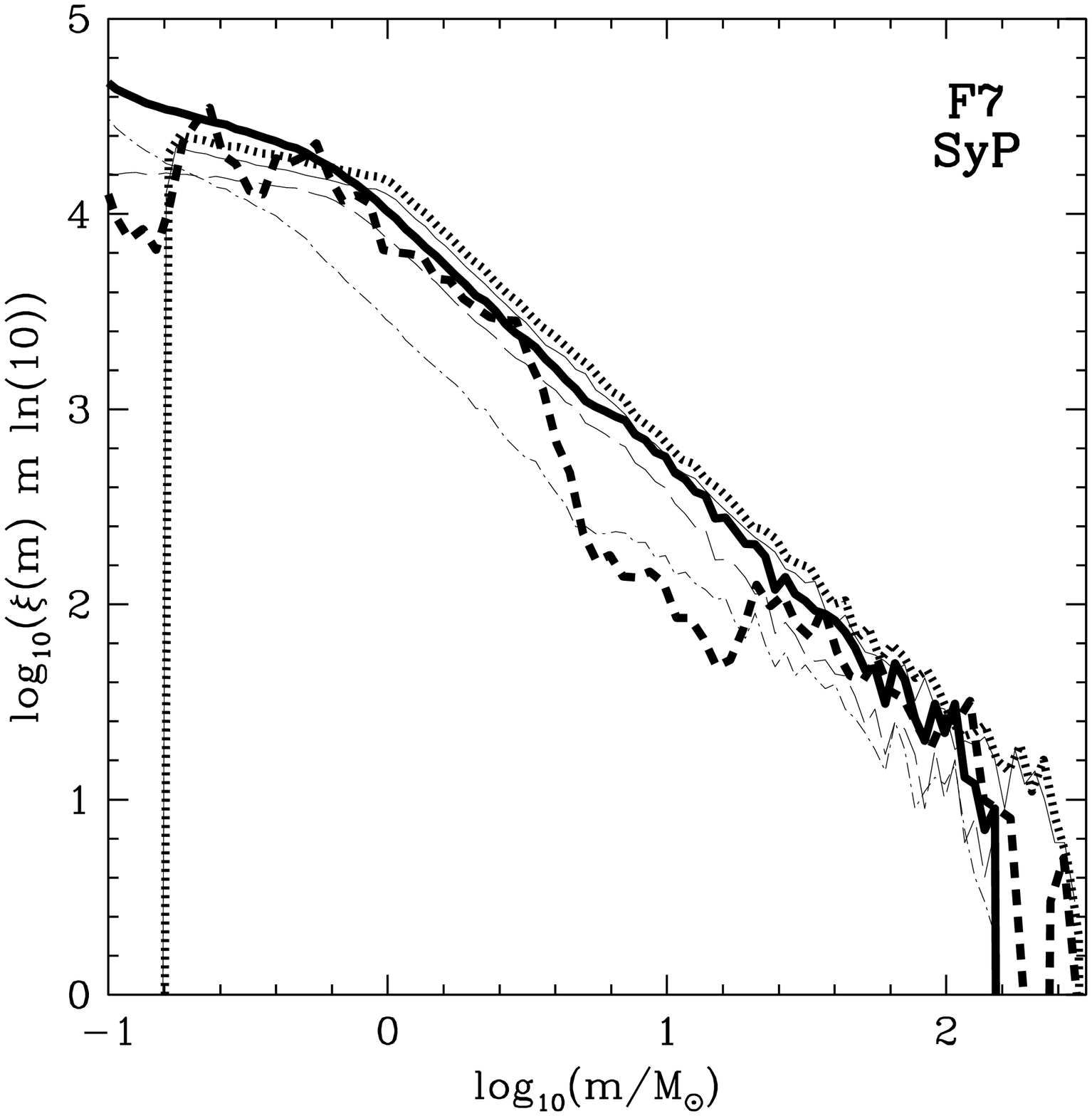}
\includegraphics[width=8cm]{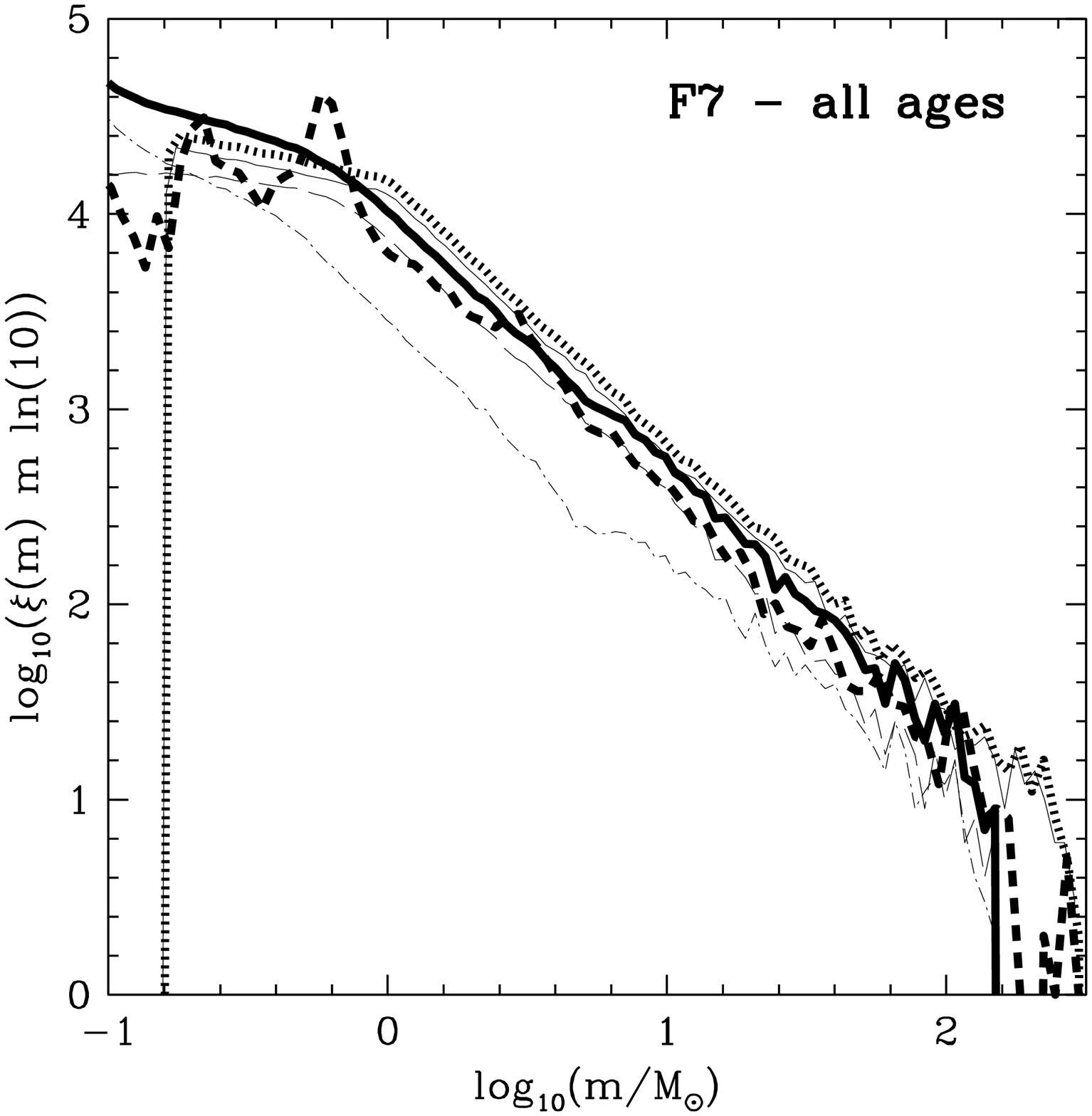}

\vspace*{-2.5cm}

\includegraphics[width=8cm]{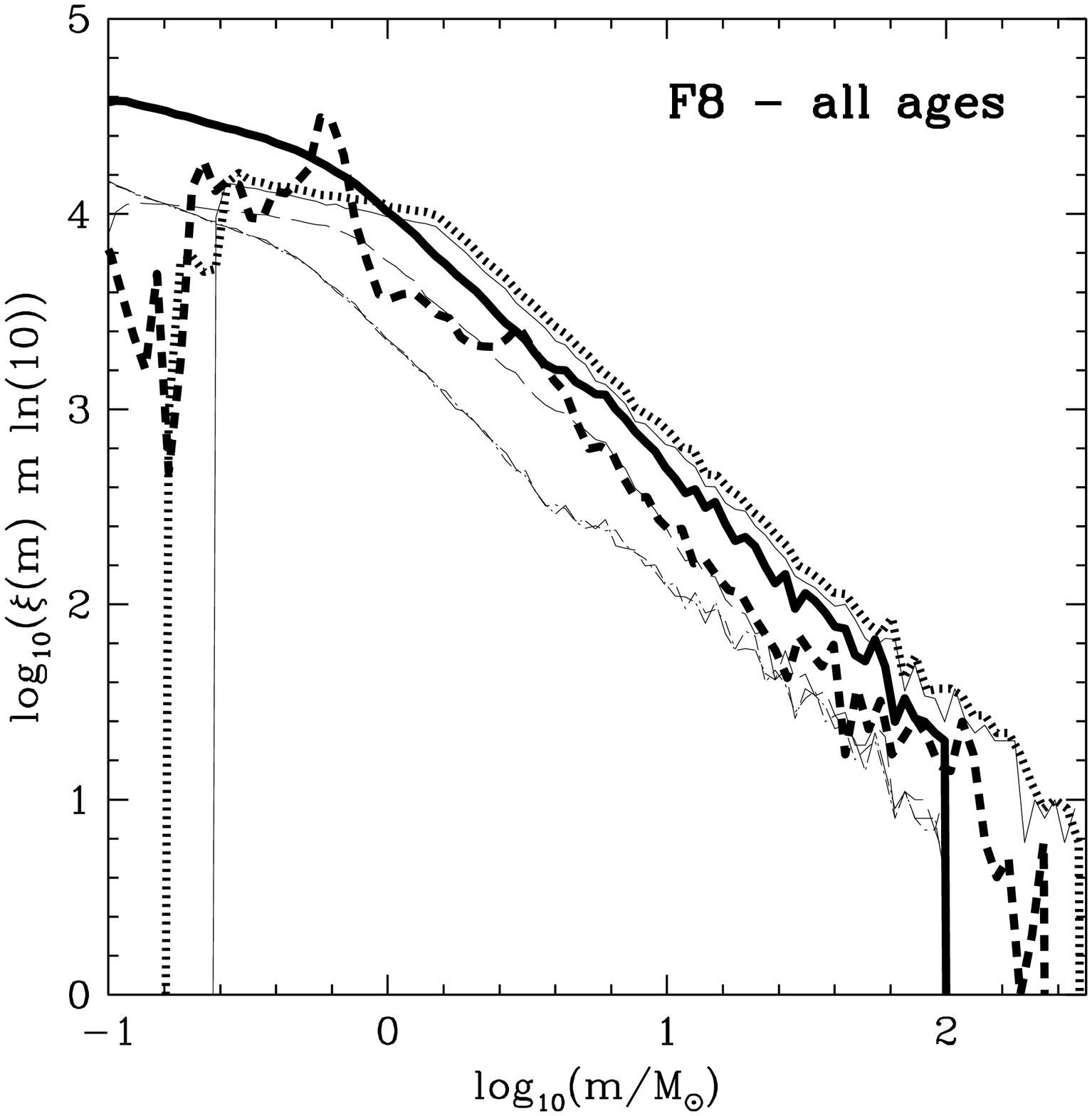}
\includegraphics[width=8cm]{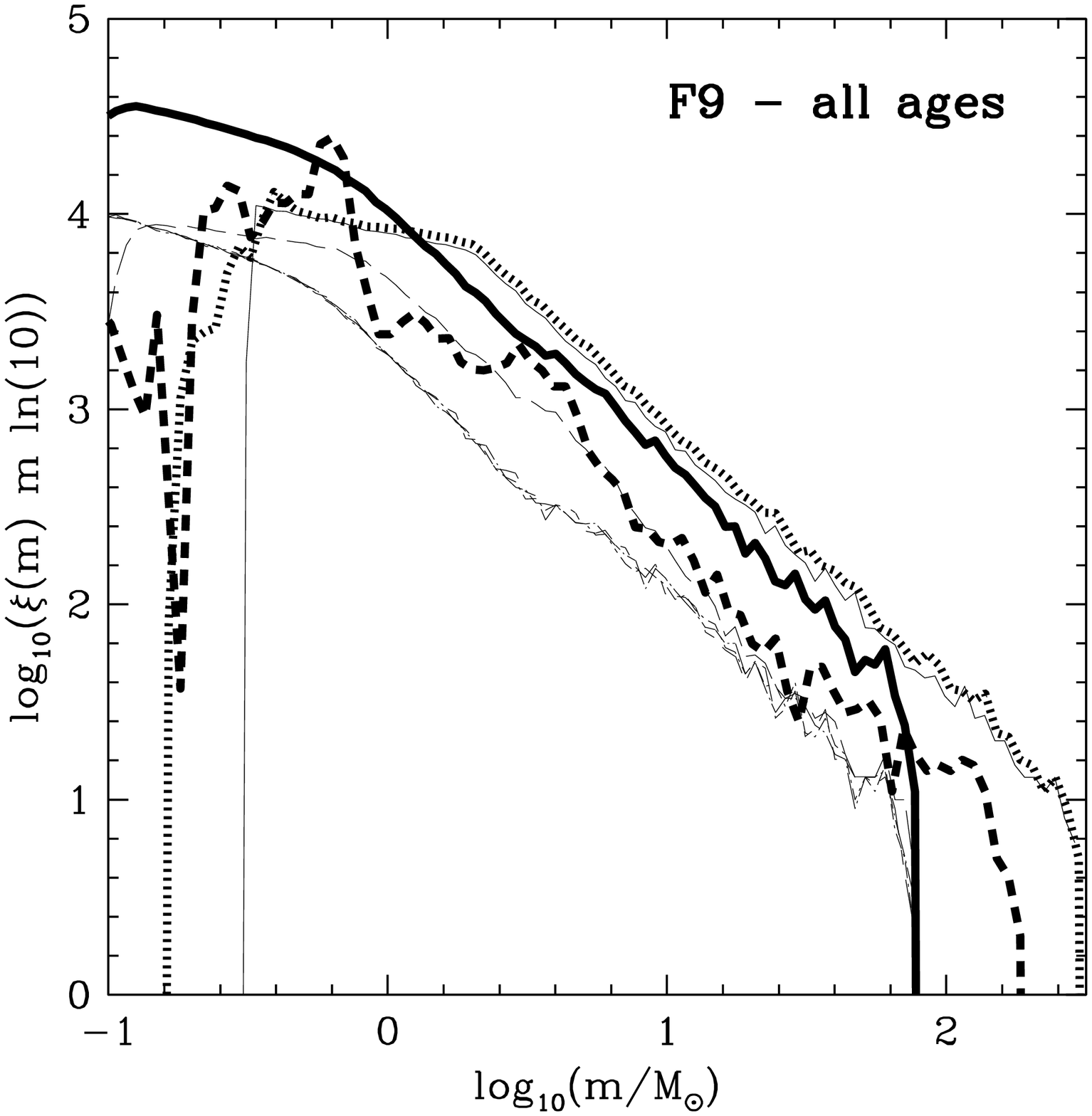}
\vspace*{-2.0cm}
\caption{Like Fig.~\ref{fig:Fit1to4} but for the Models Fit 7 to Fit
  9 from Table~\ref{tab:evo}.}
\label{fig:Fit3to4}
\end{center}
\end{figure*}

\begin{figure*}
\begin{center}
\includegraphics[width=8cm]{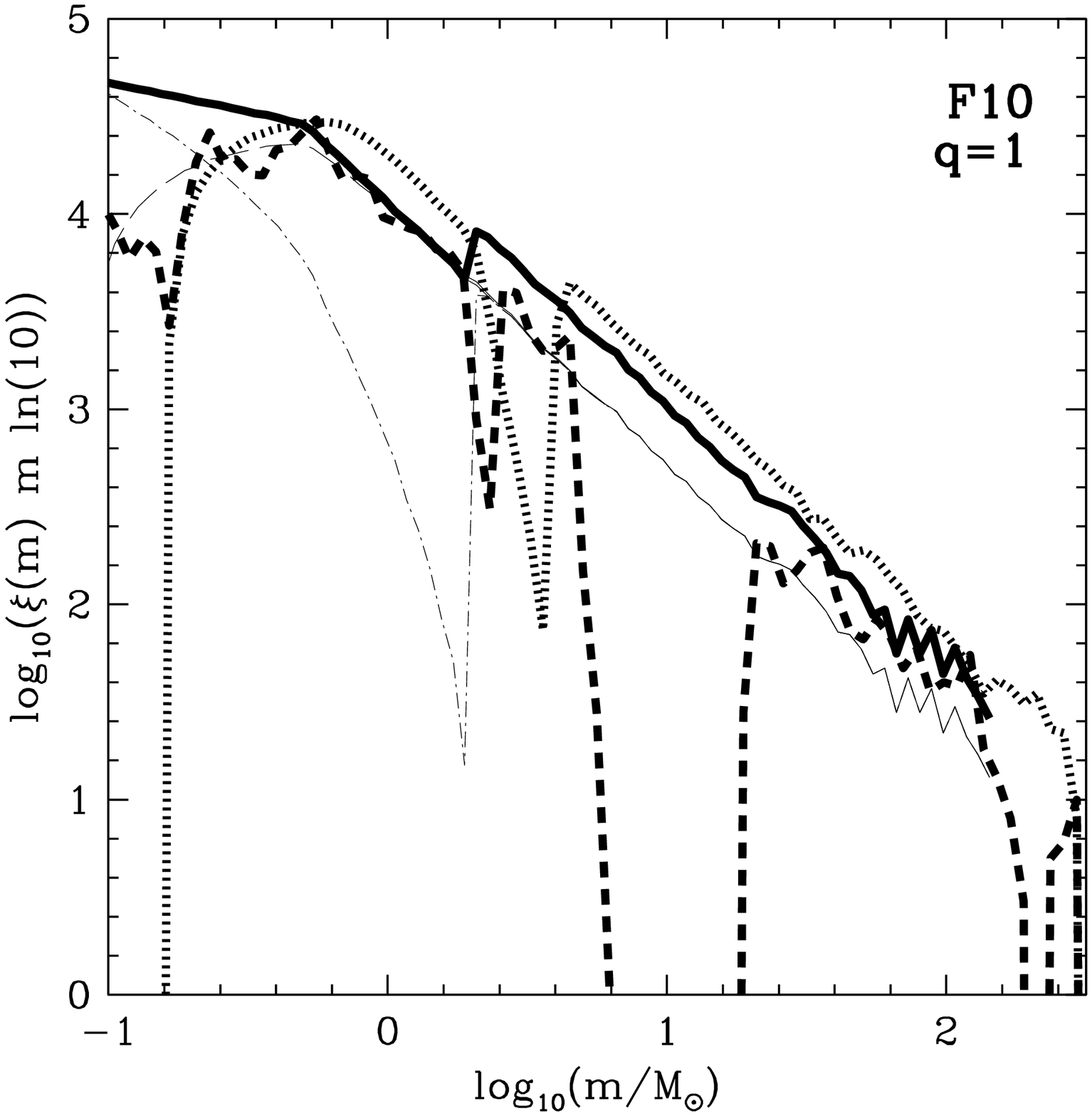}
\includegraphics[width=8cm]{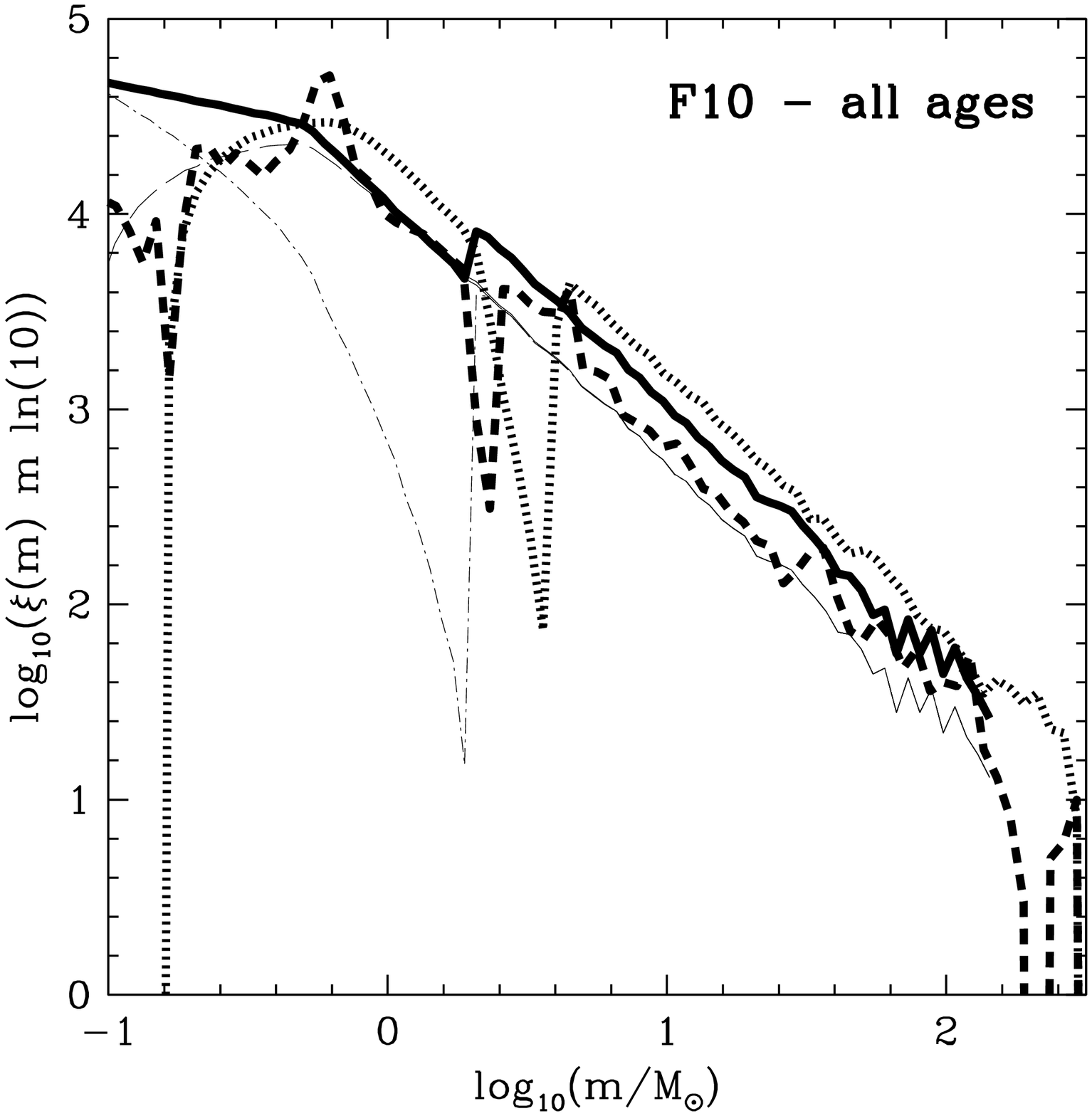}
\vspace*{-2.0cm}
\caption{Like Fig.~\ref{fig:Fit1to4} but for the Model Fit 10 from
  Table~\ref{tab:evo}.}
\label{fig:Fit10}
\end{center}
\end{figure*}

\end{appendix}

\bsp
\label{lastpage}
\end{document}